\documentclass[review,12pt]{elsarticle}

\usepackage{mathptmx}       
\usepackage{helvet}         
\usepackage{courier}        
\usepackage{type1cm}        
\usepackage{makeidx}         
\usepackage{graphicx}        
\usepackage{multicol}        
\usepackage[bottom]{footmisc}
\usepackage{amsmath,amssymb,amscd}
\usepackage{amsfonts}
\usepackage{enumitem}
\usepackage{setspace}
\usepackage{adjustbox}
\usepackage{tikz}
\usepackage{algorithm}
\usepackage{algpseudocode}

\usepackage[
colorlinks=true,
citecolor=blue,
urlcolor=blue,
linktocpage,
pagebackref
]{hyperref}

\usepackage{comment}
\usepackage{pgfplots}

\usepackage{color}
\usepackage{txfonts}
\usepackage{fontawesome}

\def\bx{{\bf x}}   
\def\ubx{\underline{\bx}}
\def\lim{{\xi}}    
\def\dd{\mathrm d}

\def\br{\mathbf{r}}   

\def\bq{{\mathbf{q}}}

\def\um{\underline{m}}
\def\ue{\underline{e}}

\def\dd{\mathrm d}
\def\ov{\bar{v}}

\def\uv{\underline{v}}
\def\ov{\bar{v}}
\def\uu{\underline{u}}
\def\ou{\bar{u}}

\def\fpt{\emph{findpts}}

\def\mm{\mathcal{M}}
\def\bigO{\mathcal{O}}
\definecolor{code}{rgb}{0.7, 0, 0.4}
\newcommand{\code}[1]{\texttt{\small\color{code} #1}}

 \journal{Journal}

\begin{document}

\begin{frontmatter}



\title{{\fontsize{15.9pt}{16pt}\selectfont General Field Evaluation in High-Order Meshes on GPUs}}

\author{Ketan Mittal \corref{cor1}\fnref{llnl}}
\author{Aditya Parik \fnref{usu}}
\author{Som Dutta \fnref{usu}}
\author{\\Paul Fischer \fnref{anl}}
\author{Tzanio Kolev \fnref{llnl}}
\author{James Lottes \fnref{google}}
\fntext[llnl]
{Lawrence Livermore National Laboratory, 7000 East Avenue, Livermore, CA 94550}
\fntext[usu]
{Utah State University, Logan, UT 84322}
\fntext[anl]
{University of Illinois at Urbana-Champaign, Urbana, IL 61801, and Argonne National Laboratory, Lemont, IL 64039}
\fntext[google]
{Google Research, Mountain View, CA, 94043}
\cortext[cor1]
{Corresponding author, mittal3@llnl.gov}

\address{}

\begin{abstract}
Robust and scalable function evaluation at any arbitrary point in the finite/spectral element mesh is required for querying the partial differential equation solution at points of interest, comparison of solution between different meshes, and Lagrangian particle tracking.
This is a challenging problem, particularly for high-order unstructured meshes partitioned in parallel with MPI, as it requires identifying the element that overlaps a given point and computing the corresponding reference space coordinates.
We present a robust and efficient technique for general field evaluation in large-scale high-order meshes with quadrilaterals and hexahedra. In the proposed method, a combination of globally partitioned and processor-local maps are used to first determine a list of candidate MPI ranks, and then locally candidate elements that could contain a given point. Next, element-wise bounding boxes further reduce the list of candidate elements. Finally, Newton's method with trust region is used to determine the overlapping element and corresponding reference space coordinates.
Since GPU-based architectures have become popular for accelerating computational analyses using meshes with tensor-product elements,
specialized kernels have been developed to utilize the proposed methodology on GPUs.
The method is also extended to enable general field evaluation on surface meshes.
The paper concludes by demonstrating the use of proposed method in various applications ranging from mesh-to-mesh transfer during r-adaptivity to Lagrangian particle tracking.
\end{abstract}

\begin{keyword}
Off-grid interpolation \sep High-Order meshes \sep Unstructured meshes \sep FEM \sep SEM
\end{keyword}
\end{frontmatter}


\section{Introduction}
\label{sec_intro}

The finite element method (FEM) and spectral element method (SEM) are powerful techniques for modeling complex physical systems described using partial differential equations (PDEs). FEM and SEM are particularly effective because
they can discretize curvilinear domains using high-order unstructured meshes.
These meshes are a union of smaller elements, each of which maps to a canonical unit-sized element in the reference space through a nonlinear transformation. Within each element, the PDE solution is then represented using a combination of coefficients and high-order bases defined on the reference element.

A well known challenge resulting from the FEM formulation is that evaluation of the solution at arbitrary physical locations in the mesh is not trivial as it requires knowledge of the element that overlaps the point and the corresponding reference space coordinates.
Efficiently finding the overlapping element is challenging as the mesh elements are  distributed across many MPI ranks in high performance computing (HPC).
Once the overlapping element is known, its map must be inverted to determine the reference space coordinates corresponding to the given point, which is computationally expensive due to the nonlinear transformation.
Thus, general function evaluation, which is critical for many applications such as Lagrangian particle tracking, overlapping-mesh based approaches, and comparison of solution between different meshes, can become a bottleneck for large scale problems when it is not done efficiently.

Most methods enable general field evaluation in three steps. First, for a given point in physical space, the element overlapping it is determined. Since this element is not usually located on the same MPI rank querying the point, some MPI communication is required.
A workaround for this problem in the context of mesh-to-mesh remap is a rendezvous algorithm to re-partition copies of the meshes such that elements in same physical region are located on same MPI ranks \cite{plimpton2004parallel,herring2021portage,ray2023efficient,slattery2013data}. While the rendezvous algorithm incurs storage overhead due to use of mesh copies, it makes the overlapping element search a processor-local problem. This search can be done for example by subdividing the local subdomain into a grid of Cartesian-aligned bins and determining element-bin intersection that can then be inversely used to determine given physical point to potential element mapping \cite{plimpton2004parallel}.

With the overlapping element determined, the reference space coordinates corresponding to that point must be computed. This requires inversion of the nonlinear transformation of that element from the reference space to physical space. Inverting the entire map for a high-order element is not trivial, so it is usually inverted locally for a given point using an optimization-based approach \cite{mittal2019nonconforming,Roca2018}. Note that applications focusing on mesh-to-mesh remap often determine intersection of overlapping elements rather than just finding the nodes of one mesh in another, to effect a conservative remap \cite{lipnikov2023conservative}. While conservative remap is desirable, determining exact intersection is virtually infeasible for curvilinear elements. The methodology proposed herein targets the more general case of point-wise interpolation that is directly leveraged in several applications. Some methods find $K$-nearest mesh nodes to the given point and then use a weighted average of the nodal values to interpolate the solution at the given point \cite{lacroix2024comparative,chandar2019overset}, but this approach cannot be used in FEM and SEM to effect spectrally accurate interpolation.
The final step of interpolating the solution is straightforward as it requires evaluation of the bases at the reference space coordinates determined in the previous step \cite{mittal2019nonconforming}.

This work describes an approach for general function evaluation at any given set of points on high-order meshes on GPUs.
The novel contributions of the current work are (i) an intra-processor map to quickly determine candidate MPI ranks that contain the element overlapping a given point ($\bx^*$), (ii) a processor-local map to determine candidate elements on that rank that overlap $\bx^*$, (iii) element-wise axis-aligned and oriented bounding boxes to determine if an element could possibly overlap a point, (iv) a Newton's method with trust region-based approach to invert the nonlinear map for an element and determine the corresponding reference space coordinates, (v) extension of this method to surface meshes, and (vi) specialized kernels to effect compute-dominated functions on GPUs.

This work builds upon our prior work on the \fpt\ library \cite{gslib-github} developed for high-order SEM. The \fpt\ library was originally authored by James Lottes to evaluate continuous SEM solutions at any given set of points on deformed 2D quadrilaterals (quads) and 3D hexahedra (hexes) on CPU-based architectures. Over the years, \fpt\ has been rigorously tested at scale through Nek5000/NekRS, an SEM-based incompressible flow solver, for applications related to Lagrangian particle tracking \cite{dutta2016large,zwick2020scalable,fabregat2021direct}, immersed boundary method \cite{yang2021scalable}, fusion and fission systems \cite{min2024exascale}, and flow visualization \cite{dutta2018visualization}.
It was eventually extended for overlapping-grid based techniques where a point can overlap multiple meshes, and the challenge then is to interpolate the solution from the mesh that has the lowest error at that point \cite{mittal2019nonconforming,mittal2020}.
More recently, \fpt\ was integrated in the general high-order FEM library MFEM \cite{mfem-github,MFEM2024} where it is used for multi-physics solution in the entire de Rham complex, even for meshes with simplices.

With advancements in High-Performance Computing (HPC), many \fpt\-based applications are leveraging GPUs to accelerate the solution of PDEs. This has necessitated the development of specialized GPU kernels for \fpt\ to prevent performance bottlenecks in simulation workflows.
This manuscript thus aims to document both the original library and its recent GPU implementation for line-, quads-, and hex-meshes.
The open-source CPU implementation of \fpt\ is available in the \emph{gslib} repository on GitHub \cite{gslib-github}, while GPU-accelerated kernels for hex meshes are accessible through NekRS \cite{lindquist2021scalable,fischer2022nekrs,nekrs-github}, the GPU-optimized version of Nek5000, and for quad-, hex-, and surface meshes through MFEM \cite{MFEM2024}.

The remainder of the paper is organized as follows. Section \ref{sec_prelim} introduces the notation and concepts necessary to understand the contribution of the current work. Section \ref{sec_method} describes the methodology for general field evaluation in 2D quadrilateral and 3D hexahedra meshes, followed by its extension to surface meshes in Section \ref{sec_method_surf}. The GPU implementation of \fpt\ is described in Section \ref{sec_gpu}, and its impact on reducing time for evaluating a field at a given set of point in comparison to a purely CPU-based approach is demonstrated in Section \ref{sec_results}. Several applications such as Lagrangian particle tracking and $r$-adaptivity that are powered by \fpt\ are also shown in Section \ref{sec_results}. Finally, Section \ref{sec_conclusion} discusses potential avenues for future work.

\section{Preliminaries} \label{sec_prelim}

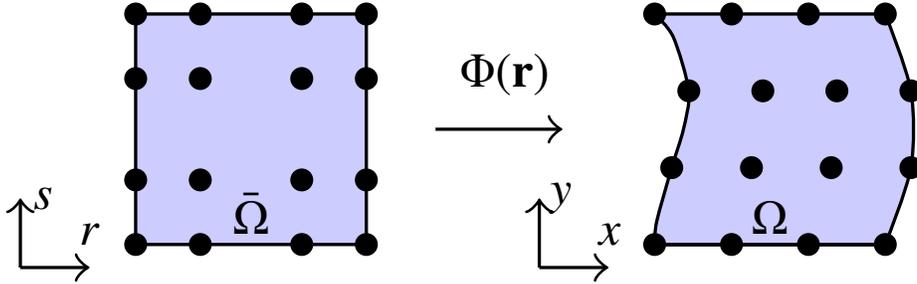
\begin{figure}[tb!]
  \centering
  \adjustbox{width=0.9\linewidth,valign=b}{\begin{tikzpicture}[scale=1]
\coordinate (0) at (0,0);
\coordinate (1) at (0.3333333333, 0);
\coordinate (2) at (0.66666666667,0);
\coordinate (3) at (1.0,0);

\coordinate (4) at (0.074074,0.33333333333);
\coordinate (5) at (0.4198, 0.33333333333);
\coordinate (6) at (0.7654,0.33333333333);
\coordinate (7) at (1.111,0.33333333333);

\coordinate (8) at (0.1481,0.66666666667);
\coordinate (9) at (0.4691, 0.66666666667);
\coordinate (10) at (0.7901,0.66666666667);
\coordinate (11) at (1.111,0.66666666667);

\coordinate (12) at (0,1);
\coordinate (13) at (0.33333333333,1);
\coordinate (14) at (0.66666666667,1);
\coordinate (15) at (1.0,1);

\coordinate (32) at (0.009,0.1);
\coordinate (33) at (0.125,0.5);
\coordinate (34) at (0.081,0.9);

\draw [blue!20!white, fill=blue!20!white] plot [smooth] coordinates {(0) (1) (2) (3)} -- (6) -- cycle;
\draw [blue!20!white, fill=blue!20!white] plot [smooth] coordinates {(3) (7) (11) (15)} -- (6) -- cycle;
\draw [blue!20!white, fill=blue!20!white] plot [smooth] coordinates {(15) (14) (13) (12)} -- (6) -- cycle;
\draw [blue!20!white, fill=blue!20!white] plot [smooth] coordinates {(12) (34) (8) (33) (4) (32) (0)} -- (6) -- cycle;
\draw [blue!20!white, fill=blue!20!white] plot [smooth] coordinates {(6) (10) (9)} -- (5) -- cycle;

\draw[black] plot [smooth cycle] coordinates {(0) (1) (2) (3)} ;
\draw[black] plot [smooth] coordinates {(15) (14) (13) (12)} ;
\draw[black] plot [smooth] coordinates {(12) (34) (8) (33) (4) (32) (0)} ;

\coordinate (16) at (-2-0.25,0);
\coordinate (17) at (-1.72-0.25,0);
\coordinate (18) at (-1.28-0.25,0);
\coordinate (19) at (-1-0.25,0);

\coordinate (20) at (-2-0.25,0.28);
\coordinate (21) at (-1.72-0.25,0.28);
\coordinate (22) at (-1.28-0.25,0.28);
\coordinate (23) at (-1-0.25,0.28);

\coordinate (24) at (-2-0.25,0.72);
\coordinate (25) at (-1.72-0.25,0.72);
\coordinate (26) at (-1.28-0.25,0.72);
\coordinate (27) at (-1-0.25,0.72);

\coordinate (28) at (-2-0.25,1);
\coordinate (29) at (-1.72-0.25,1);
\coordinate (30) at (-1.28-0.25,1);
\coordinate (31) at (-1-0.25,1);

\draw [blue!20!white, fill=blue!20!white] plot [smooth] coordinates {(16) (19)} -- (22) -- cycle;
\draw [blue!20!white, fill=blue!20!white] plot [smooth] coordinates {(19) (31)} -- (22) -- cycle;
\draw [blue!20!white, fill=blue!20!white] plot [smooth] coordinates {(28) (31)} -- (22) -- cycle;
\draw [blue!20!white, fill=blue!20!white] plot [smooth] coordinates {(16) (28)} -- (22) -- cycle;

\draw[black] plot [smooth] coordinates {(3) (7) (11) (15)};

\draw[black] plot [smooth] coordinates {(16) (19)};
\draw[black] plot [smooth] coordinates {(19) (31)};
\draw[black] plot [smooth] coordinates {(31) (28)};
\draw[black] plot [smooth] coordinates {(16) (28)};

\foreach \i in {0,...,31}
{
    \draw (\i) node[circle, fill=black, inner sep=1pt] {} ;
}

\draw[->] (-0.5,-0.1)--(-0.2,-0.1) node[above]{};
\draw[] (-0.2,-0.15) node[above]{\tiny{$x$}};
\draw[->] (-0.5,-0.1)--(-0.5,0.2) node[right]{\hspace{-1mm}\tiny{$y$}};

\draw[->] (-0.7-0.25,0.5)--(-0.4,0.5) node[above]{\hspace{-5mm}\tiny{$\Phi(\mathbf{r})$}};

\draw[->] (-2.5-0.25,-0.1)--(-2.2-0.25,-0.1) node[above]{};
\draw[] (-2.2-0.25,-0.15) node[above]{\tiny{$r$}};
\draw[->] (-2.5-0.25,-0.1)--(-2.5-0.25,0.2) node[right]{\hspace{-1mm}\tiny{$s$}};


\draw[] (-1.75,-0.1) node[above]{\tiny{$\bar{\Omega}$}};
\draw[] (0.5,-0.1) node[above]{\tiny{${\Omega}$}};

\end{tikzpicture}}
  \vspace{-3mm}
  \caption{Schematic showing a finite element ($p=3$) mapped from the reference space to physical space.}
  \label{fig_2d_mapped_element}
  \vspace{-3mm}
\end{figure}

In the finite element framework \cite{MFEM2021,MFEM2024}, the domain $\Omega \subset \mathbb{R}^d$, $d=\{2,3\}$, is
discretized as a union of $N_E$ curved mesh elements, $\Omega^e$, $e=1\dots N_E$, each of order $p$. To obtain a discrete representation of these elements, we select a set of scalar basis
functions $\{ \phi_i \}_{i=1}^{N^d}$, $N=p+1$, on the reference element $\bar{\Omega}^e \in [-1,1]^d$.
These bases are typically chosen to be Lagrange
interpolation polynomials at the Gauss-Lobatto nodes, $\{\mathbf{\zeta}_j \}_{j=1}^{N^d}$, of the reference element such that $\phi_i(\mathbf{\zeta}_j) = \delta_{ij}$.
The position of an element $\Omega^e$ in the mesh $\mathcal{M}$ is then
fully described by a matrix $\bx^e$, of size $d \times N^d$,
which contains all nodal coordinates.
Given $\bx^e$ and the choice of Lagrange interpolants as the bases, we introduce the map $\Phi_e:\bar{\Omega}^e \to \mathbb{R}^d$ whose image is the geometry of the physical element $\Omega^e$:
\begin{equation}
\label{eq_x}
\bx(\br) \vert_{\Omega^e} =
   \Phi_e(\br) \equiv
   \sum_{i=1}^{N^d} \mathbf{x}^e_{i} \phi_i(\br),
   \qquad \br \in \bar{\Omega}^e, ~~ \bx=\bx(\br) \in \Omega^e,
\end{equation}
where $\mathbf{x}^e_{i}$ denotes the coordinates of the $i$-th node of $\Omega^e$. Hereafter, $\bx(\br)$ will denote the element-wise position function defined by \eqref{eq_x} with $\Omega^e$ omitted for brevity.
Figure \ref{fig_2d_mapped_element} shows an example of a cubic element ($d=2,\,p=3,\,N=4$) mapped from the reference space to physical space.

For tensor-product elements (quadrilaterals and hexahedra), basis function evaluation at a given reference point $\br=\{r,s,t\}$ is simplified using a product of 1D bases in each direction
\begin{equation}
\label{eq_x_tensor}
  \bx(\br) \vert_{\Omega^e} = \sum_{i=1}^{N} \sum_{j=1}^{N} \sum_{k=1}^{N} {\bx}^e_{ijk} \phi_i(r) \phi_j(s) \phi_k(t).
\end{equation}
This reduces the computational cost and memory footprint for evaluating $\bx(\br)$ \cite{deville2002high}, among other advantages, and is well suited for GPU-based architectures \cite{MFEM2021}.

Similar to \eqref{eq_x_tensor}, any discrete solution (e.g., velocity or temperature) $u(\bx)$ is defined on $\Omega^e$ as:
\begin{equation}
  \label{eq_u_tensor}
  u(\bx(\br)) = u(\br)\vert_{\Omega^e} = \sum_{i=1}^{\tilde{N}} \sum_{j=1}^{\tilde{N}} \sum_{k=1}^{\tilde{N}} {u}^e_{ijk} \phi_i(r) \phi_j(s) \phi_k(t),
\end{equation}
where $\tilde{N}$ depends on the polynomial order used to represent the solution, and is often not the same as that used for mesh nodal coordinates \eqref{eq_x_tensor}.

Note that in the case of an area/volume mesh, the reference space is the same dimension as the physical space, i.e. $\Omega \subset \mathbb{R}^d$ and $\bar{\Omega}^e \in [-1,1]^d$. For manifolds, the reference space dimension ($d_r$) is lower than that of the physical space, i.e. $\bar{\Omega}^e \in [-1,1]^{d_r}$, $d_r < d$. For example, the nodal position function for a curved line element embedded in 2D space is
\begin{equation}
\label{eq_x_tensor_surf}
  x(r) \vert_{\Omega^e} = \sum_{i=1}^{N} x^e_{i} \phi_i(r), \qquad
  y(r) \vert_{\Omega^e} = \sum_{i=1}^{N} y^e_{i} \phi_i(r).
\end{equation}
Similarly, a curved quadrilateral element embedded in 3D space $\bx \subset \mathbb{R}^3$ is mapped from a 2D reference element $\bar{\Omega}^e \in [-1,1]^2$.

In the HPC setting, a mesh with $N_E$ elements is partitioned across $N_P$ processors such that each MPI rank $m=1\dots N_P$ is mainly responsible for storage and computation related to the $N_{m,E} \approx N_E/N_P$ elements it owns. This mesh partitioning allows distribution of the work for solving a PDE. Figure \ref{fig_2d_meshpartition2}(a) shows a simple example of a 2D mesh partitioned onto 4 processors.

\section{Methodology for General Field Evaluation in Volume/Area Meshes} \label{sec_method}

This section first describes the \emph{Setup} step of the proposed framework (Section \ref{sec_method_precomp}) that computes the data structures to map a given point to first, the candidate MPI ranks $\um^*$, and then the candidate elements $\ue^*$ on those ranks, that could overlap the point.
Next, the \emph{Find} step determines which of the candidate elements actually contains the point $\bx^*$, and inverts its map to determine the corresponding reference space coordinates $\br^*$. The \emph{Find} step is described in Section \ref{sec_method_find}.
Finally, the function evaluation step (\emph{Interpolate}) is discussed in Section \ref{sec_method_eval}.

\subsection{Computing data structures to map a point $\bx^*$ to MPI ranks $\um^*$ and elements $\ue^*$} \label{sec_method_precomp}

To compute the maps from $\bx^*$ to $\um^*$ and $\ue^*$, the global and processor-local domain are first partitioned into Cartesian-aligned meshes.
Next, we identify the elements of the processor-local Cartesian-aligned mesh $\mm_L$ that intersect with each of the $N_{m,E}$ elements on rank $m$. This information is stored in a process-local map and maps $\bx^*$ to $\ue^*$ in $\bigO(1)$ time (Section \ref{sec_method_precomp_local_map}).
Since computation of high-order element intersections with $\mm_L$ is not trivial, \emph{linear} axis-aligned bounding boxes (AABBs) for each element of $\mm$ are used. This AABB computation is described in Section \ref{sec_method_precomp_aabb}.
Similar to the local map, we identify the elements of the global Cartesian-aligned mesh $\mm_G$ that intersect with each of the $N_{m,E}$ elements. The global map is partitioned across the $N_P$ ranks, and maps $\bx^*$ to candidate ranks $\um^*$ (Section \ref{sec_method_precomp_global_map}).
In addition to AABBs, oriented bounding boxes (OBB) for each element are computed as they provide relatively tighter bounds around each element  (Section \ref{sec_method_precomp_obb}). AABBs and OBBs together are used to eliminate elements from $\ue^*$ that cannot contain $\bx^*$.

\begin{figure}[bt!]
  \begin{center}
  $\begin{array}{ccc}
  \includegraphics[width=0.3\linewidth]{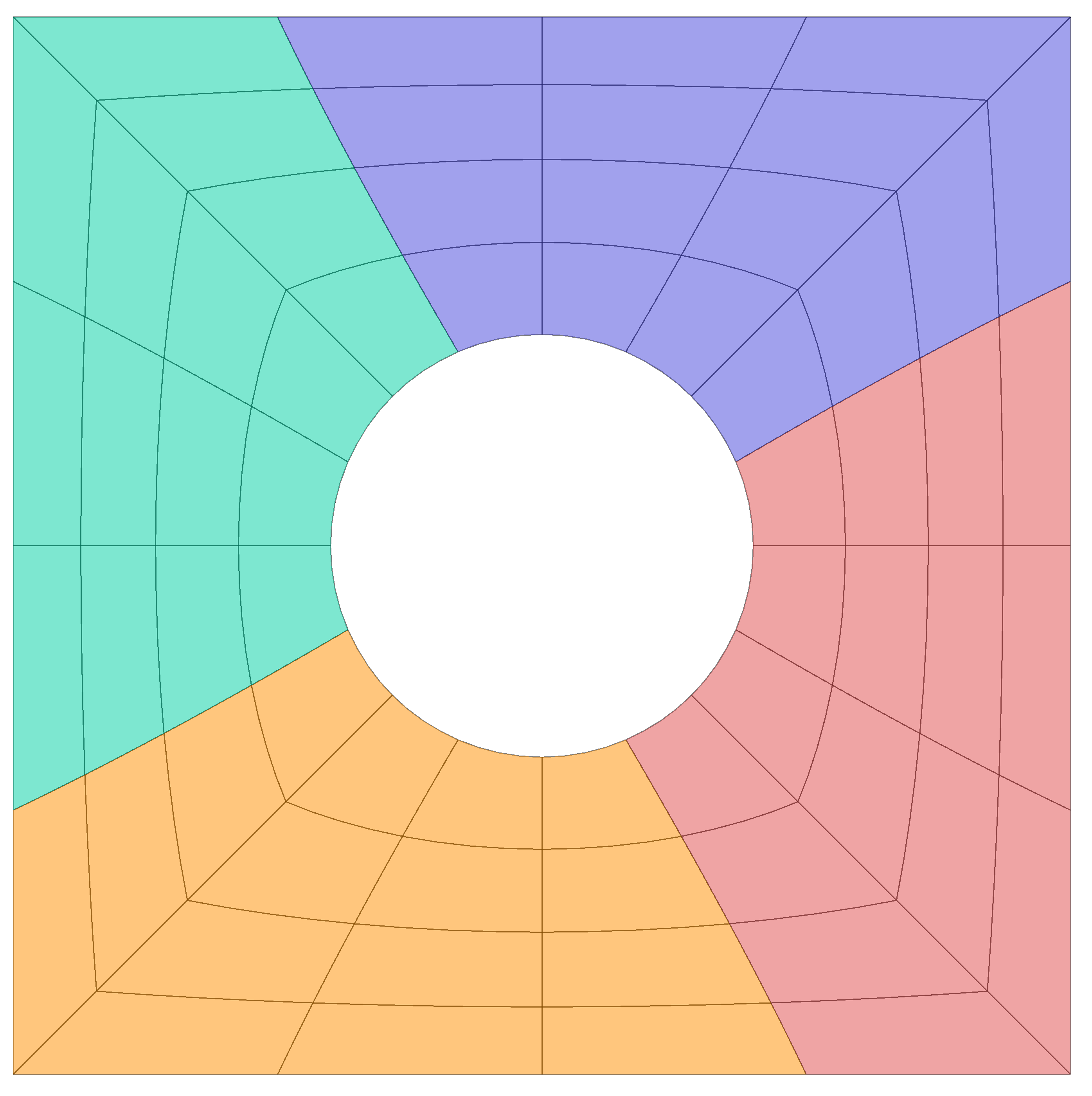} &
  \includegraphics[width=0.3\linewidth]{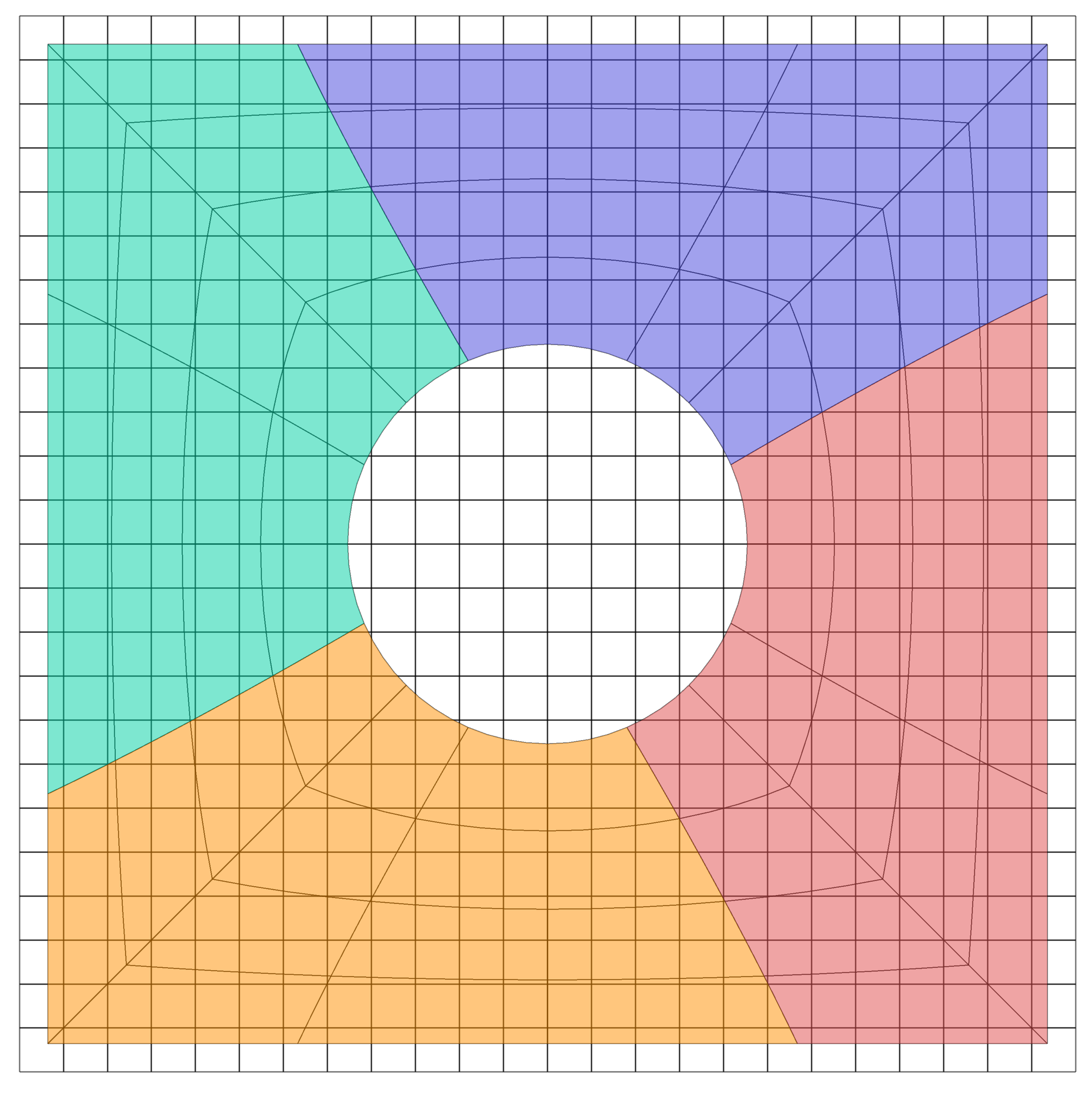} &
  \includegraphics[width=0.3\linewidth]{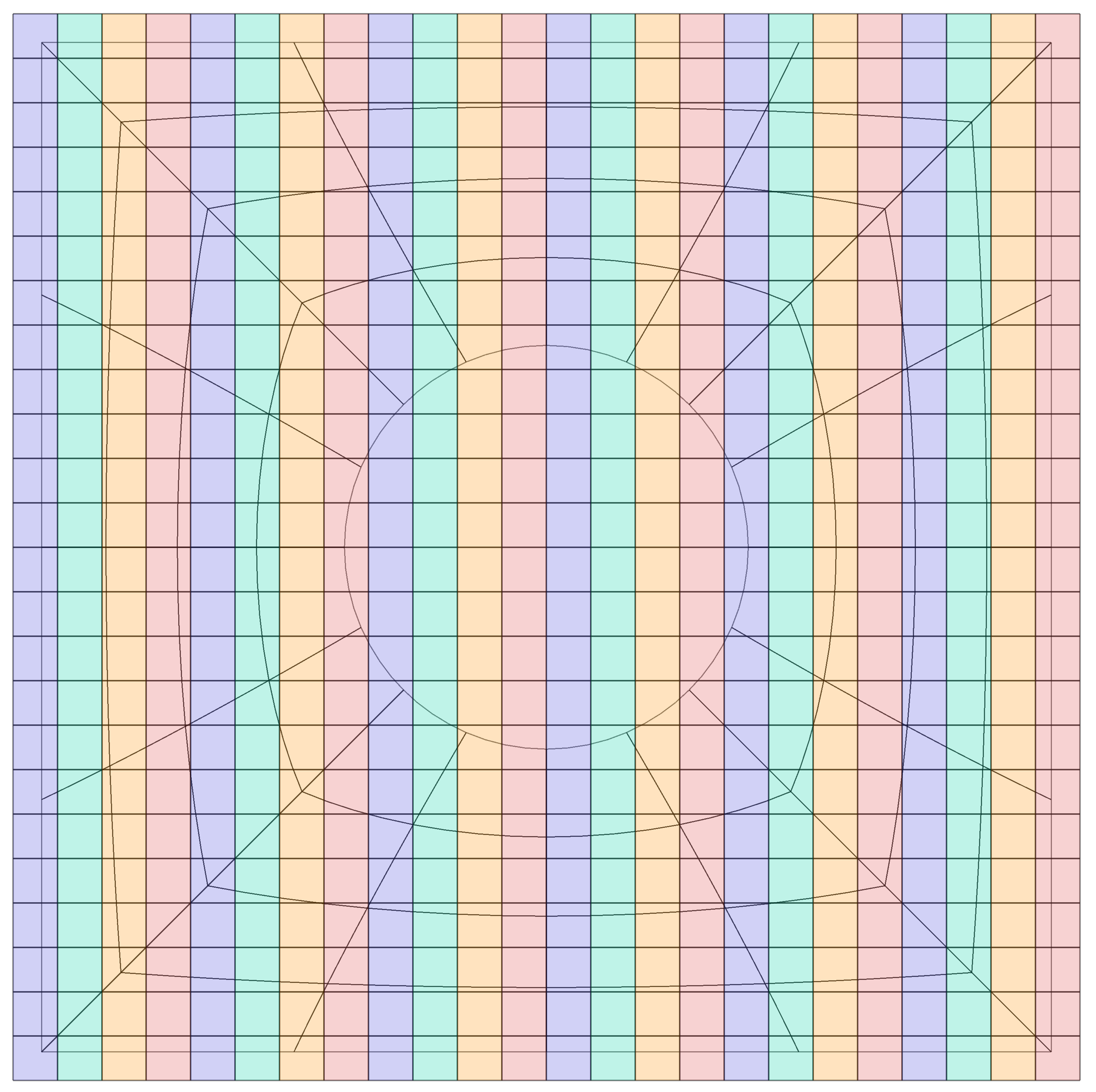} \\
  \textrm{(a)} & \textrm{(b)} & \textrm{(c)} \\
  \includegraphics[width=0.3\linewidth]{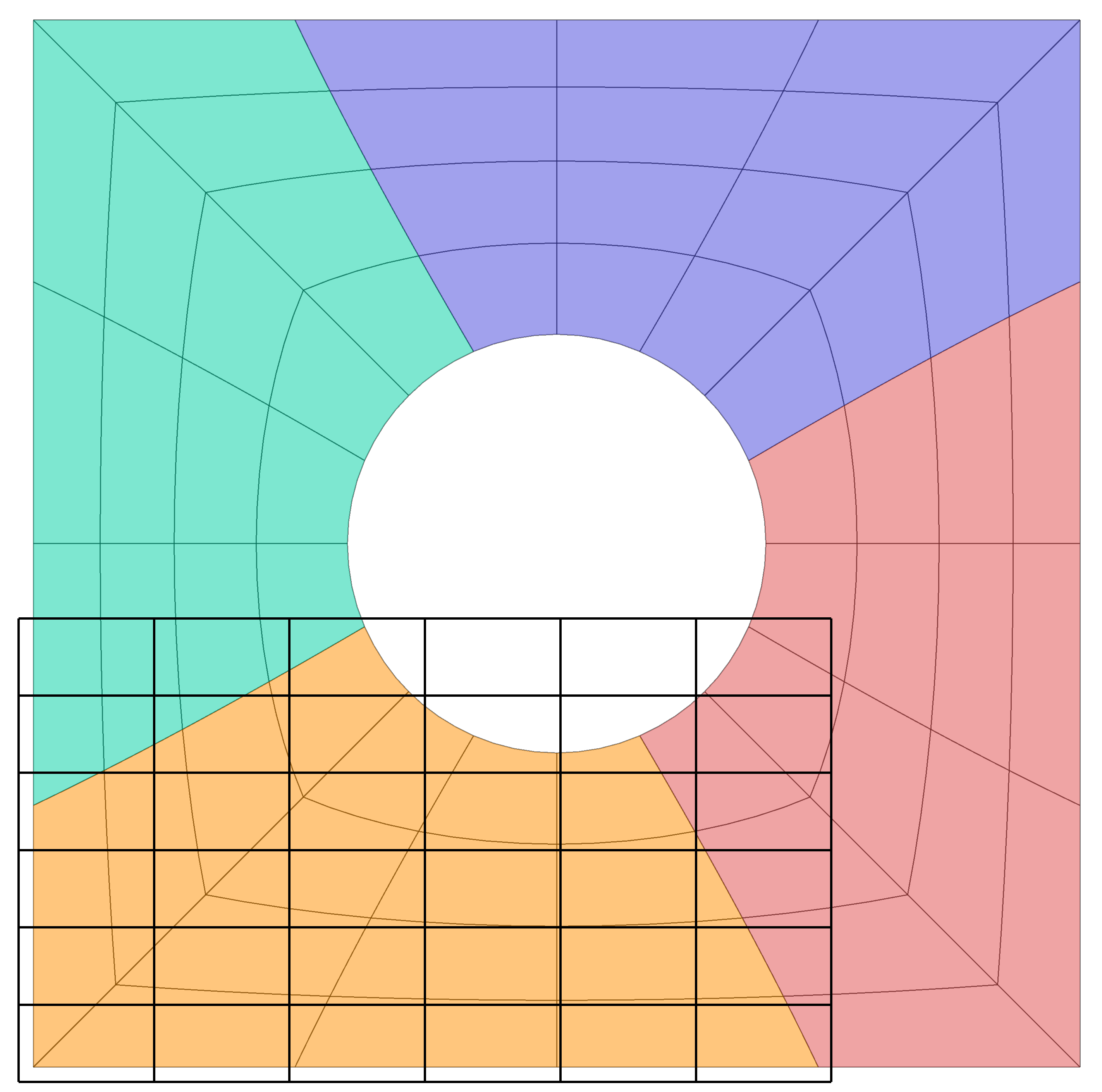} &
  \includegraphics[width=0.3\linewidth]{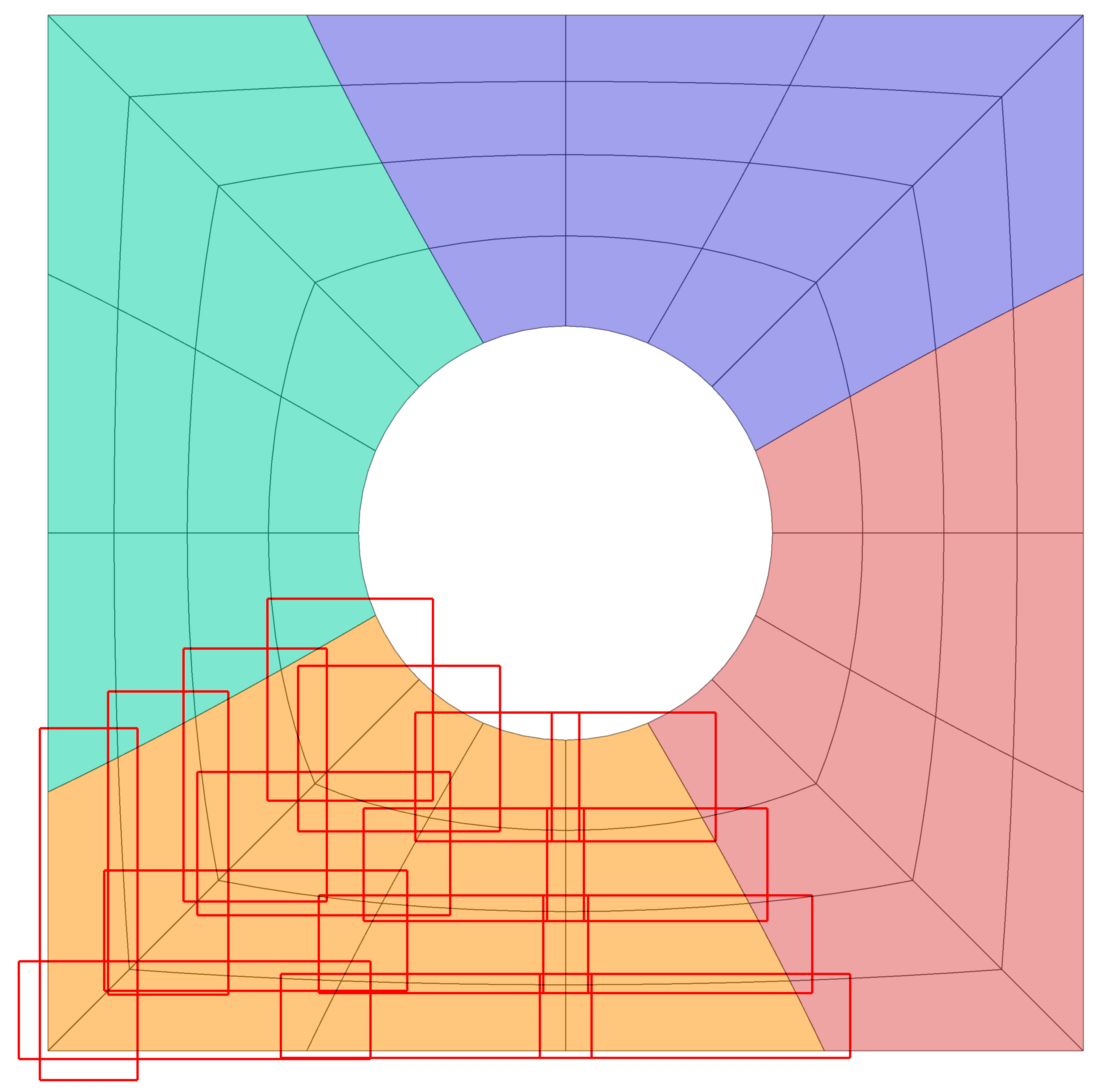} &
  \includegraphics[width=0.3\linewidth]{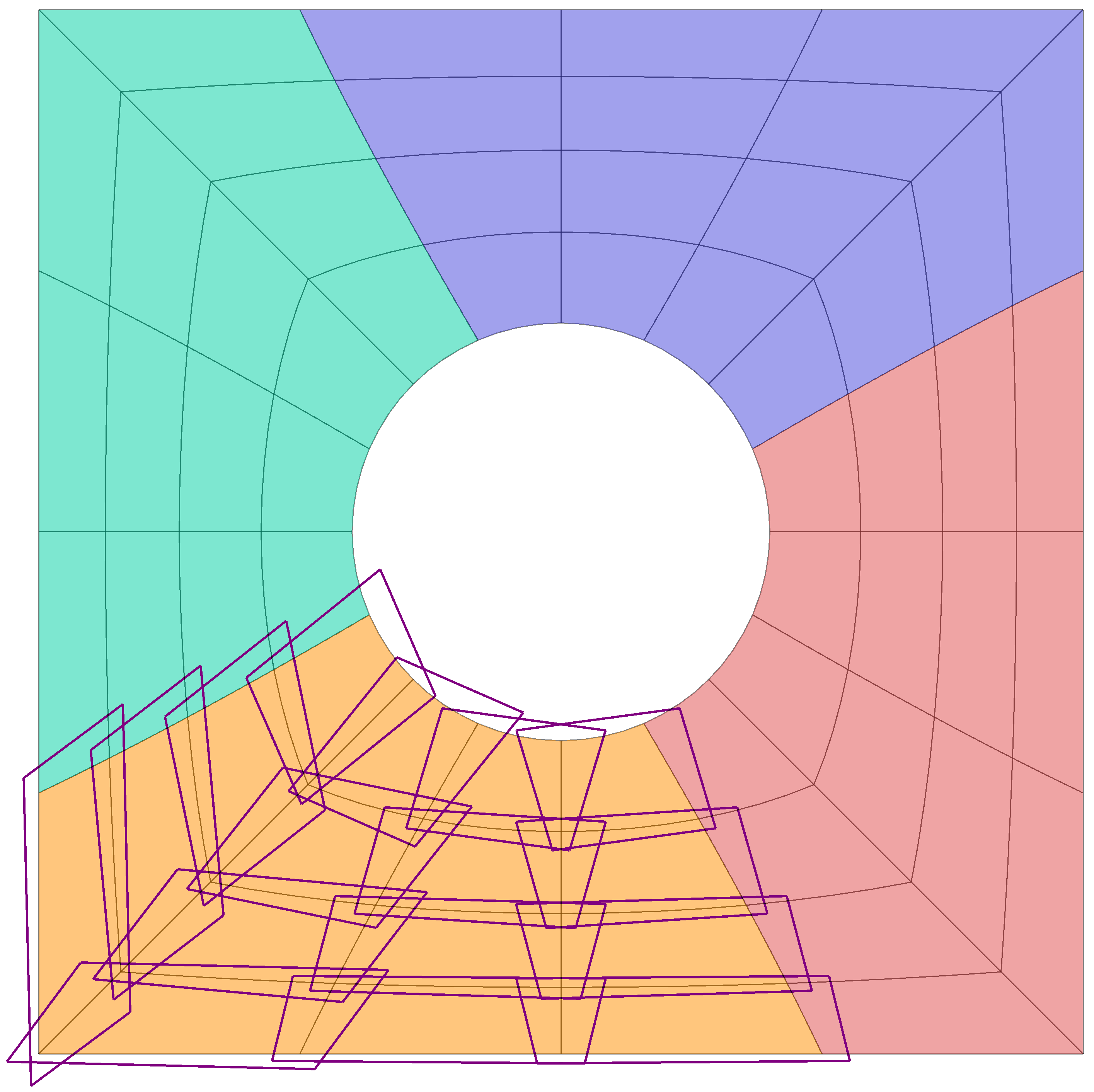} \\
  \textrm{(d)} & \textrm{(e)} & \textrm{(f)}
  \end{array}$
  \end{center}
  \vspace{-4mm}
  \caption{Key ingredients used to determine candidate elements that overlap a given point $\bx^*$. (a) $p=3$ mesh with $N_E=64$ partitioned on to 4 MPI ranks. Different colors represent different ranks. (b) Global Cartesian mesh $\mm_G$ spanning the domain of $\mm$, (c) partitioned on to 4 MPI ranks. $\mm_G$ is used to construct a map from its elements to MPI ranks corresponding to intersecting elements of $\mm$. (d) The process local Cartesian mesh $\mm_L$ on one of the ranks, which is used to map elements of $\mm_L$ to corresponding intersecting elements of $\mm$ on that rank. (e) Axis-aligned bounding boxes (AABBs) and (f) oriented bounding boxes (OBBs) for the elements of $\mm_L$.}
\label{fig_2d_meshpartition2}
\end{figure}

Figure \ref{fig_2d_meshpartition2} shows the key ingredients of the pre-processing step. The global Cartesian mesh is shown in Figure \ref{fig_2d_meshpartition2}(b) along with its partitioning in Figure \ref{fig_2d_meshpartition2}(c). The local Cartesian mesh for one of the ranks is shown in Figure \ref{fig_2d_meshpartition2}(d), along with the AABBs and OBBs for the elements on that rank in Figure \ref{fig_2d_meshpartition2}(e) and (f), respectively.

\subsubsection{Axis-aligned bounding boxes (AABBs)} \label{sec_method_precomp_aabb}

The axis-aligned bounding box for each element $\Omega^e$ is determined by computing the lower and upper bounds on the nodal position function $\bx(\br)\vert_{\Omega^e}$ \eqref{eq_x_tensor}. This is not a trivial task in the case of high-order elements as the function extrema can be located anywhere on the element.

\begin{figure}[bt!]
  \begin{center}
  $\begin{array}{cc}
  \includegraphics[width=0.4\linewidth]{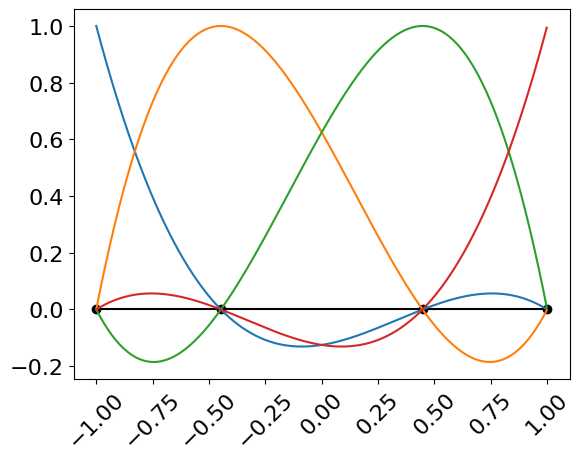} &
  \includegraphics[width=0.4\linewidth]{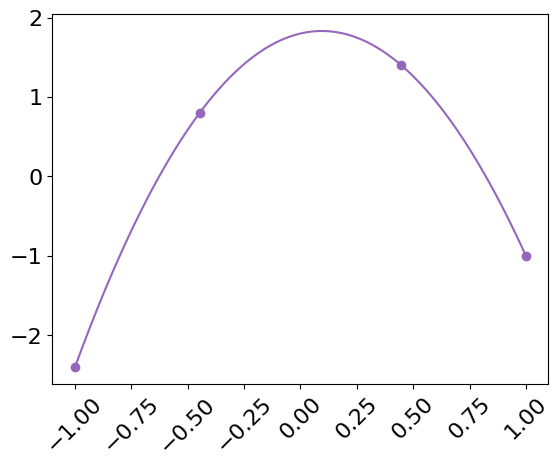} \\
  \textrm{(a)} & \textrm{(b)}
  \end{array}$
  \end{center}
  \vspace{-4mm}
  \caption{(a) $\phi_i(r)$, $i=1 \dots 4$, Lagrange interpolants. (b) Cubic 1D function $u(r)$ with nodal coefficients $u_i=\{-2.4,0.8,1.4,-1\}$ in \eqref{eq_1D_function}.}
\label{fig_1D_function}
\end{figure}

Consider a 1D function,
\begin{equation}
  \label{eq_1D_function}
  u(r) = \sum_{i=1}^{N} {u}_i \phi_i(r),\,\,\qquad r\in[-1,1].
\end{equation}
Linear bases, $\phi_i(r)$, would result in a linear function $u(r)$ and its bounds would depend on the end-points $u(r=\pm 1)$.
Similarly, piecewise linear bases would result in a piecewise linear $u(r)$ and its bounds would depend on the pointwise values, $u(\zeta_i), i=1\dots N$ at the GLL points.
Since the bases are high-order polynomials instead, pointwise values cannot be used.
See for example Figure \ref{fig_1D_function} for a cubic 1D function ($p=3,N=4$) with nodal coefficients $u_i=\{-2.4,0.8,1.4,-1\}$. The bases $\phi_i(r)$ are Lagrange interpolants at the GLL points, shown in Figure \ref{fig_1D_function}(a), and the high-order function $u(r)$ is shown in Figure \ref{fig_1D_function}(b).


\begin{figure}[bt!]
  \begin{center}
  $\begin{array}{cc}
  \includegraphics[width=0.4\linewidth]{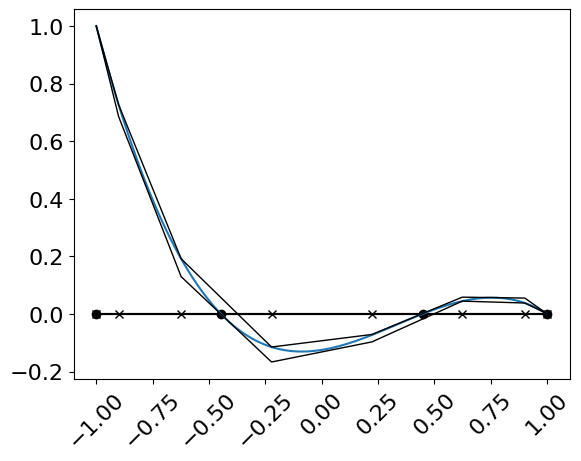} &
  \includegraphics[width=0.4\linewidth]{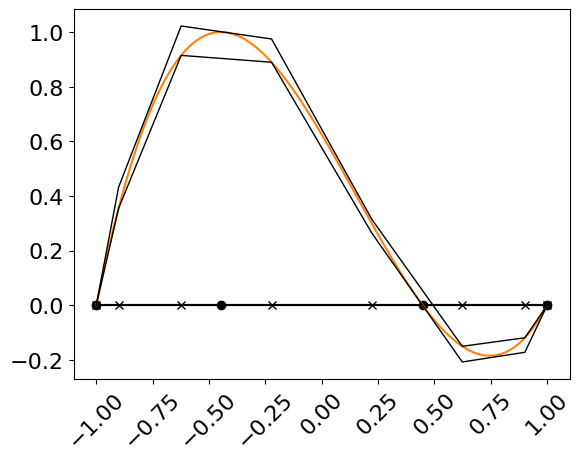} \\
  \textrm{(a)} & \textrm{(b)} \\
  \includegraphics[width=0.4\linewidth]{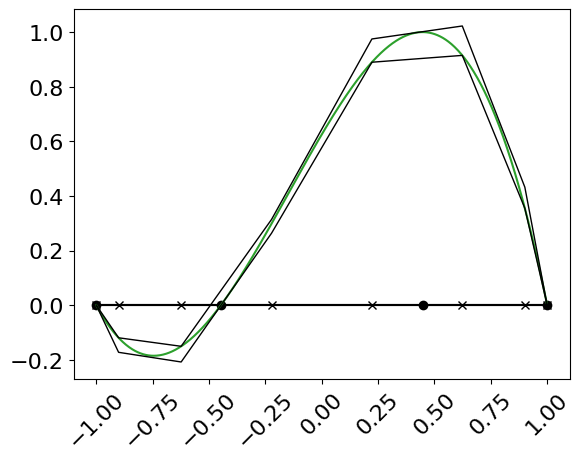} &
  \includegraphics[width=0.4\linewidth]{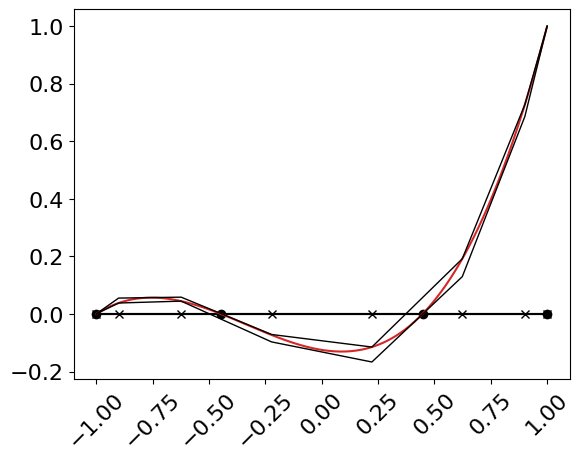} \\
  \textrm{(c)} & \textrm{(d)}
  \end{array}$
  \end{center}
  \vspace{-4mm}
  \caption{Lagrange interpolants $\phi_i(r)$, $i=1\dots 4$ at GLL points (\textbullet) along with the piecewise linear bounds constructed with $M=8$ Chebyshev interval points (x). (a) $\phi_1(r)$, (b) $\phi_2(r)$, (c) $\phi_3(r)$, and (d) $\phi_4(r)$.}
\label{fig_1D_basis_bound}
\end{figure}

We simplify the bound computation problem for \eqref{eq_1D_function} by  determining piecewise linear bounds $\uv_i (r)$ and $\ov_i (r)$ of the bases such that $\phi_i(r) \in [\uv_i (r),\ov_i (r)]$.
These bounding functions are represented using their values at Chebyshev interval points $\uv_i(\eta_j)$ and $\ov_i(\eta_j)$, where
\begin{equation}
  \label{eq_chebyshev}
  \eta_j = -\cos\bigg(\frac{j-1}{M-1}\pi\bigg), \qquad j=1\dots M.
\end{equation}
Then, using subscripts $()_{ij}$ to represent quantities associated with the $i$th basis function and the $j$th interval point, superscript $()^{'}$ to represent the derivative of the basis function, and $j^+$ and $j^-$ to represent the midpoint between the $j$th interval point and the interval point to its right- and left-hand side, respectively, the bounding functions are:
\begin{eqnarray}
  \label{eq_1D_bases_bounds}
  \underline{v}_{ij}=\min V_{ij},\,\,\bar{v}_{ij}=\max V_{ij},\qquad i = 1 \dots N, j = 1 \dots M \\
  \label{eq_1D_bases_bounds_2}
  V_{ij} =
  \begin{cases}
  {\delta_{ij}} & \text{if j=1 or j=M} \\
  \bigg\{\phi_{ij},\,\,
  \phi_{ij^-} + (\eta_j - \eta_{j-1})\cdot\phi_{ij^-}^{'}, \,\,\,
  \phi_{ij^+} - (\eta_{j+1} - \eta_{j})\cdot \phi_{ij^+}^{'} \bigg\}  & \text{otherwise},
\end{cases} \\
  \label{eq_1D_bases_bounds_3}
\eta_{j^{+}} = \frac{\eta_j + \eta_{j+1}}{2},\,\,\,\eta_{j^{-}}= \frac{\eta_j + \eta_{j-1}}{2}\end{eqnarray}
The idea in \eqref{eq_1D_bases_bounds}-\eqref{eq_1D_bases_bounds_3} is to bound the basis function in each interval between the line segment connecting the function evaluated at interval end points and the line segment that is tangent to the curve at the interval midpoint. Bounding each interval independently would result in piecewise discontinuous bounds so the more conservative of the incident bounds at each interval point is chosen to obtain piecewise continuous linear bounds.
This approach is guaranteed to work for a function that is strictly convex/concave in each interval.
For sufficiently large $M > N$, the Lagrange interpolants are \emph{predominantly} convex/concave in each interval with at-most one change in sign of their first derivatives, and we have empirically verified that the proposed approach works with $M=2N$ points for at-least $N \leq 30$ ($p \leq 29$)\footnote{The minimum number of points needed to bound the bases using \eqref{eq_1D_bases_bounds}-\eqref{eq_1D_bases_bounds_3} for $N=2 \dots 12$ represented as $(N,M_{\min})$ pairs are: (2,2), (3,4), (4,7), (5,9), (6,11), (7,12), (8,14), (9, 16), (10, 18), (11, 20), and (12, 21).}.
Figure \ref{fig_1D_basis_bound} shows the Lagrange interpolants $\phi_i(r)$ at $N=4$ GLL points along with the piecewise linear bounds constructed with $M=8$ Chebyshev interval points.

With piecewise linear bounds for the bases, $\uv_{ij}$ and $\ov_{ij}$, the bounding functions for $u(r)$ are:
\begin{eqnarray}
  \underline{u}(\eta_j) = \sum_{i=1}^{N} \min \{u_i \underline{v}_{ij}, u_i \bar{v}_{ij} \}, \\
  \bar{u}(\eta_j) = \sum_{i=1}^{N} \max \{u_i \underline{v}_{ij}, u_i \bar{v}_{ij} \}
\end{eqnarray}
and the overall bounds over the domain $r \in [-1,1]$ are just the pointwise minimum and maximum of $\underline{u}(\eta_j)$ and $\bar{u}(\eta_j)$, respectively.

We further compact these bounds by rewriting
\begin{eqnarray}
  u(r) = \sum_{i=1}^{N} u_i \phi_i(r) = a_0 + a_1 r  + \sum_{i=1}^{N} w_i\, \phi_i(r),
\end{eqnarray}
where $w_{i} = u_i - a_0 - a_1 \zeta_i$, and $a_0$ and $a_1$ are chosen based on the truncated Legendre expansion of $u(r)$ with two terms to make $w_i$ small:
\begin{eqnarray}
  a_l = \frac{2l+1}{2} \int_{-1}^{1} u(r) P_l(r) \dd r, \qquad l=0,1.
\end{eqnarray}
Here, $P_l(r)$ represents the $l$th Legendre polynomial, and the Legendre expansion ensures that the obtained bounds are exact up to a linear function.
The piecewise bounds on $u(r)$ are then:
\begin{eqnarray}
  \label{eq_1D_bound_final_1}
  \underline{u}(\eta_j) = a_0 + a_1\eta_j + \sum_{i=1}^{N} \min \{w_i \underline{v}_{ij}, w_i \bar{v}_{ij} \}, \\
  \label{eq_1D_bound_final_2}
  \bar{u}(\eta_j) = a_0 + a_1\eta_j + \sum_{i=1}^{N} \max \{w_i \underline{v}_{ij}, w_i \bar{v}_{ij} \}
\end{eqnarray}

\begin{figure}[bt!]
  \begin{center}
  $\begin{array}{ccc}
  \includegraphics[width=0.3\linewidth]{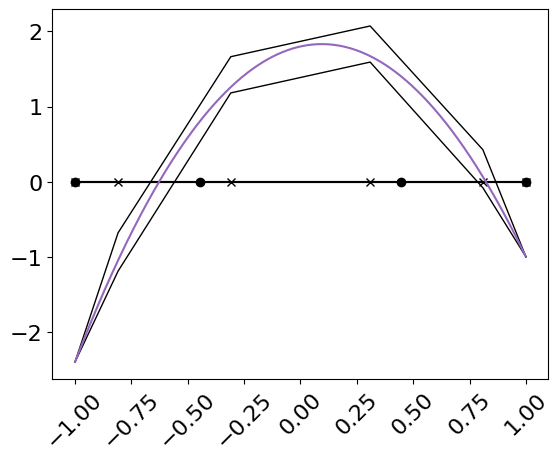} &
  \includegraphics[width=0.3\linewidth]{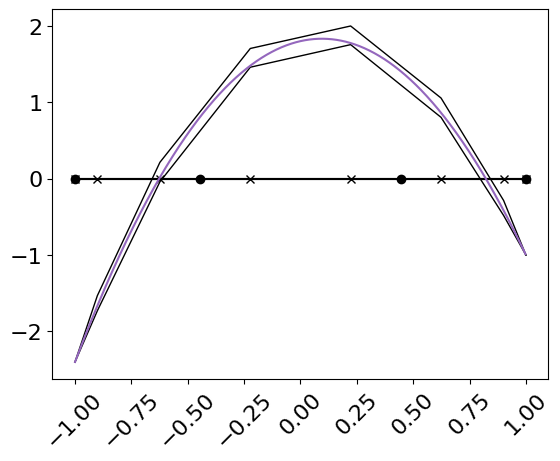} &
  \includegraphics[width=0.3\linewidth]{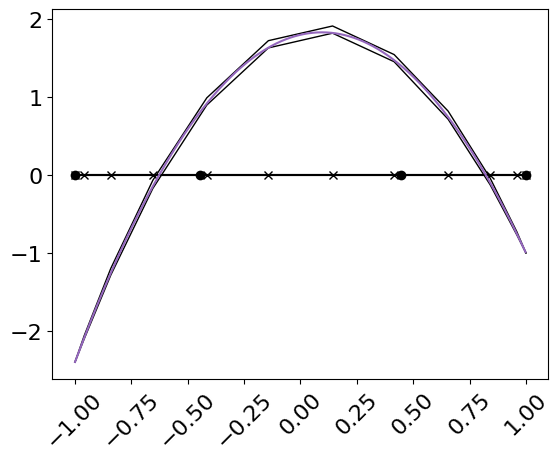} \\
  \textrm{(a)} & \textrm{(b)} & \textrm{(c)}
  \end{array}$
  \end{center}
  \vspace{-4mm}
  \caption{Piecewise linear bounds on $u(r)$ constructed using (a) $M=6$, (b) $M=8$, and (c) $M=12$ Chebyshev interval points. Higher number of interval points lead to tighter bounds around $u(r)$.}
\label{fig_1D_u_bound}
\end{figure}

Figure \ref{fig_1D_u_bound} shows the bounding functions for the cubic function in Figure \ref{fig_1D_function}(a) for $M = 6,\,8,$ and $12$. As expected, higher $M$ results in tighter bounds around $u(r)$. The computational cost of the 1D bounding procedure in \eqref{eq_1D_bound_final_1}-\eqref{eq_1D_bound_final_2} is $\bigO(NM)$, and we set $M=2N$ by default.

For each 2D quadrilateral element, we compute the bounds on each of their edges using \eqref{eq_1D_bound_final_1}-\eqref{eq_1D_bound_final_2} as the nodal position function is 1D along each edge. We then take the union of the bounds for the four edges to get the final bounding box for the quad element, $\breve{\bx}=\{\bx_{min},\bx_{max}\}$. In the remainder of the text, $(\breve{\,\,\,\,})$ will denote the quantities related to the axis-aligned bounding box of an element (e.g., $\breve{\Omega}^e$ and $\breve{\bx}(\br)$).

This approach directly extends to hexahedron elements as we bound the 2D nodal position function along each of the element faces. Consider the 2D function,
\begin{equation}
  \label{eq_2D_function}
  u(r,s) = \sum_{i=1}^{N} \sum_{j=1}^{N} {u}_{ij} \phi_i(r) \phi_j(s),\,\,\qquad r,s\in[-1,1].
\end{equation}
Using piecewise linear bounds for $\phi_i(r)$ and $\phi_j(s)$, the bounding functions for $u(r,s)$ are:
\begin{eqnarray}
  \label{eq_2D_function_bound_1}
  \uu(\eta_k,\eta_l) &=& \sum_{i=1}^{N} \sum_{j=1}^{N}
  \min\{{u}_{ij} \uv_{ik} \uv_{jl},\,
       {u}_{ij} \uv_{ik} \ov_{jl},\,
       {u}_{ij} \ov_{ik} \uv_{jl},\,
  {u}_{ij} \ov_{ik} \ov_{jl}\}, \\
  \label{eq_2D_function_bound_2}
  \ou(\eta_k,\eta_l) &=& \sum_{i=1}^{N} \sum_{j=1}^{N}
  \max\{{u}_{ij} \uv_{ik} \uv_{jl},\,
       {u}_{ij} \uv_{ik} \ov_{jl},\,
       {u}_{ij} \ov_{ik} \uv_{jl},\,
  {u}_{ij} \ov_{ik} \ov_{jl}\},
\end{eqnarray}
where $\{k,l\}=1\dots M$. The computational complexity of executing the double sum in \eqref{eq_2D_function_bound_1}-\eqref{eq_2D_function_bound_2}  for $M^2$ interval points is reduced from $\bigO(N^2M^2)$ to $\bigO(NM^2)$ using the tensor-product property of the bases (see e.g., \cite{deville2002high}).
Since the polynomial order is same for all the elements in the mesh, the piecewise linear bounds on the bases $\phi_i$ are computed only once and reused.
Using \eqref{eq_2D_function_bound_1}-\eqref{eq_2D_function_bound_2}, the bounds are computed for the nodal position function of each face of a hex element, and then combined to obtain the AABB for that element.

The AABB is represented using $2D$ values per element ($\breve{\bx}_{\tt min}$ and $\breve{\bx}_{\tt max}$) and a given point $\bx^*$ overlaps this AABB if
\begin{eqnarray}
  \label{eq_AABB_test}
(x^*_i - \breve{x}_{i,\tt min})(\breve{x}_{i,\tt max} - x^*_i) \geq 0\,\,,\qquad i=1\dots D.
\end{eqnarray}
From an implementation perspective, the bounding boxes are expanded by a nominal percentage (10\% by default) to avoid false negatives due to floating-point errors resulting from points located on the bounding-box boundary.

\begin{figure}[bt!]
  \begin{center}
  $\begin{array}{cc}
  \includegraphics[height=0.35\linewidth]{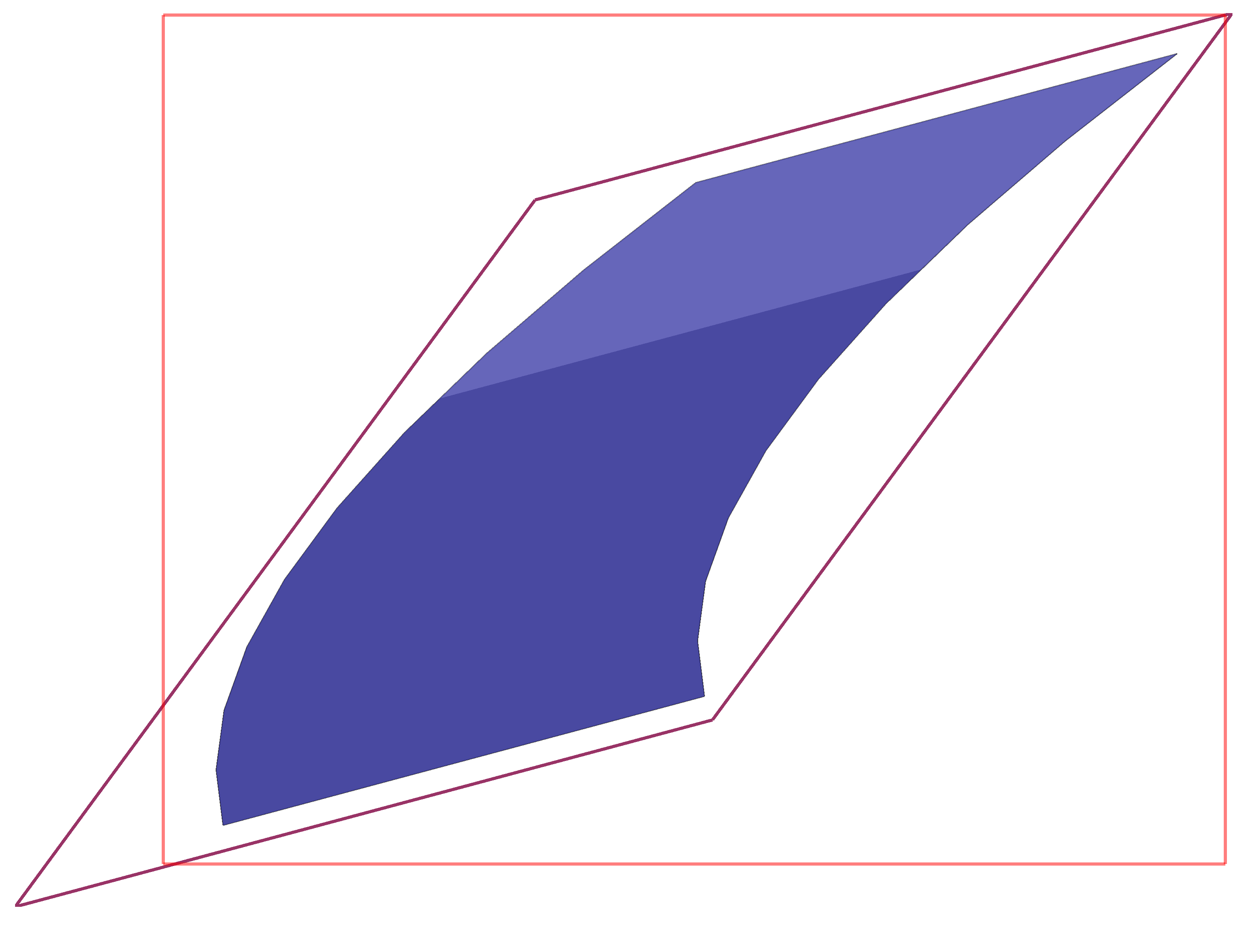} &
  \includegraphics[height=0.35\linewidth]{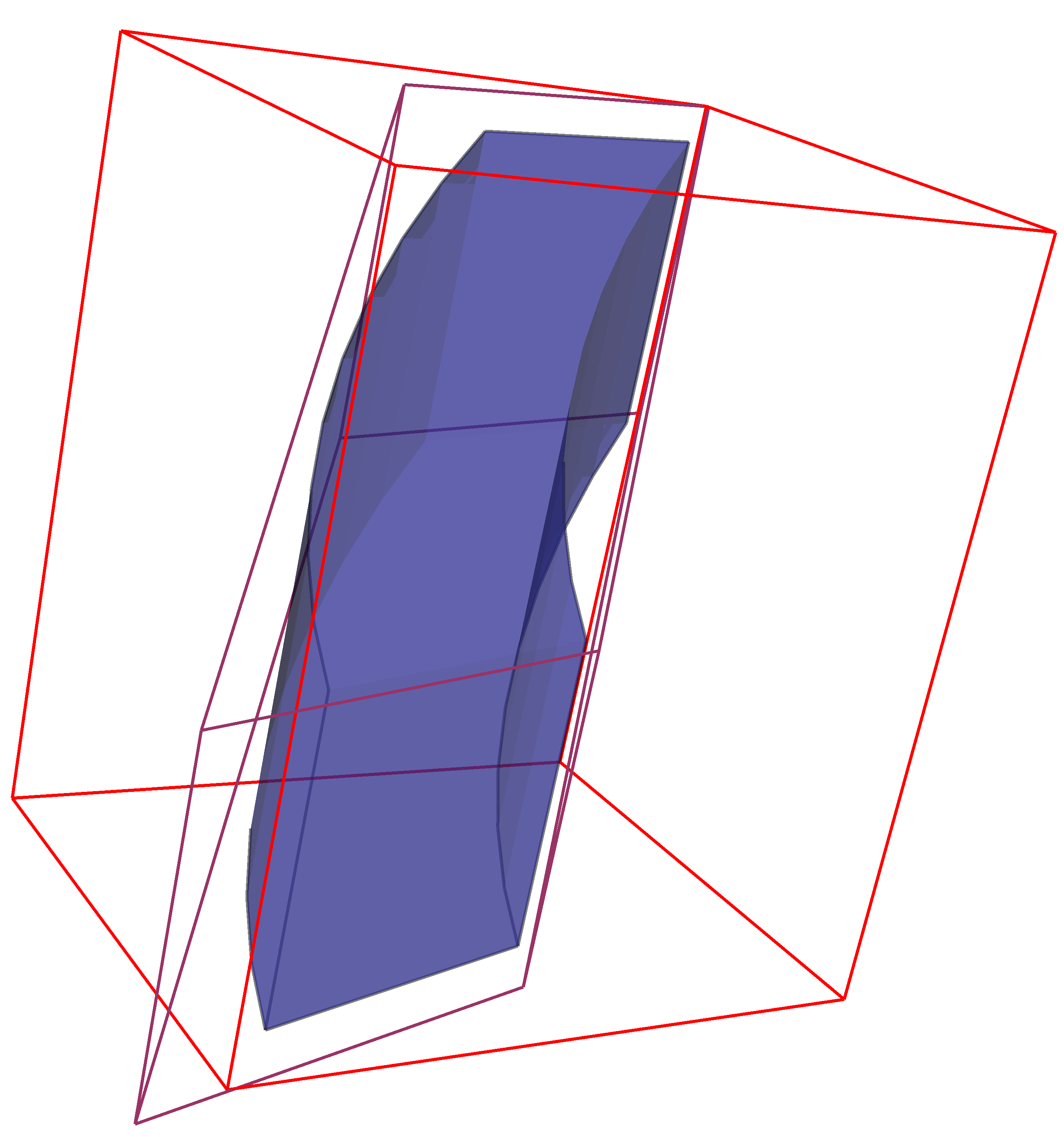} \\
  \textrm{(a)} & \textrm{(b)}
  \end{array}$
  \end{center}
  \vspace{-4mm}
  \caption{Axis-aligned bounding box and oriented bounding box around a unit-sized cubic element in (a) 2D and (b) 3D. For the 2D problem, the area of the AABB is 3.6 versus 1.6 for the OBB, and for the 3D problem, the volume of the AABB is 9 versus 1.76 for the OBB.}
\label{fig_single_el_AABB}
\end{figure}

Figure \ref{fig_single_el_AABB} shows the bounding boxes for a cubic 2D quadrilateral and 3D hexahedron element. In each case, the initial cubic element is unit-sized, but the AABB area is 3.6 for the 2D case and volume is 9 for the 3D case. While an AABB can be used to quickly determine if a candidate element contains a given point, this example demonstrates that it can occupy a much larger area/volume than the curvilinear element it bounds.

\subsubsection{Oriented bounding boxes (OBBs)} \label{sec_method_precomp_obb}

\begin{figure}[bt!]
  \begin{center}
  $\begin{array}{c}
  \includegraphics[width=0.8\linewidth]{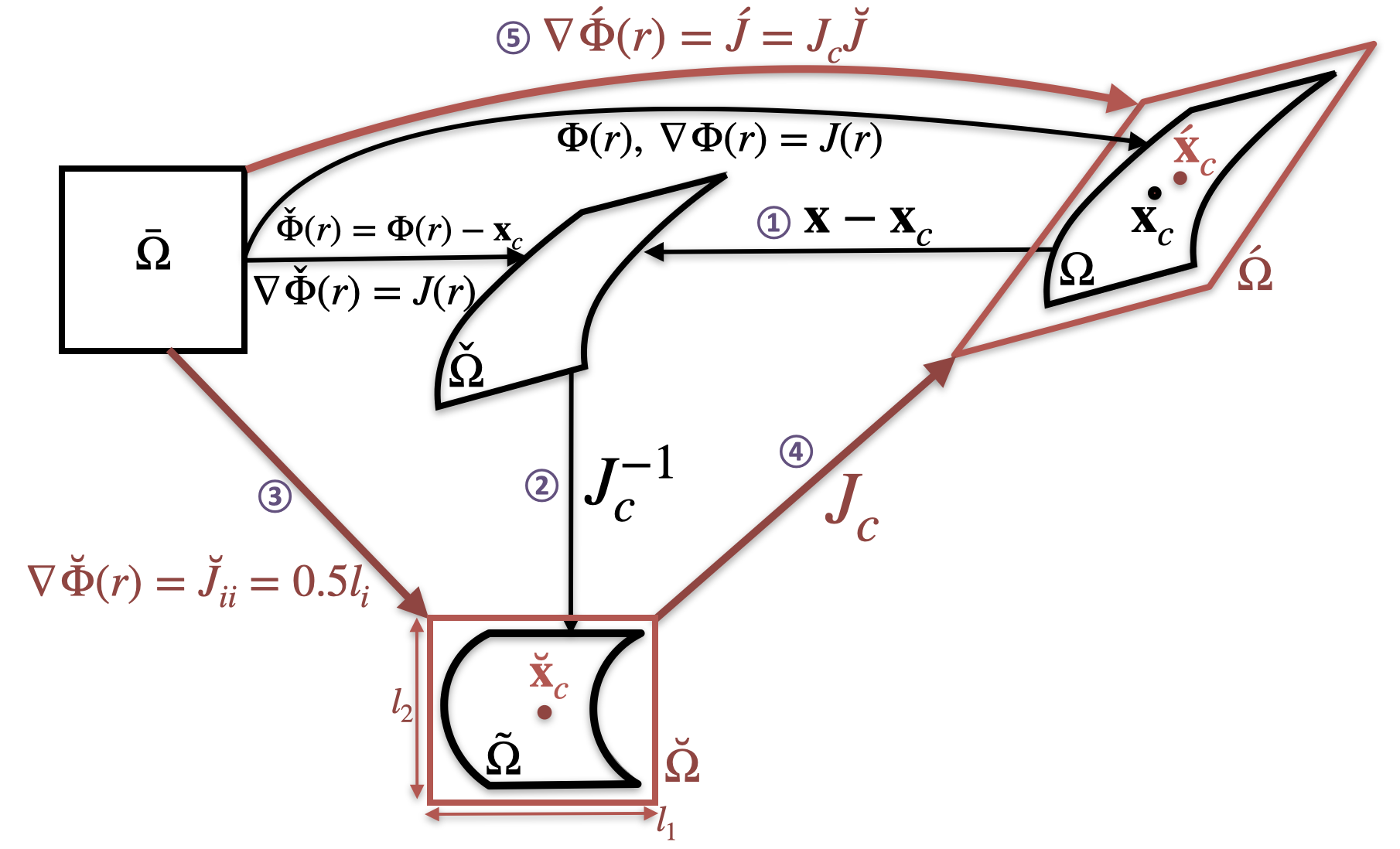}
  \end{array}$
  \end{center}
  \vspace{-4mm}
  \caption{Jacobian of the transformations used to compute the oriented bounding box around an element. The original element $\Omega$ is shown in top-right, the modified element $\check{\Omega}$ with its center at the origin is in top-center, the reference element $\bar{\Omega}$ is in top-left, and the element $\tilde{\Omega}$ transformed to align with the Cartesian axes is at the bottom. The AABB around $\tilde{\Omega}$, the OBB around the original element $\Omega$, and relevant transformations are shown in red.}
\label{fig_OBB_transformation}
\end{figure}

We obtain a compact oriented bounding box for an element by first transforming it to the reference space based on the Jacobian of the transformation at its center. Next, the AABB for this transformed element is computed. Finally, the AABB is transformed back to the physical space using the Jacobian from the first step. This procedure is summarized with relevant transformations in Figure \ref{fig_OBB_transformation}.


Consider an element $\Omega \in \mathbb{R}^d$ with nodal positions $\bx(\br)$ defined via a transformation $\mathbf{\Phi}(\br)$ of the reference element $\bar{\Omega}$.
The element $\Omega$ is first linearly transformed using its center \big($\bx_c = \bx(\br=0)$\big) to obtain the element $\check{\Omega}$ with its center at the origin, i.e., $\check{\mathbf{\Phi}}(\br)=\mathbf{\Phi}(\br)-\bx_c$.
Note that since these two transformations vary by a constant, their Jacobian is the same, i.e.,  $\nabla_\br \mathbf{\Phi}=\nabla_\br \check{\mathbf{\Phi}} = J(\br)$.
Then, $\check{\Omega}$ is transformed using $J_c = J(\br=0)$ to obtain the element $\tilde{\Omega}$ that aligns with the axes in the reference space:
\begin{eqnarray}
  \label{eq_OBB_transformation}
  \tilde{\bx}(\br)\vert_{\tilde{\Omega}^e} = J^{-1}_{c} \bigg(\bx(\br)\vert_{{\Omega^e}}-\bx_{c} \bigg).
\end{eqnarray}

Next, bounds of the nodal positions $\tilde{\bx}(\br)$ of $\tilde{\Omega}$ are computed using the strategy described in Section \ref{sec_method_precomp_aabb} to obtain its AABB, $\breve{\Omega}$. The Jacobian of the transformation of $\breve{\Omega}$ with respect to the reference-element $\bar{\Omega}$ is the diagonal matrix, $\breve{J}_{ii} = L_i/2$, where $L_i=\breve{x}_{i,\tt max}-\breve{x}_{i,\tt min}$.
The AABB is shown in red at the bottom center of Figure \ref{fig_OBB_transformation}.
Finally, $\breve{\Omega}$ is transformed back using $J_c$ to align with the original element and obtain its OBB. The center of the OBB is at $\Acute{\bx}_{c} = \bx_{c} + J_c \breve{{\bx}}_{c}$, where ${\breve{{\bx}}}_{c}$ is the center of $\breve{\Omega}$, and the Jacobian of its transformation with respect to the reference space is $\Acute{J} = J_c \breve{J}$.
Note that $\Omega$ and $\acute{\Omega}$ are obtained from $\tilde{\Omega}$ and $\breve{\Omega}$ via the same linear transformation, $\bx = J_c \tilde{\bx} + \bx_c$, and so  $\tilde{\Omega} \subset \breve{\Omega}$ implies $\Omega \subset \acute{\Omega}$.
 The OBB $\acute{\Omega}$ is shown in red around the original element $\Omega$ in Figure \ref{fig_OBB_transformation}.

Each element's OBB is stored using $D^2$ values for the inverse transformation $\Acute{J}^{-1}$ and $D$ values for the center $\Acute{\bx}_{c}$, and is used to determine if a given point $\bx^*$ is potentially inside an element by checking if
\begin{eqnarray}
  \label{eq_OBB_test}
\acute{J}^{-1}(\bx^*-\Acute{\bx}_{c}) \in [-1, 1]^D.
\end{eqnarray}
Figure \ref{fig_single_el_AABB} shows the oriented bounding boxes for the cubic elements. As expected, the OBB provide much tighter bounds around the element: the area of the 2D OBB is 1.6 vs 3.6 for AABB, and the volume of the 3D OBB is 1.76 versus 9 for the AABB.

Note that in the proposed approach, the orientation of the element is essentially inferred from the Jacobian of the transformation at its center ($\br=0$). In future work, we will explore other techniques to determine the orientation such that the resulting OBB provides even tighter bounds.

\subsubsection{Process-local map from $\bx^*$ to $\ue^*$} \label{sec_method_precomp_local_map}

\begin{figure}[bt!]
  \begin{center}
  $\begin{array}{cc}
  \includegraphics[width=0.45\linewidth]{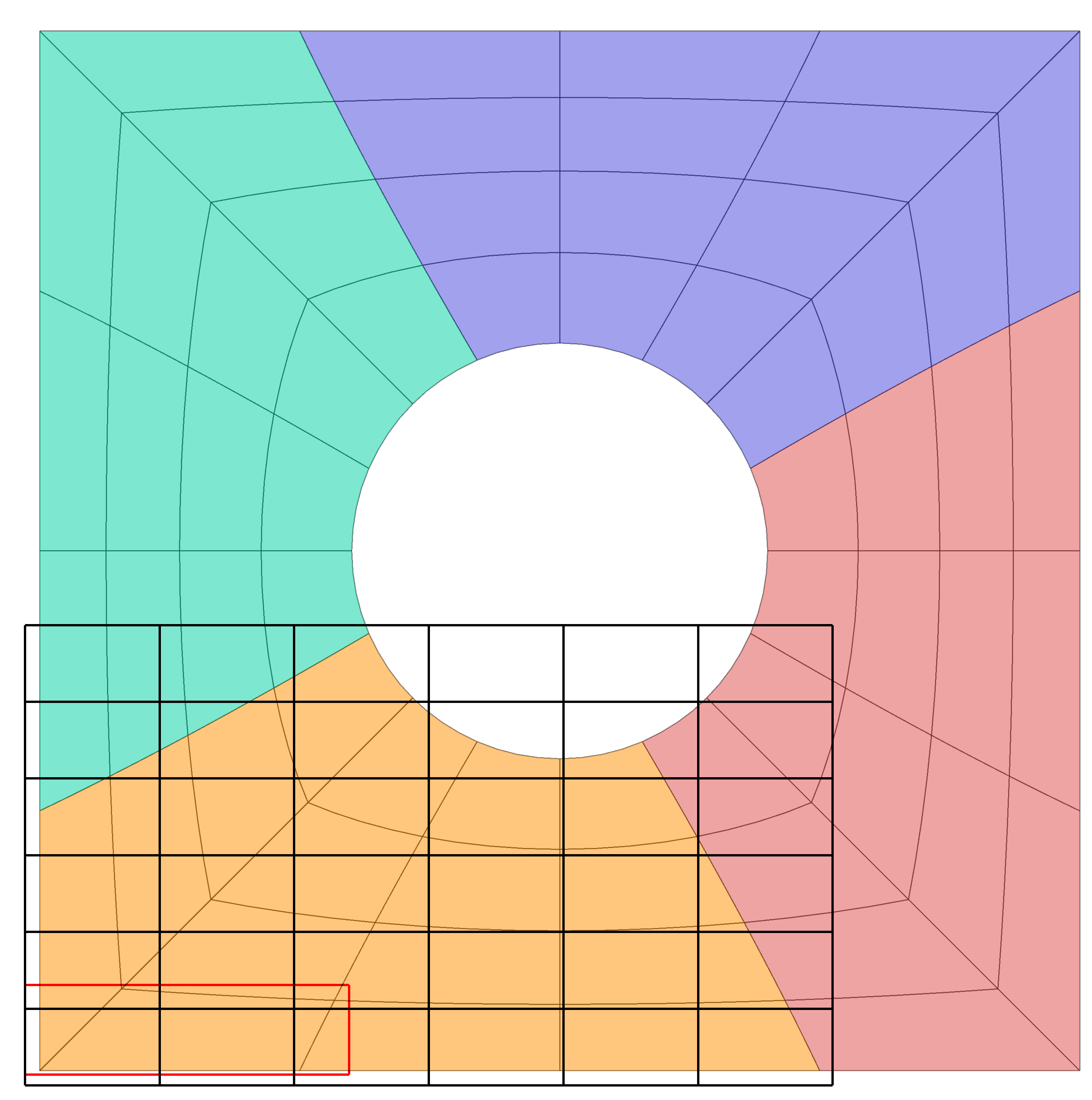} &
  \includegraphics[width=0.45\linewidth]{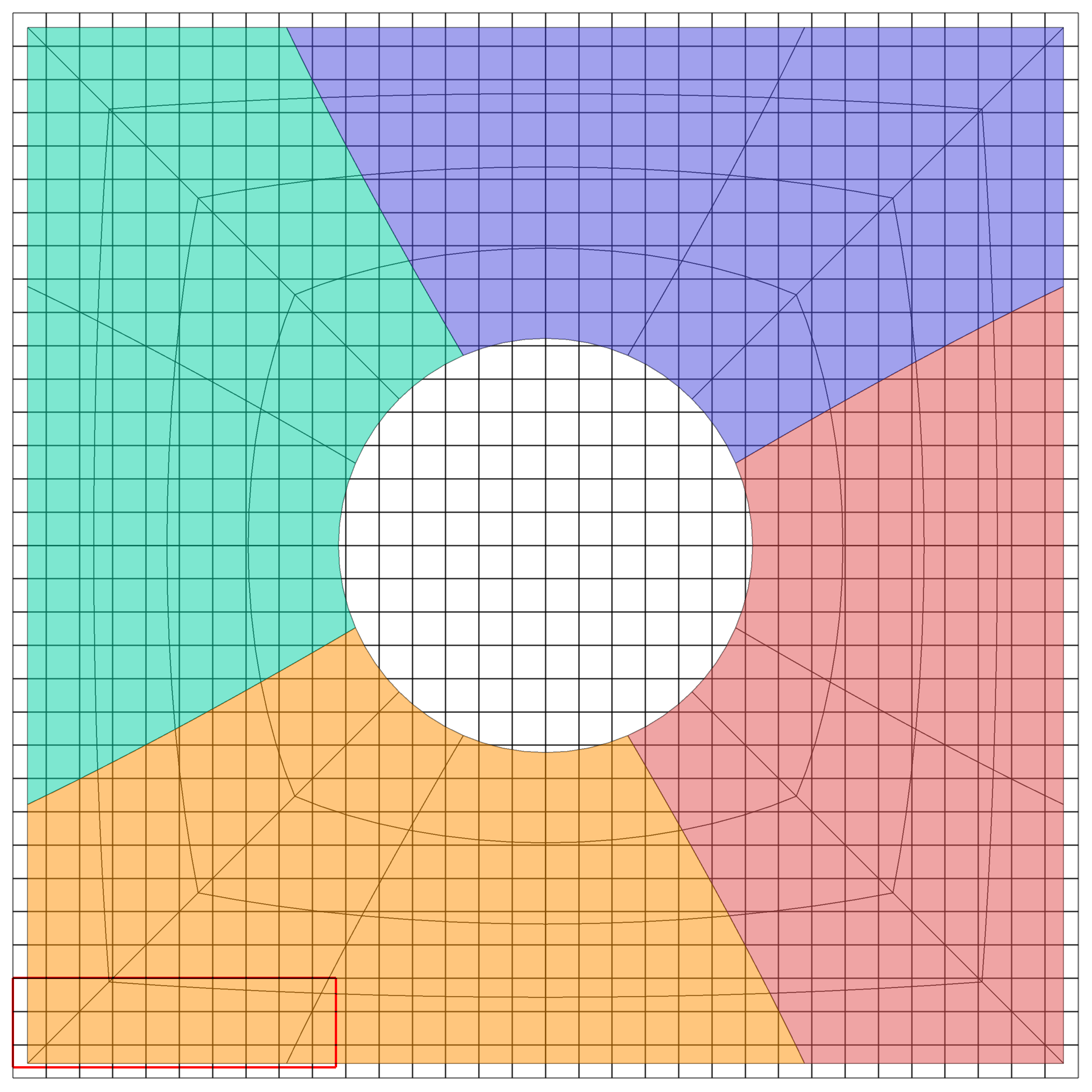}
  \end{array}$
  \end{center}
  \vspace{-4mm}
  \caption{Axis-aligned bounding box of one of the elements of $\mm$ intersecting with (left) 8 elements of the process-local Cartesian mesh $\mm_L$ and (right) 30 elements of the global Cartesian mesh $\mm_G$.}
\label{fig_local_hash_map}
\end{figure}

We use a process-local Cartesian mesh $\mm_L$ to determine the candidate elements $\ue^*$ on a given rank $m^*$ that could contain the point $\bx^*$. $\mm_L$ has $N_L$ cells in each direction, and spans the union of the bounding boxes of all the elements on that rank. We first identify the elements of $\mathcal{M}_L$ that intersect the AABB of each of the $N_{m^*,E}$ elements of $\mathcal{M}$. The computation of this intersection is straightforward: determine the cells of $\mm_L$ that contain the diagonally opposite corners of each AABB, and save all the elements between those cells in a multi-valued map, $\Psi_L : e_{\mathcal{M}_L} \mapsto \ue_{\mathcal{M}}$.  Here, $e_{\mm_L}$ is the element index corresponding to $\mm_L$ and $\ue_{\mathcal{M}}$ is the list of element indices corresponding to $\mm$. For example, the AABB of one of the elements from Figure \ref{fig_2d_meshpartition2} is shown to intersect 6 elements of $\mm_L$ in Figure \ref{fig_local_hash_map}(left).

During the \emph{Find} step, we determine the cell of $\mm_L$ that contains a given point $\bx^*$ and use $\Psi_L$ to retrieve the candidate elements $\ue^*$.
Note that as $N_L$ increases, each cell of $\mm_L$ maps to a decreasing number of elements of $\mm$. For the example shown in Figure \ref{fig_2d_meshpartition2}(d), $N_L=6$ and each of the 36 cells map to at-most 6 cells of $\mm$.
We set $N_L$ such that the resolution of $\mm_L$ is similar to that of $\mm$ by default, though the user can adjust this parameter based on their storage and computational constraints.

Due to the Cartesian-aligned structure of $\mm_L$, the mesh is never explicitly constructed and is represented using $2D$ values for the lower and upper bounds of the mesh, 1 value for the mesh resolution $(N_L$), and $D$ values for the size of the elements in each direction.

\subsubsection{Global map from $\bx^*$ to $\um^*$} \label{sec_method_precomp_global_map}

Similar to the local map, we use a Cartesian mesh $\mm_G$ to determine the candidate MPI ranks $\um^*$ that could contain a given point $\bx^*$. This mesh spans the union of the bounding boxes of all the elements across all the MPI ranks, and is divided into $N_G$ cells in each direction. $\mm_G$ is implicitly partitioned across the $N_P$ ranks such that each cell with lexicographic index $i$ is owned by rank $i\%N_P$ and is assigned a local index $i \backslash N_P$ on that rank.

Next, we identify the cells of $\mathcal{M}_G$ that intersect with the AABB of each of the $N_{m,E}$ elements of $\mathcal{M}$ on rank $m$. Each process $m$ then communicates its MPI rank to the ranks (e.g., $i\%N_P$) that own the corresponding cells of $\mm_G$ along with corresponding element indices (e.g., $i \backslash N_P$).
Upon receipt, this information is stored in a multi-valued map $\Psi_G: e_{\mathcal{M}_G} \mapsto \um_{\mathcal{M}}$, where $e_{\mm_G}$ is the element index corresponding  to $\mm_G$ and  $\um_{\mathcal{M}}$ is the list of ranks corresponding to $\mm$.
For example, the AABB of one of the elements from Figure \ref{fig_2d_meshpartition2} is shown to intersect 30 elements of $\mm_G$ in Figure \ref{fig_local_hash_map}(right). The owning process for this element, will thus communicate its MPI rank and the overlapping element indices corresponding to $\mm_G$ to the ranks that own each of those 30 cells.

During the \emph{Find} step, we determine the cell of $\mm_G$ that contains a given point $\bx^*$, forward it to the owning rank based on the lexicographic index of that cell, look up the list of ranks using $\Psi_G$, and forward the point to those ranks. The process-local map, $\Psi_L$ is then locally used to determine candidate elements that could contain the point.
The resolution of $\mm_G$ is set based on the mesh size by default, similar to $\mm_L$, but can also be specified by the user as needed.

\subsection{Computing reference space coordinates for a given point} \label{sec_method_find}
After the \emph{Setup} step is completed for a given mesh, the \emph{Find} step determines the computational coordinates required to evaluate the solution for each point $\ubx^*_i$, $i=1\dots N_{pt}$.
These computational coordinates are $\bq_i^*=\{m^*,e^*,\br^*,d^*, c^* \}_i$ that indicate the MPI rank $m_i^*$ that element $e_i^*$ is on, and the reference space coordinates $\br_i^*$ inside $e_i^*$ corresponding to $\bx_i^*$. $\bq_i^*$ also include the distance $d_i^*$ between the input point $\bx_i^*$ and the point $\bx(\br_i^*)\vert_{\Omega^{e_i^*}}$ found using the Newton's method, and an enum $c_i^*=\{{\tt INTERIOR\,|\,BORDER\,|\,NOT\_FOUND}\}$ that indicates the location of the converged solution with respect to the reference element.

Typically each process issues the \emph{Find} query for a unique set of points that are first searched locally on that rank.
For each $\bx_i$, the process-local map $\Psi_L$ and bounding boxes determine a list of candidate overlapping elements $\ue_i$. If a point does not overlap the bounding boxes of any of the elements returned by $\Psi_L$, $c_i^*$ is set to ${\tt NOT\_FOUND}$.
For remainder of the points, we sequentially go through the elements in $\ue_i$ and use the Newton's method described in Section \ref{sec_method_newton} to determine the reference space coordinates in each element.
When a point is found inside an element ($\br^*_i \in (-1,1)^D$), the remaining candidate elements are discarded and $c_i^*$ is set to ${\tt INTERIOR}$.
If a point is found on an element's boundary ($\br^*_i \in \partial \bar{\Omega}^{e^*}$) or outside it (because it was located within its bounding boxes), the point is searched in the remaining elements until it is found in the interior of an element. When a point is not interior to any of the elements on that rank, the element returning the smallest $d^*$ is chosen. For such points, $c_i^*$ is set to ${\tt BORDER}$

After the process-local search, all points with $c^* = \{\tt BORDER\,|\,NOT\_FOUND\}$ are sent to other ranks determined from the global map described in Section \ref{sec_method_precomp_global_map}. The process-local search is then repeated, and all the processes return the computational coordinates of the points found with $c^* = \{\tt INTERIOR\,|\,BORDER\}$, to the process that had originally issued the query for the point. These returned computational coordinates are then compared with the process-local results from the first step and preference is given to the element that returns $c^* = {\tt INTERIOR}$, followed by the element that returns $c^* = {\tt BORDER}$ with the smallest $d^*$.

\subsubsection{Newton's method with trust-region} \label{sec_method_newton}
For a given $\bx^*$ and $\Omega^e$, the Newton's method is used to determine the corresponding reference space coordinates by minimizing the functional $f(\br) = 0.5\|\bx^*-\bx(\br)\|_2^2=0.5\|\Delta \bx\|_2^2$. The initial guess, $\br_0$, is based on the closest mesh node and it is updated iteratively:
\begin{eqnarray}
  \label{eq_newton}
  \br_{l+1} = \br_{l} -  \underbrace{\mathcal{H}^{-1}_{ji} \mathcal{J}_{j}}_{\Delta \br_l}:\,\,\,\, \br_{l+1} \in [-1, 1]^D, \Delta \br_l \in \alpha_l [-1, 1]^D,
\end{eqnarray}
where the subscript $()_l$ denotes the $l$th Newton iteration, $\mathcal{J}_{j}$ and $\mathcal{H}_{jk}$ are the Jacobian and Hessian of the functional $f(\br_l)$, and $\alpha_l$ is the trust-region factor. The Jacobian and Hessian are:
\begin{eqnarray}
  \label{eq_newton_jacobian}
  \mathcal{J}_j &=& \frac{\partial f}{\partial r_j} = \sum_i \Delta x_i (-\frac{\partial x_i}{\partial r_j}) = - G_{ij}^T \Delta x_i, \\
  \label{eq_newton_hessian}
  \mathcal{H}_{jk} &=& \frac{\partial^2f}{\partial r_j \partial r_k} = G_{ji} G_{ik} - {\beta} \Delta x_i \frac{\partial^2 x_i}{\partial r_j \partial r_k},\,\,\, \beta={0,1}, \\
  \label{eq_jacobian}
  G_{ij} &=& \frac{\partial x_i}{\partial r_j}.
\end{eqnarray}
The second derivative term in the Hessian, $\frac{\partial^2 x_i}{\partial r_j \partial r_k}$, $\{i,j,k\}=1\dots D$, is the most expensive to compute as it has $D^3$ components ($6$ terms in 2D and $18$ terms in 3D after accounting for symmetries). It is thus included ($\beta=1$) only after the first Newton iteration ($l>0$) when the current approximation is on the element edge/face; $\beta = 0$ otherwise. Since one or more reference space coordinates are fixed at element boundaries, the number of components required for the $\frac{\partial^2 x_i}{\partial r_j \partial r_k}$ term reduce to 9 for a hexahedron face and $D$ for a quadrilateral/hexahedron edge.

The trust-region factor $\alpha_l$ is initialized to 1 and then updated after each Newton iteration based on the actual decrease in the distance, i.e. $\Delta f_{\tt decr} = \|\Delta \bx_l\|_2^2 - \|\Delta \bx_{l+1}\|_2^2$ and the predicted decrease in distance, i.e. $\Delta f_{\tt pred} = \|\Delta \bx_l\|_2^2 - \|\Delta \tilde{\bx}_{l}\|_2^2$ between the point to find and the current approximation. Here,  $\|\Delta \bx_l\|_2^2$ and $\|\Delta \bx_{l+1}\|_2^2$ are computed using the nodal position function at $\br_l$ and $\br_{l+1}$, respectively, and $\Delta \tilde{\bx}_{l}$ is the anticipated decrease in distance based on the derivatives of the functional at $\br_l$. We use a linear approximation when $\beta=0$:
\begin{eqnarray}
  \Delta \tilde{\bx}_{l} = \bx^* - \tilde{\bx}(\br_l), \\
  \tilde{\bx}(\br_l) = \bx(\br_l) + G \Delta \br_l,
\end{eqnarray}
and a quadratic approximation when $\beta=1$:
\begin{eqnarray}
  \Delta f_{\tt pred} = (\mathcal{J}^T\Delta \bx_l)\cdot \Delta \br_l +  \frac{\Delta \br^T_l \mathcal{H} \Delta \br_l}{2}.
\end{eqnarray}
The trust region is updated as:
\begin{equation}
  \alpha_{l+1} =
  \begin{cases}
    2 \alpha_l, & \text{if $\Delta f_{\tt decr} >= 0.9 \Delta f_{\tt pred}$}\\
    \alpha_l, & \text{else if $\Delta f_{\tt decr} >= 0.01 \Delta f_{\tt pred}$} \\
    \frac{\alpha_l}{4} & \text{otherwise}
   \end{cases}
\end{equation}
In the last case when $\Delta f_{\tt decr} < 0.01 \Delta f_{\tt pred}$, we discard the Newton update and repeat \eqref{eq_newton} using the updated trust region factor. For a given point, at-most 50 Newton iterations are done until $\|\Delta \br_l\|_\infty < \epsilon = 10^{-10}$. As we will demonstrate in Section \ref{sec_results}, even 5 Newton iterations are sufficient to accurately determine the reference space coordinates in very high-order elements.

Note that a consequence of the constraint $\br_{l+1} \in [-1,1]^D$ in \eqref{eq_newton} is that when a point is located outside the element but within its bounding boxes, the converged reference space coordinates indicate the closest point to $\bx^*$ on the element boundary. This constraint is crucial to enable tangential relaxation during $r$-adaptivity where \fpt\ is used for closest point projection (Section \ref{sec_results_sliding}).


\subsection{General field evaluation using computational coordinates} \label{sec_method_eval}
Solution evaluation is straightforward after the computational coordinates $\bq_i$, $i=1\dots N_{pt}$, have been determined for each point $\bx_i$.
The solution $u(\bx(\br))$ in \eqref{eq_u_tensor} is first evaluated for each point that overlaps an element on the same MPI rank that queried the point, $u(\br^*_i)\vert_{\Omega^{e_i^*}}$.
Next, points that overlap elements on other ranks are exchanged using MPI. The solution is then evaluated locally for all the received points, and these values are returned.

Due to the proposed methodology, the \emph{Setup} step in Section \ref{sec_method_precomp} needs to be done only once for a given mesh. Then, the \emph{Find} step in Section \ref{sec_method_find} is used to find the computational coordinates of any given set of points. Similarly, if multiple fields (e.g., velocity and temperature for Navier-Stokes) need to be evaluated at a given set of points, the \emph{Find} step is done only once and the returned computational coordinates are re-used.

\section{Extension to Surface Meshes} \label{sec_method_surf}

\begin{figure}[bt!]
  \begin{center}
  $\begin{array}{cc}
  \includegraphics[width=0.3\linewidth]{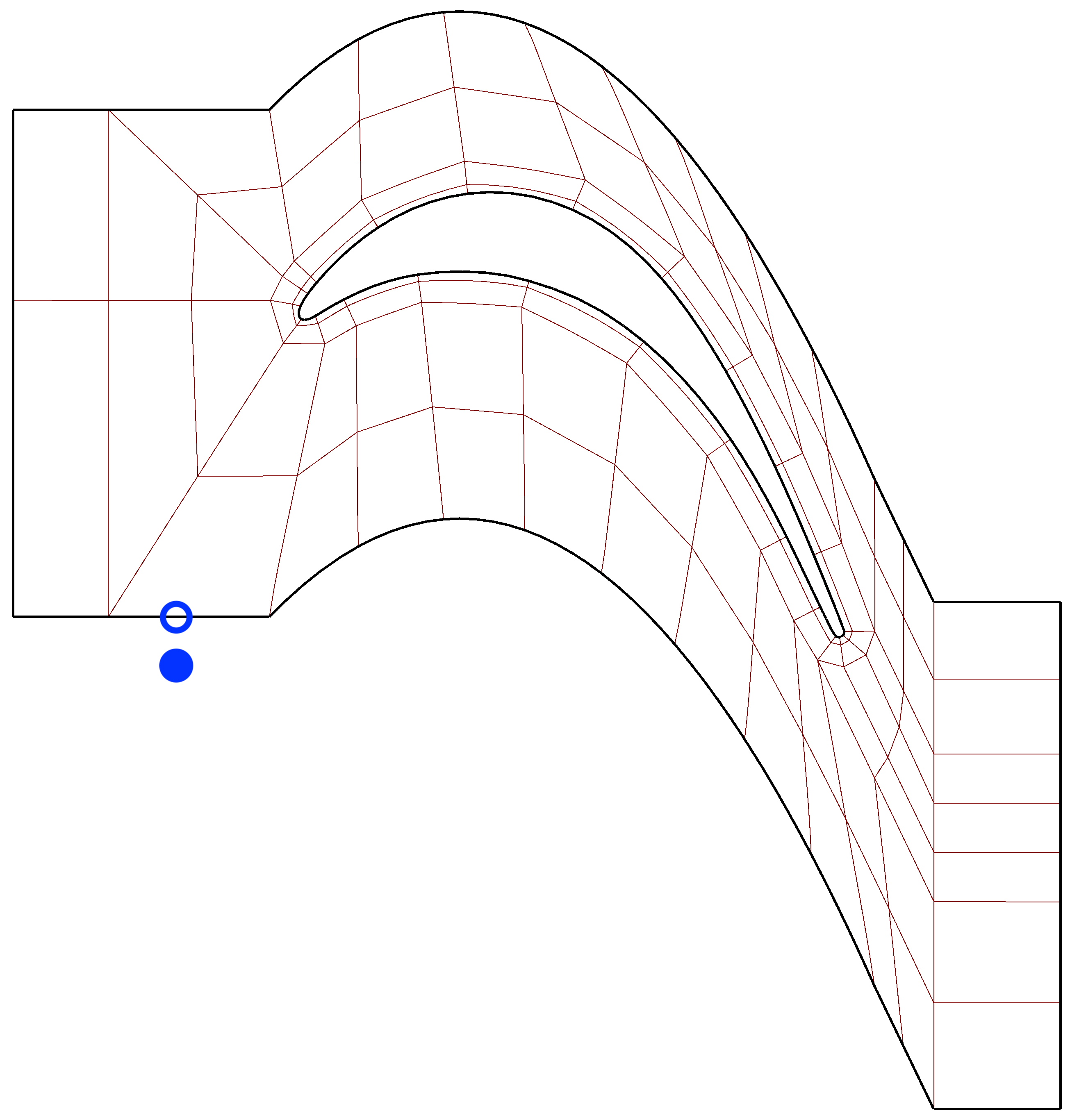} &
  \includegraphics[width=0.3\linewidth]{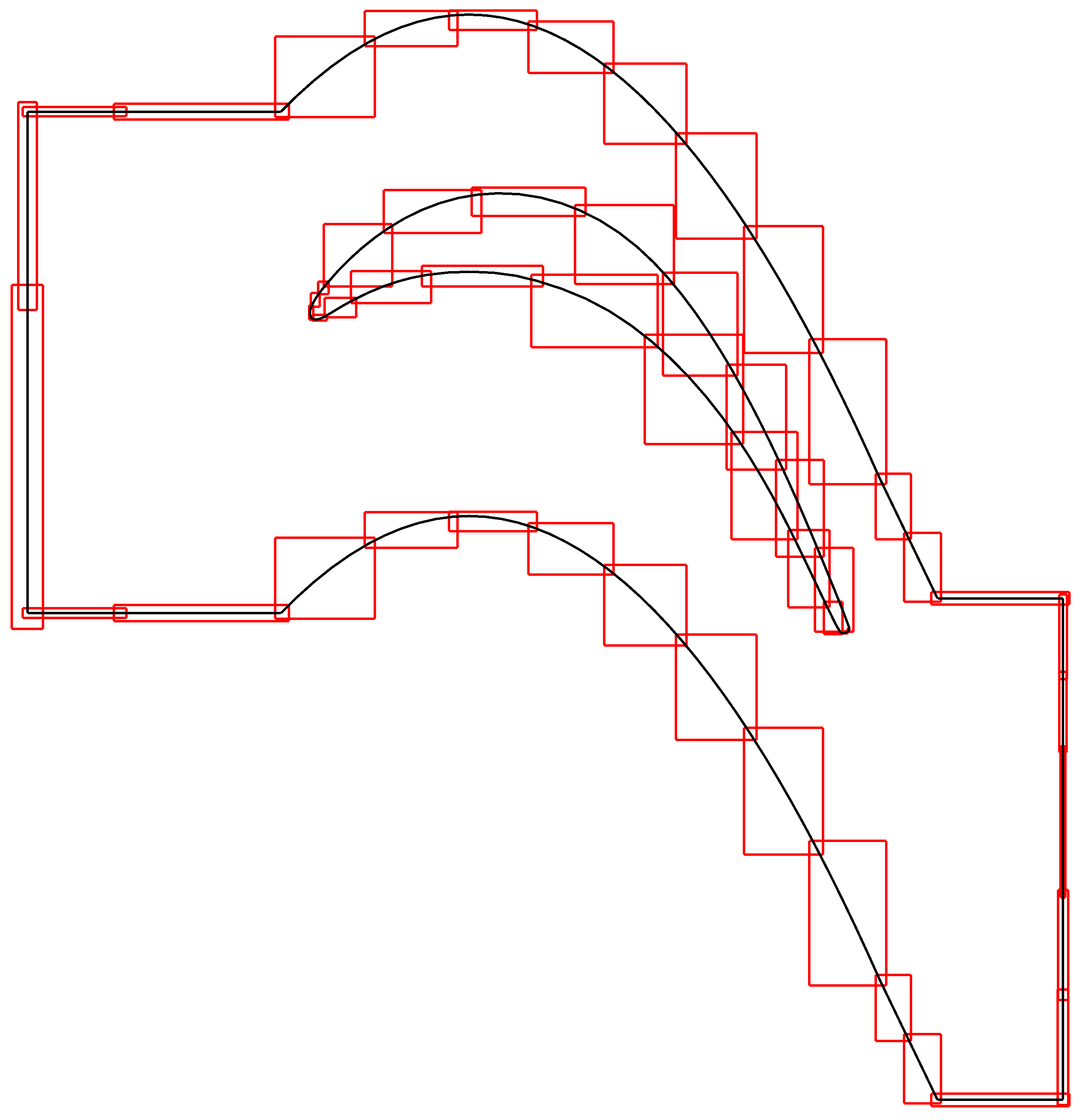} \\
  \textrm{(a)} & \textrm{(b)} \\
  \includegraphics[width=0.3\linewidth]{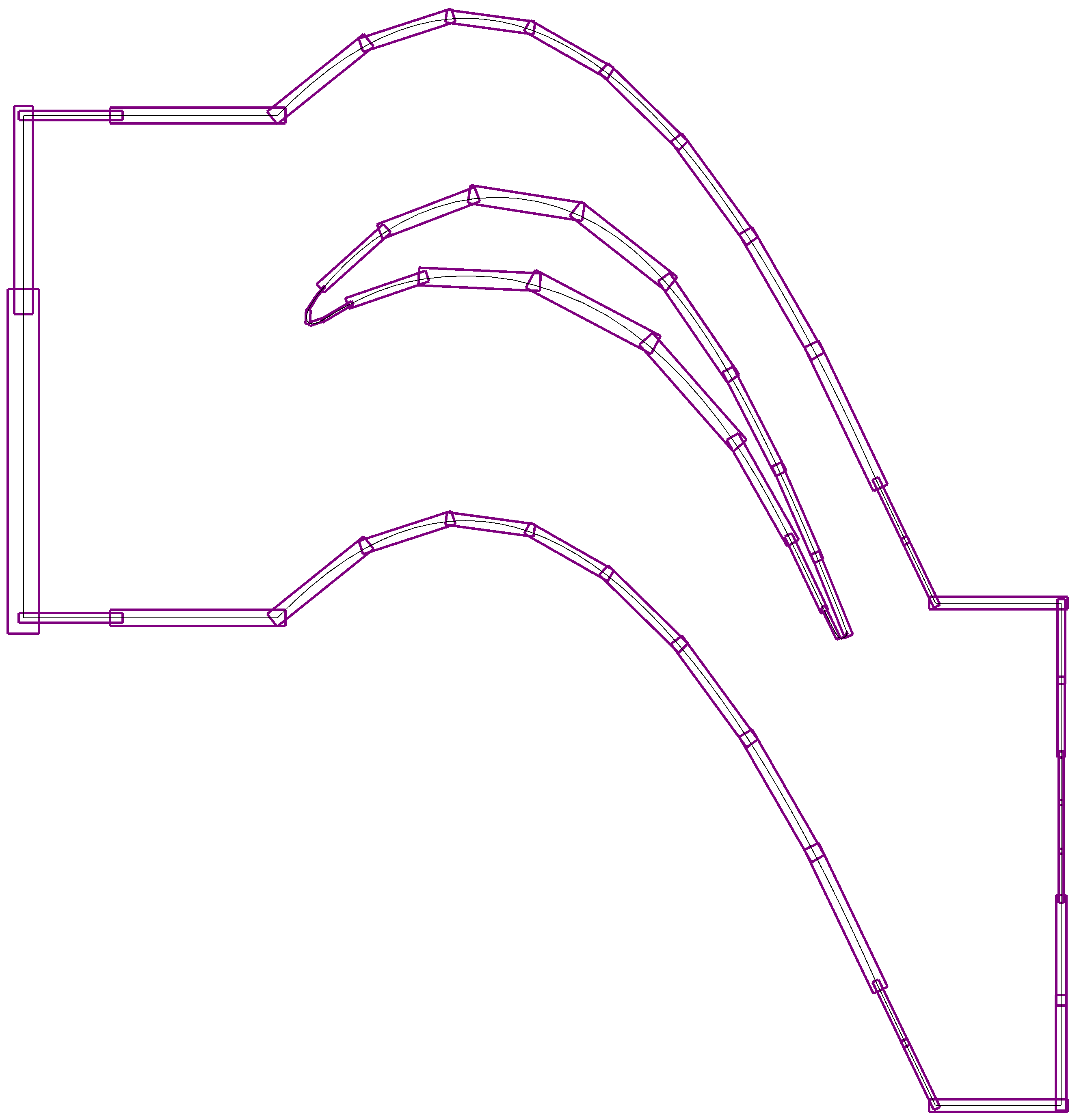} &
  \includegraphics[width=0.3\linewidth]{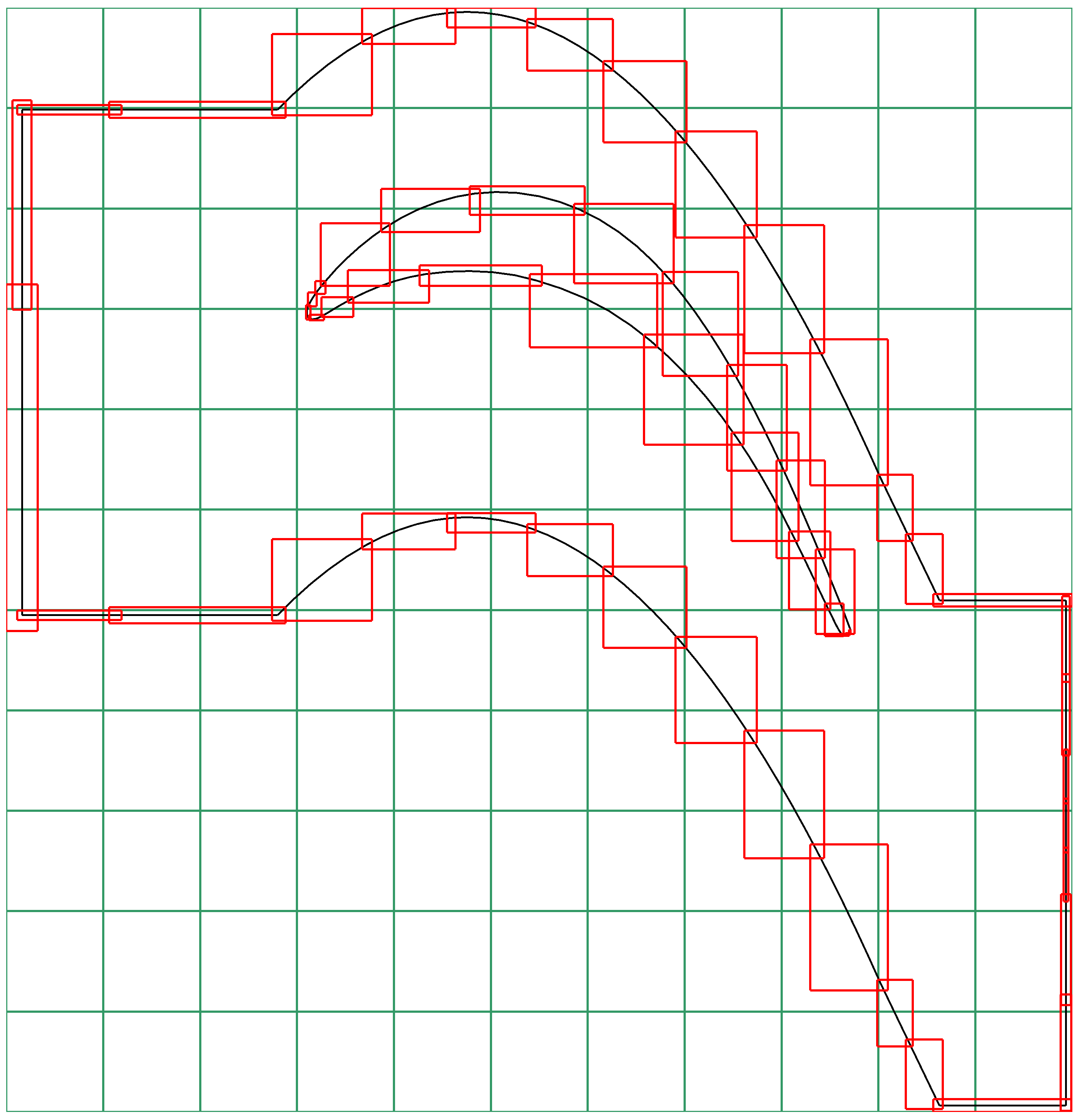} \\
  \textrm{(c)} & \textrm{(d)}
  \end{array}$
  \end{center}
  \vspace{-4mm}
  \caption{(a) Quartic surface mesh (solid black) extracted from a mesh for analyzing fluid flow over a turbine blade. (b) AABB and (c) OBB of the surface mesh elements. (d) Cartesian aligned mesh used to generate the local map $\Psi_L$. In (a), \scalebox{0.5}{{\color{blue}\faCircle}} represents a point outside the mesh and \scalebox{0.5}{{\color{blue}\faCircleO}} is the closest point found in the \emph{interior} of the surface mesh element.}
\label{fig_surface_blade}
\end{figure}

The extension of the proposed methodology to surface meshes presents some interesting challenges.
First, the axis-aligned bounding box of a surface element can have a zero-extent in one of the directions, which makes the proposed AABB test \eqref{eq_AABB_test} susceptible to false negatives due to rounding errors.
Second, the inverse of the Jacobian of the transformation at the center of the element is used in the computation of OBB of a volume element. This approach does not readily extend to surface meshes.
Finally, the way the code $c^*$ is set to indicate whether $\bx^*$ was found inside an element or on its boundary, does not directly extend from volume meshes. We will address these challenges in this section and demonstrate the proposed solution using a quartic surface mesh in Figure \ref{fig_surface_blade}.
This example mesh is extracted from a 2D mesh constructed to analyze flow over the GE LPT-106 turbine blade.

\subsection{Axis-aligned bounding box for surface meshes} \label{sec_method_surf_aabb}

For a surface element, the AABB is determined as for volume elements (Section \ref{sec_method_precomp_aabb}) by computing the bounds on the nodal position function (e.g., $x(\br)$ and $y(\br)$).
When a geometrically linear surface element is aligned with the Cartesian axes, one of the nodal position function will have the same minima and maxima. For example, the surface elements at the left inflow boundary of the turbine blade are aligned with the $y$-axis, and thus the bound on $x(\br)$ is constant, i.e. $x_{\tt min} = x_{\tt max}$.
In such cases, the bounding box in extended ($10\%$ by default) in the zero-extent direction proportional to the smallest extent in other directions. Figure \ref{fig_surface_blade}(b) shows the AABB obtained for the elements of the surface mesh, with AABBs extended for Cartesian-aligned elements in the orthogonal direction.

\subsection{Oriented bounding box for surface meshes} \label{sec_method_surf_obb}

For surface elements, the Jacobian of the transformation gives the tangent vectors to the surface at a given point. We use these tangent vectors at the center of the element ($\br=0$) to infer the orientation of the element and align it with the Cartesian axes. Consequently, the transformation $J_c$ used to transform $\check{\Omega}$ to $\tilde{\Omega}$ in Section \ref{sec_method_precomp_obb} and Figure \ref{fig_OBB_transformation} is replaced with a composite transformation $R$ that aligns the tangent vectors with the Cartesian axes.

For line elements embedded in 2D and 3D space, $R$ is simply determined using Rodrigues' formula based on the angle between the tangent vector ($\partial x_i/\partial r$, $i=1\dots D$) and the $x$-axis.
For quadrilateral elements in 3D space, we first compute the normal vector to the surface using the cross-product of the two tangent vectors: $t_1 = \partial \bx/\partial r$, $t_2 = \partial \bx/\partial s$, $n = t_1 \times t_2$. Then, $R=R_2 R_1$ is a combination of two transformations: $R_1$ is the rotation matrix that aligns the normal vector $n$ with the $z$-axis, and $R_2$ is a planar transformation that aligns the rotated tangent vectors ($R_1t_1$ and $R_1t_2$) with the $xy$-axes,
$R_2 = [(R_1 t_1)^T \,\,(R_1 t_2)^T\,\, e_3]^{-1}$, where $e_3=[0,0,1]$.
Figure \ref{fig_surface_blade}(c) shows the OBB obtained using the proposed methodology for the quartic surface mesh.

\subsection{Local and global maps for $\bx^*$ to $\ue^*$ and $\um^*$} \label{sec_method_surf_maps}

Figure \ref{fig_surface_blade}(d) shows the Cartesian mesh $\mm_L$ used to generate the process-local map $\Psi_L$ when the entire surface mesh is located on a single mpi rank.
The procedure described for the maps $\Psi_L$ (Section \ref{sec_method_precomp_local_map}) and $\Psi_G$ (Section \ref{sec_method_precomp_global_map}) for volume meshes is directly applicable to surface meshes as it relies only on AABBs.
Using this existing procedure allows us to complete the \emph{Setup} step for surface meshes with minimal changes, but it is clear from Figure \ref{fig_surface_blade}(d) that the resulting maps can be suboptimal with a lot of elements of $\mm_L$ and $\mm_G$ not intersecting any of the surface elements. In future work, we will explore ways to construct more efficient maps for surface meshes, for example using techniques such as bounding volume hierarchies \cite{novak2012rasterized}.

\subsection{Criterion for setting $c^*$} \label{sec_method_surf_newton}

The Newton's method described in Section \ref{sec_method_newton} remains virtually unchanged for surface meshes. The only change required is in the criterion for setting $c^*$ to indicate whether a point is inside an element or on its boundary. For volume elements, we set $c^*={\tt INTERIOR}$ only when $\br^* \in (-1, 1)^D$. This approach does not work for surface elements; see Figure \ref{fig_surface_blade}(a) for an example of a point that is located outside the mesh and Newton's method finds the closest point in the \emph{interior} of a surface element. To address this issue, we set $c^*={\tt INTERIOR}$ only when $\br^* \in (-1, 1)^{D_r}$ and the distance $d^*=\|\bx^*-\bx(\br^*)\|_2^2$ is less than a threshold $\epsilon_{d}$. Here, $\epsilon_d$ can optionally be an absolute threshold (e.g., $\epsilon_d = 10^{-10}$) or based on the element size (e.g., $\epsilon_d = 10^{-10} |\Omega^{e^*}|$).

The proposed changes in this section enable field evaluation at arbitrary points on high-order surface meshes. The effectiveness of this approach is demonstrated in Section \ref{sec_results_surf} and Section \ref{sec_results_sliding}.

\section{GPU Implementation} \label{sec_gpu}

In the proposed implementation \cite{fischer2022nekrs,MFEM2024}, the MPI communication intensive \emph{Setup} step is executed on CPUs and the compute intensive \emph{Find} and function evaluation steps are on GPUs. In this section, we describe the key concepts of the GPU implementation in the context of volume meshes as they similarly extend to surface meshes.

At the beginning of the \emph{Find} step, element-wise bounding boxes, the local map based on $\mm_L$, and physical space coordinates of points to find, are moved from the host (CPU) to device (GPU).
Then, 1 thread block is allocated per $N_{pt}$ point to find, with $N\cdot D$ threads per thread block. This approach maps well to the tensor product decomposition of the bases \eqref{eq_x_tensor} as computation of the Newton functional and its derivative typically requires summation over $N$ terms in each of the $D$ space dimensions.

In 2D for example, the gradient of the transformation \eqref{eq_jacobian} required for the Jacobian and Hessian of the Newton functional is:
\begin{eqnarray}
  \label{eq_jacobian_2D_first_row}
  \frac{\partial x}{\partial r} = \sum_{i=1}^{N} \sum_{j=1}^{N} x_{ij}^e \frac{\partial \phi_i(r)}{\partial r} \phi_j(s),\,\,\,\,
\frac{\partial y}{\partial r} = \sum_{i=1}^{N} \sum_{j=1}^{N} y_{ij}^e \frac{\partial \phi_i(r)}{\partial r} \phi_j(s) \\
\label{eq_jacobian_2D_second_row}
\frac{\partial x}{\partial s} = \sum_{i=1}^{N} \sum_{j=1}^{N} x_{ij}^e \phi_i(r) \frac{\partial \phi_j(s)}{\partial s},\,\,\,\,
\frac{\partial y}{\partial s} = \sum_{i=1}^{N} \sum_{j=1}^{N} y_{ij}^e \phi_i(r) \frac{\partial \phi_j(s)}{\partial s},
\end{eqnarray}
where the basis function and its derivatives are computed at $\br_l$.
From the available $2N$ threads, each of the first $N$ threads compute $\phi_i(r)$ and $\partial \phi_i(r)/\partial r$ for a single $i$, and each of the next $N$ threads compute $\phi_j(s)$ and $\partial \phi_j(s)/\partial s$ for a single $j$. Next, the first $N$ threads effect the inner summation corresponding to a fixed $i$ for both terms of \eqref{eq_jacobian_2D_first_row}, and the next $N$ threads similarly effect the inner summation for \eqref{eq_jacobian_2D_second_row}. Finally, one of the threads accumulates the outer summation required to determine $\partial x_i/\partial r_j$.
Similarly, the work required for computation of the quadratic term in the Hessian ($\partial^2 x_i/\partial r_j \partial r_k$) is distributed across the $N \cdot D$ threads.

Where possible, \emph{shared} memory with coalesced access is used to perform all the work efficiently. After the \emph{Find} step, the computational coordinates stored in an array of length $(4+d)N_{pt}$ are moved back from the device to host.

The function evaluation step similarly utilizes a thread block for each of the $N_{pt}$ points with $N^d$ threads. Every thread then evaluates one of $N^d$ bases at $\br^*$ in $\Omega^{e^*}$, followed by aggregation by thread 0 (see e.g., \eqref{eq_u_tensor}) to determine the interpolated value $u(\bx^*)$. The GPU kernels are available open source in the folder \code{fem/gslib/} of MFEM \cite{mfem-github} and in \code{kernels/findpts} folder of NekRS \cite{nekrs-github,fischer2022nekrs}.

\noindent
\textbf{Remark on CPU implementation:} An important distinction between the GPU and CPU implementation of the \emph{Find} and \emph{Interpolate} step is that the GPU implementation leverages multithreading where each thread block is responsible for the computation and memory read/writes associated with a single point. In contrast, the CPU implementation does not use multithreading, and we first group the points by candidate elements and then perform the computation for all the points in an element simultaneously. This approach significantly reduces the memory bandwidth as the geometry of an element is only fetched into memory once. The reader is referred to the open-source implementation of \emph{findpts} on Github for further details \cite{gslib-github}.

\section{Results \& Applications} \label{sec_results}

This section demonstrates the robustness of the proposed approach using various examples and applications. The first three numerical experiments are done on the Lassen supercomputer at the Lawrence Livermore National Laboratory, which has IBM Power9 CPUs (792 nodes with 44 CPU cores per node) and NVIDIA V100 GPUs (4 per node). All reported computations utilize a single machine node. The CPU runs use 4 CPU cores, while the GPU runs utilize 4 CPU cores with 1 GPU per core, unless otherwise stated.  Note that prior to this work, applications leveraging \fpt\ did their PDE-related computation on GPU followed by \fpt\ on CPU. The following examples demonstrate that the specialized GPU kernels for \fpt\ significantly reduce time for the \emph{Find} and \emph{Interpolate} steps.

\subsection{General field evaluation on a 9th order spiral element} \label{sec_results_spiral}

In this first example, we consider a single 9th order spiral element in 3D on 1 CPU core and 1 GPU. Note that 9th order elements are rarely used in practice, and the purpose of this example is to highlight the robustness and speed of the proposed method for a high-order element.

\begin{figure}[bt!]
  \begin{center}
  $\begin{array}{cc}
  \includegraphics[width=0.4\linewidth]{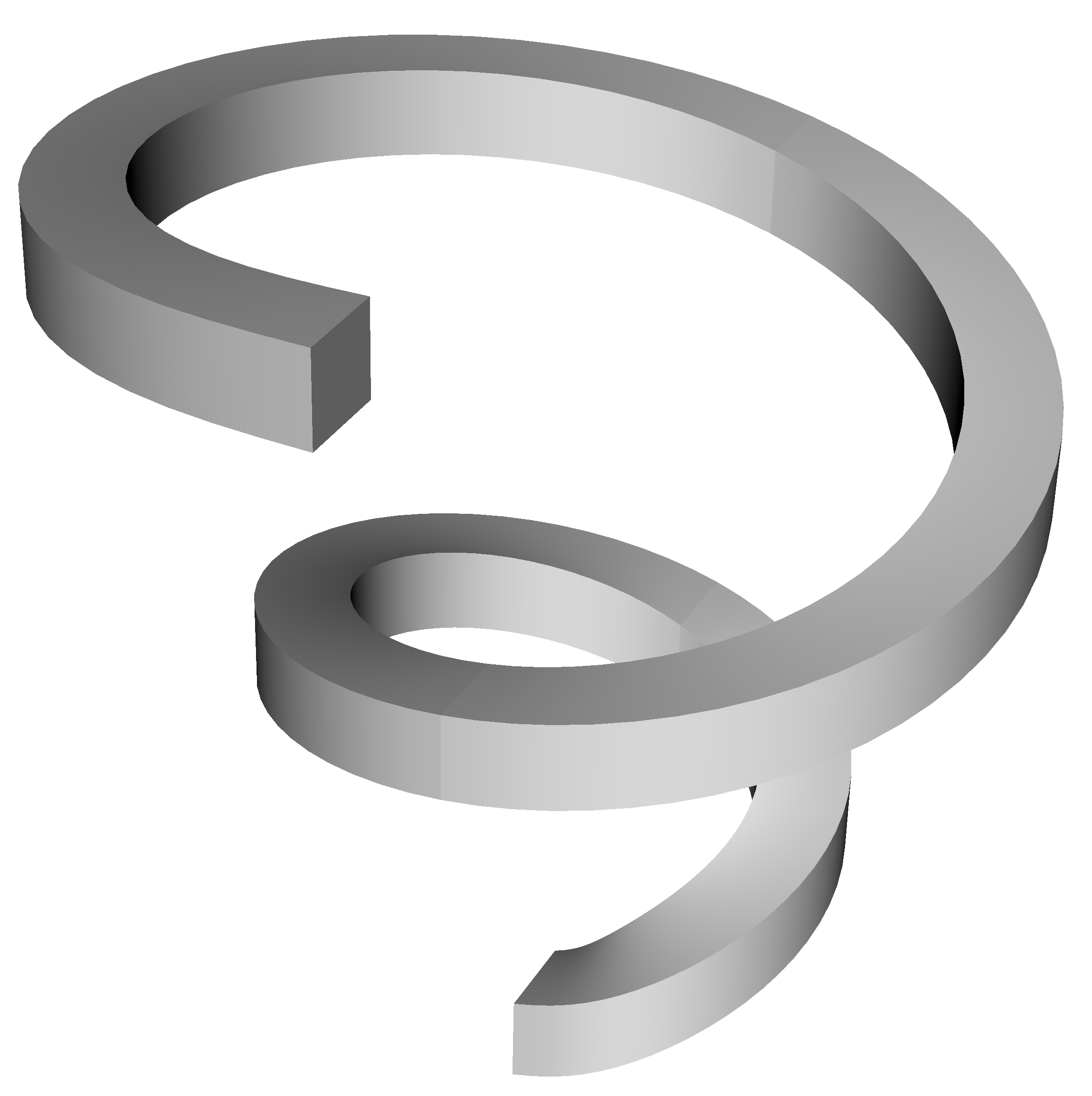} &
  \includegraphics[width=0.5\linewidth]{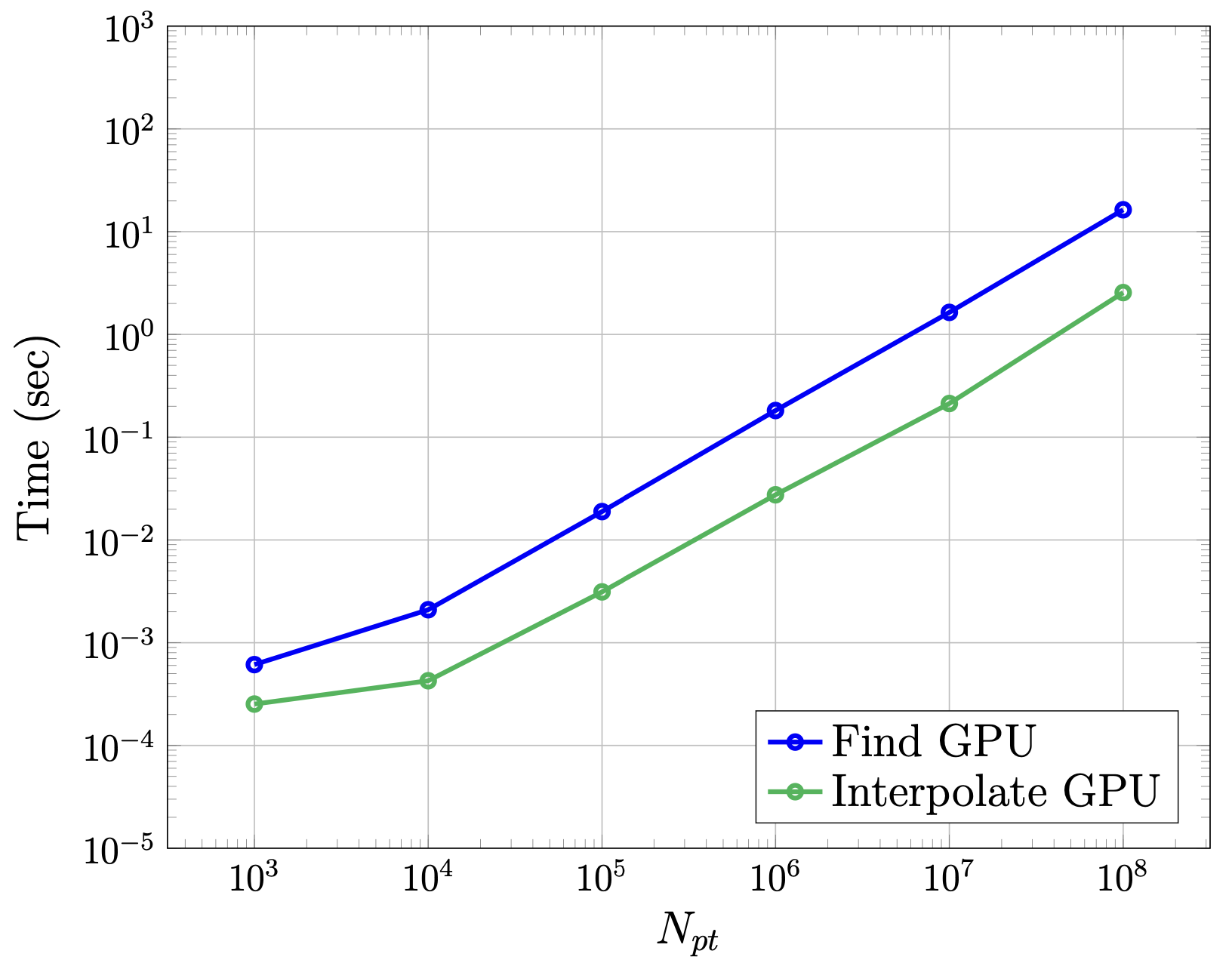}
  \end{array}$
  \end{center}
  \vspace{-4mm}
  \caption{(left) 9th-order spiral element and (right) time to find and interpolate a known function at $N_{pt}=10^3$ to $10^8$ points.}
\label{fig_spiral}
\end{figure}

We randomly distribute $N_{pt}$ points in the element and use the Newton's method to determine the reference space coordinates for each point. Then, a known function is evaluated at the returned coordinates to verify the accuracy of the method.
Figure \ref{fig_spiral} shows the spiral element and the time to determine the reference space coordinates for $N_{pt}=10^3$ up to $10^8$ points. We observe the efficiency of the proposed method as it takes less than 1 millisecond to find a 1000 points, about 0.1 second to find 1 million points, and about 16 seconds to find 100 millions points. On average, it takes only 5 iterations per point to find the reference space coordinates such that the function is evaluated at each point with machine precision accuracy.

\subsection{General field evaluation for the triple point problem} \label{sec_results_triplept}

\begin{figure}[bt!]
  \begin{center}
  $\begin{array}{cc}
  \includegraphics[width=0.4\linewidth]{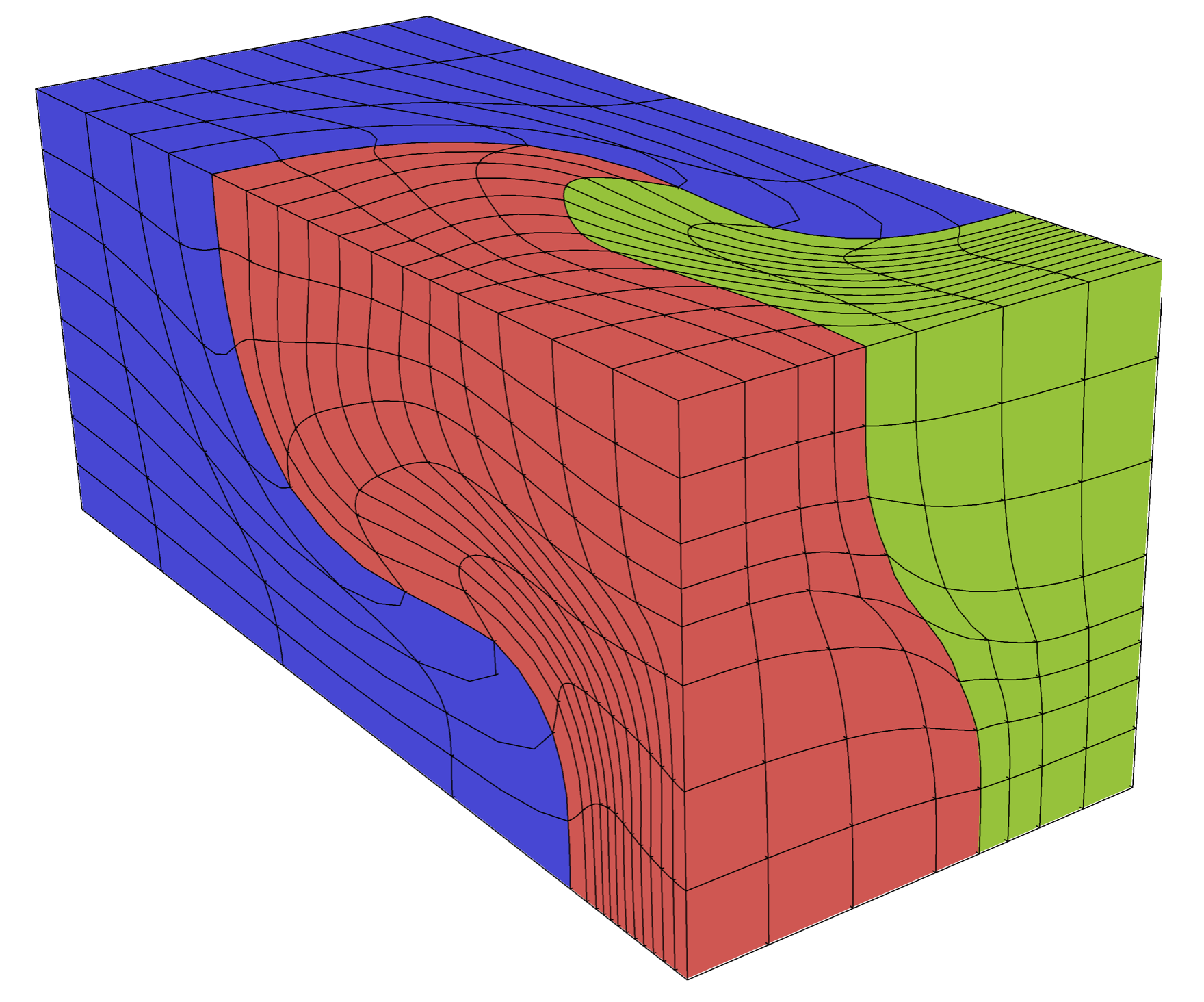} &
  \includegraphics[width=0.5\linewidth]{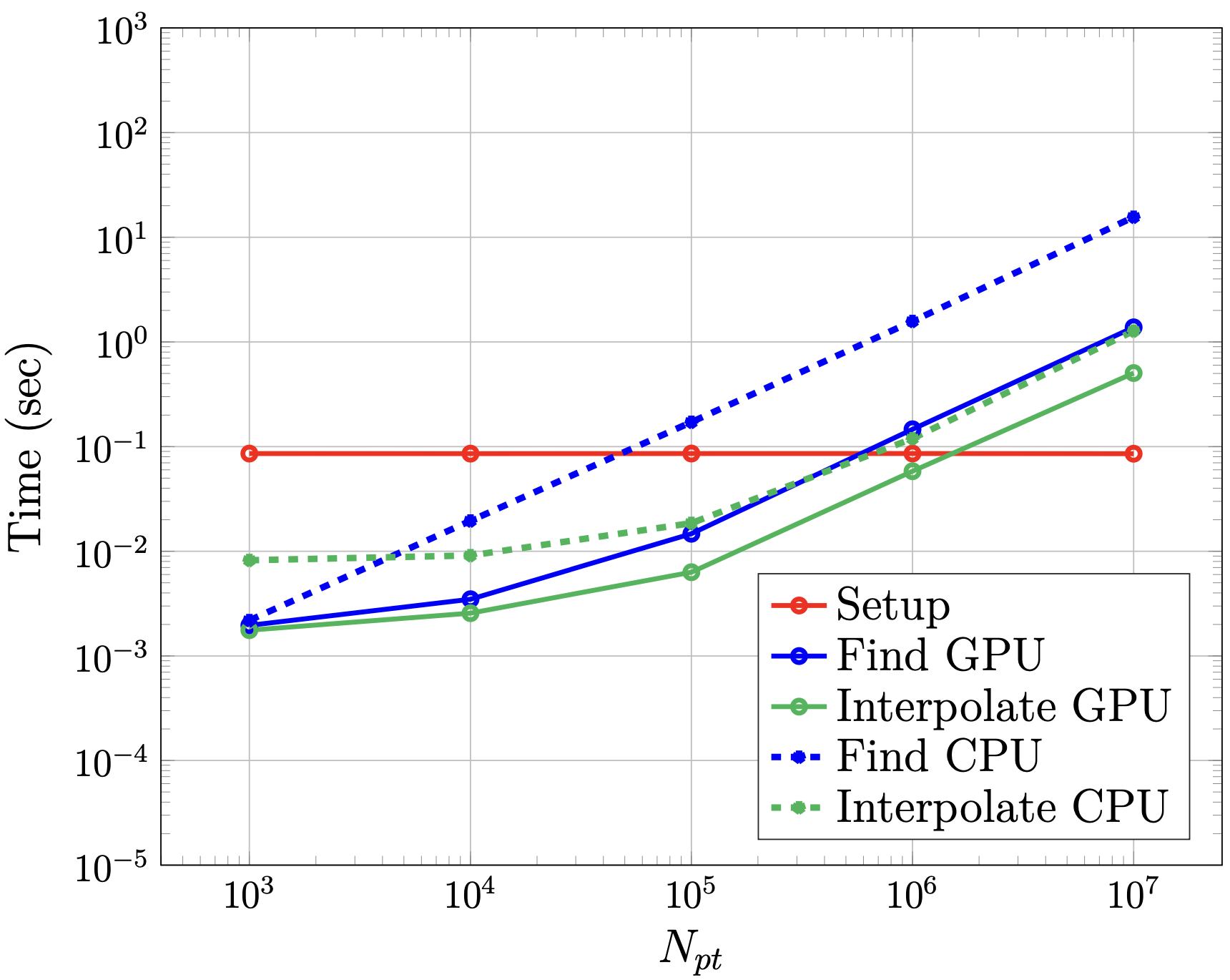}
  \end{array}$
  \end{center}
  \vspace{-4mm}
  \caption{(left) Cubic mesh ($N_E=1024$) for the triple-point problem that is oct-refined to obtain a mesh with $N_E=65,536$ elements. (right) Time to find and evaluate a known function at $N_{pt}=10^3$ to $10^8$ points on CPUs versus GPUs.}
\label{fig_triplept3d}
\end{figure}

This next example is demonstrative of our target applications where \fpt\ is needed for capabilities such as Lagrangian particle tracking, $r$-adaptivity, and mesh-to-mesh transfer.
Figure \ref{fig_triplept3d} shows a coarse version of the multi-material mesh ($N_E=1024$) used for the triple point problem that is oct-refined to obtain $N_E=65,536$ elements. We also show here the time for different components of the proposed method on CPUs and GPUs. In each case, $N_{pt}=10^3$ up to $10^8$ points are randomly distributed in the mesh and then a known function is evaluated at the returned coordinates to measure the accuracy of the method.

We observe that for low number of points (e.g., $N_{pt}=1000$) the CPU version is faster than the GPU version for the \emph{Find} step. This is due to the overhead associated with data movement between the host and the device. As the number of points increases, the GPU version proves to be an order of magnitude faster than the CPU version. For $N_{pt}=10^8$, it takes about 14.8 seconds to find all the points on the GPUs versus 188.7 seconds on CPUs, and 4.8 to interpolate a cubic function on the GPUs versus 17.4 seconds on CPUs. The GPU kernels thus provides a speed-up of about 12.7 times for finding the points and 3.6 times for interpolation.

\subsection{General field evaluation on a surface mesh} \label{sec_results_surf}

The extension of the proposed method to surface meshes is required to enable a suite of capabilities (e.g., tangential relaxation along boundaries in section \ref{sec_results_sliding}) critical for our target applications. We demonstrate the effectiveness of our method here using a cubic surface mesh generated using the 2D version of the triple point problem.

\begin{figure}[bt!]
  \begin{center}
  $\begin{array}{cc}
  \includegraphics[width=0.4\linewidth]{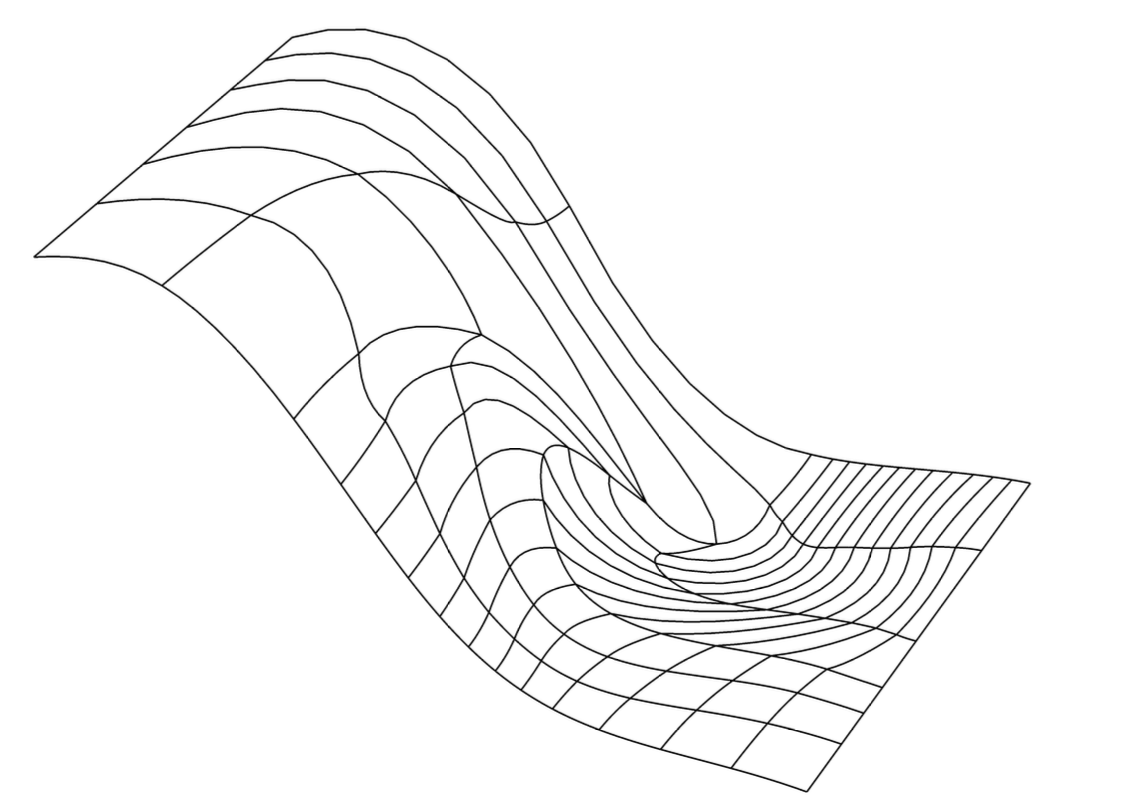} &
  \includegraphics[width=0.5\linewidth]{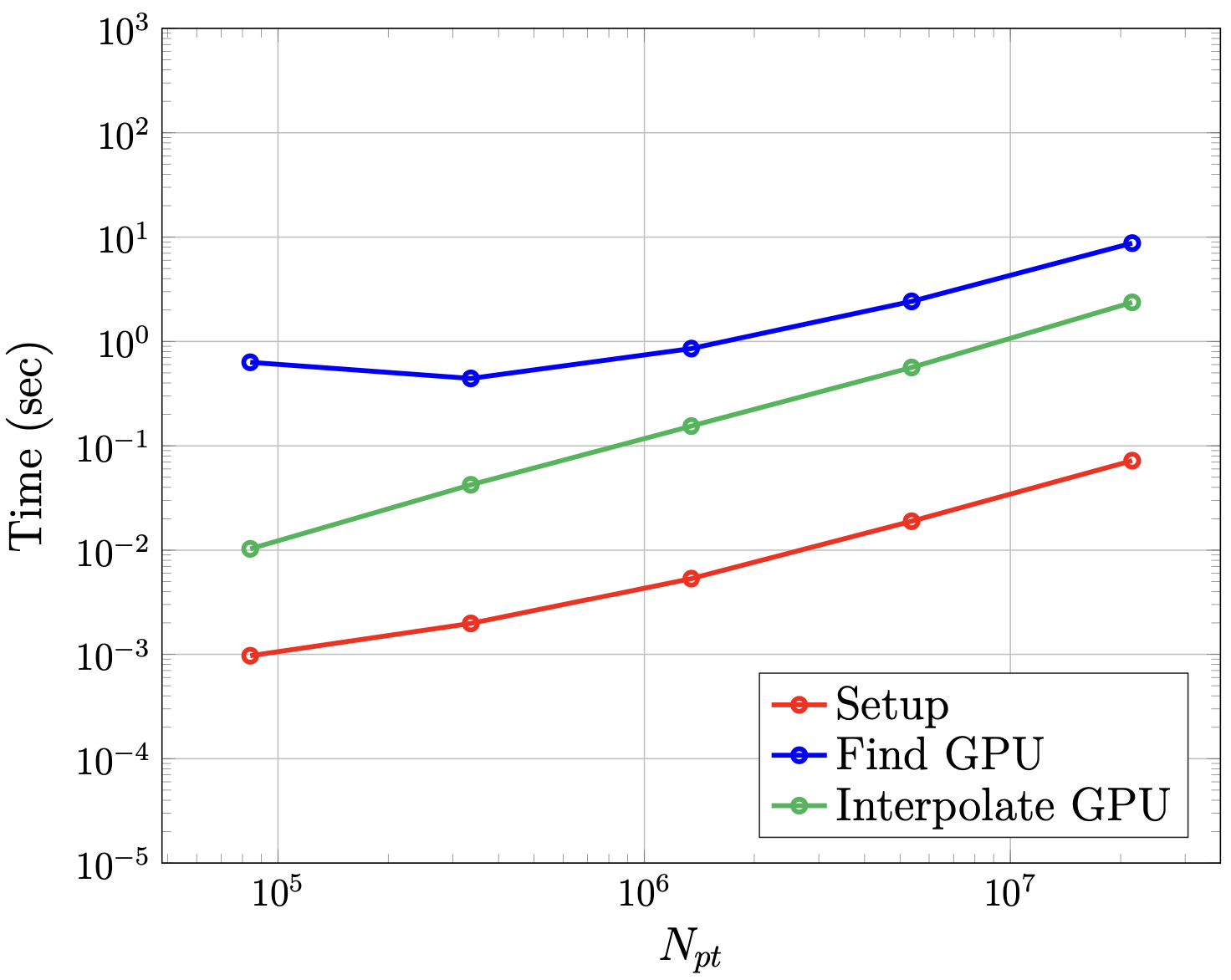}
  \end{array}$
  \end{center}
  \vspace{-4mm}
  \caption{(left) Cubic surface mesh and (right) time to find and interpolate a known function at $N_{pt}=10^3$ to $10^8$ points on GPUs.}
\label{fig_surf_triplept}
\end{figure}

Figure \ref{fig_surf_triplept}(left) shows the initial surface mesh with 84 cubic elements. We iteratively do uniform quadtree refinement on this mesh, and randomly distribute 1000 points per element for interpolation.
Figure \ref{fig_surf_triplept}(right) shows the time for different components of the proposed capability. We note that even for the case with highest resolution where $N_E=21,504$ and $N_{pt}=1000 N_E$, the proposed framework is efficient as it takes only 0.06 seconds to construct the bounding boxes and maps during the \emph{Setup} phases, 8.6 seconds to find the reference space coordinates during the \emph{Find} phase, and 2.4 seconds to evaluate a cubic function.

Note that the recent extension to surface meshes is only available through kernels designed to leverage the parallelism available on GPUs. These kernels can be executed on CPUs if needed, but are not expected to be as performant.

\subsection{Mesh-to-mesh transfer during $r$-adaptivity} \label{sec_results_radapt}

In prior work, we have proposed simulation-driven $r$-adaptivity to optimize a given mesh to the solution of the PDE and reduce the resulting discretization error \cite{TMOP2020}.
This $r$-adaptivity approach is based on the Target-Matrix Optimization Paradigm (TMOP) where a nonlinear functional dependent on the current and target geometric parameters (e.g., element size and aspect-ratio) of the mesh is minimized \cite{TMOP2019}.
For simulation-driven adaptivity, the target geometric parameters  are typically derived at each mesh node using the gradient of the discrete solution.
Since these geometric parameters are defined on the initial mesh, they must be remapped after each optimization iteration as the nodes move.
This remapping is enabled using general field evaluation capability described in the current work.

\begin{figure}[bt!]
  \begin{center}
  $\begin{array}{cccc}
  \includegraphics[width=0.25\linewidth]{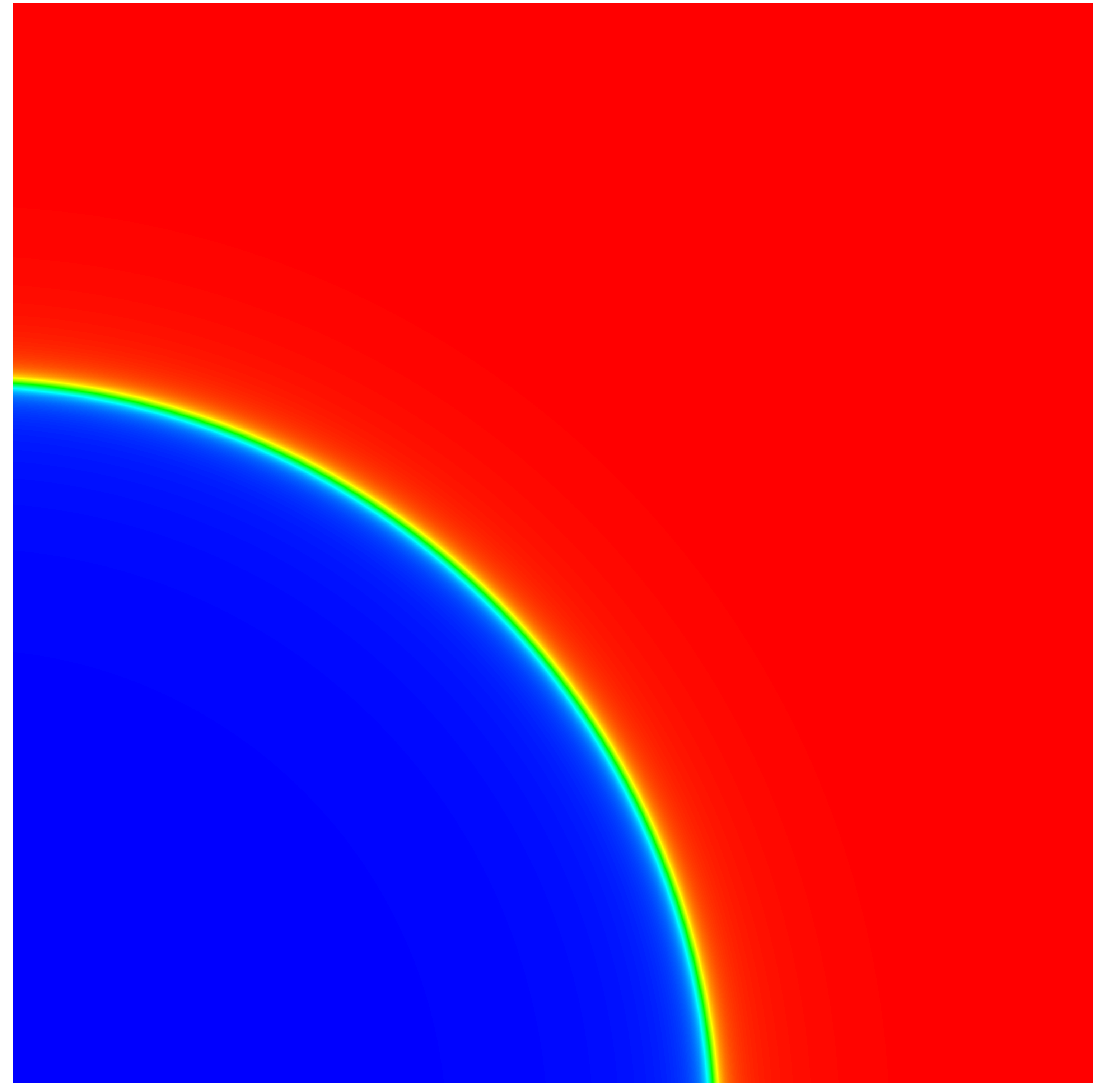} &
  \includegraphics[width=0.25\linewidth]{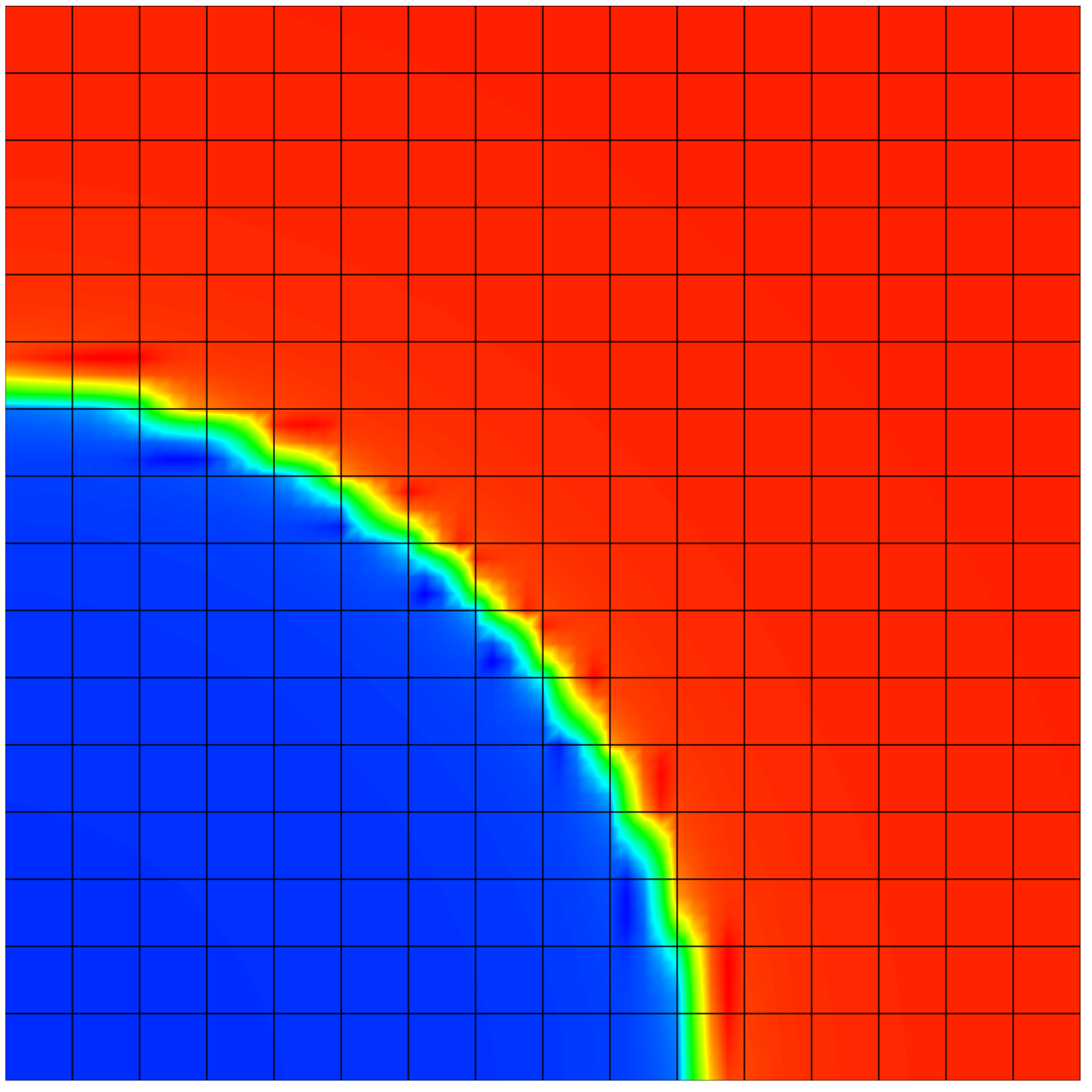} &
  \includegraphics[width=0.25\linewidth]{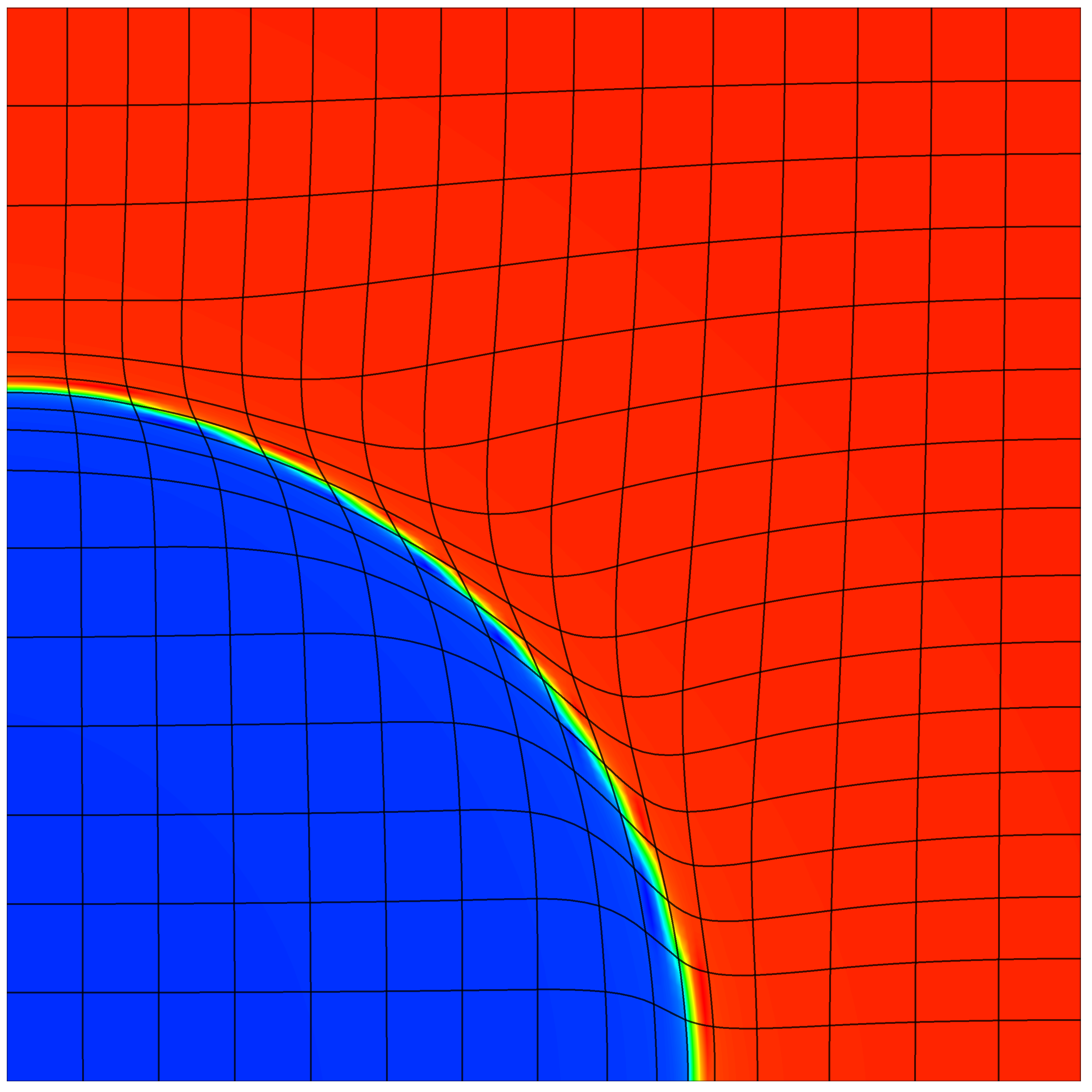} &
  \includegraphics[width=0.1\linewidth]{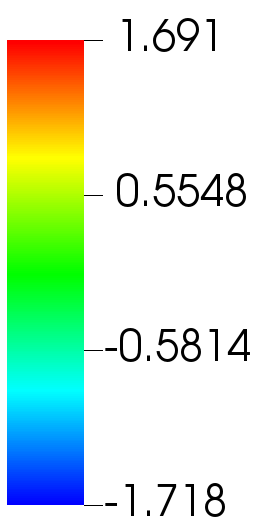}
  \end{array}$
  \end{center}
  \vspace{-4mm}
  \caption{(left) Solution mimicing a sharp circular wave front \eqref{eq_pois_u} projected on the (center) original quadratic mesh, and (right) the optimized mesh.}
\label{fig_radapt}
\end{figure}

Figure \ref{fig_radapt} shows an example of a uniform Cartesian-aligned mesh optimized to reduce the error in representing a function that mimics a sharp circular wave front,
\begin{equation}
\label{eq_pois_u}
  u(\bx) = \arctan\bigg[\alpha\bigg(\sqrt{(x-x_c)^2 + (y-y_c)^2}-r\bigg)\bigg],\,\, \bx \in [0,1]^2,
\end{equation}
where $r = 0.7$, $(x_c,y_c) = (-0.05, -0.05)$, and  $\alpha = 200$.
As evident, the optimized quadratic mesh is able to capture the sharp wave front better than the original Cartesian-aligned mesh.

\begin{figure}[bt!]
  \begin{center}
  $\begin{array}{cccc}
  \includegraphics[width=0.25\linewidth]{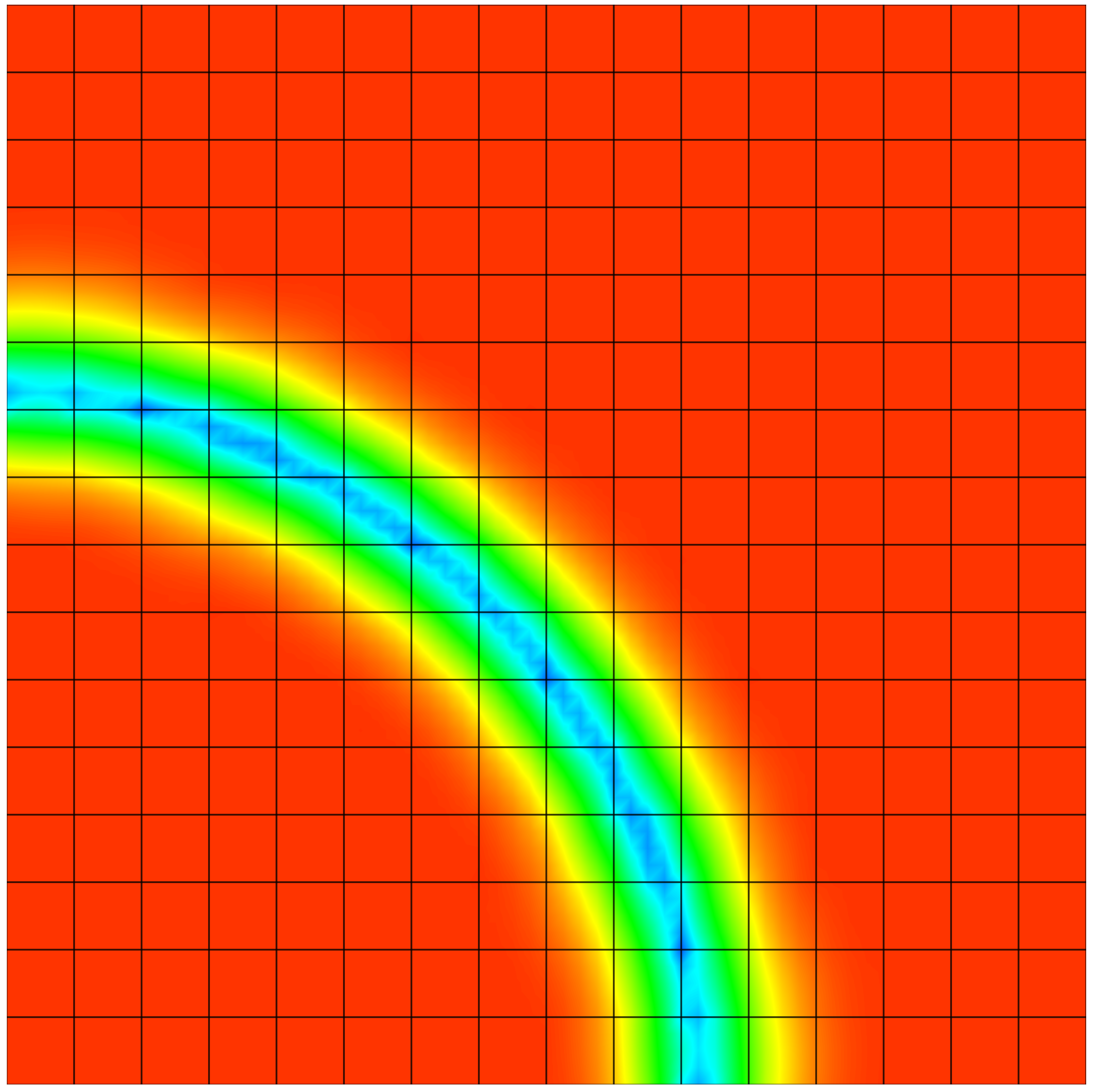} &
  \includegraphics[width=0.25\linewidth]{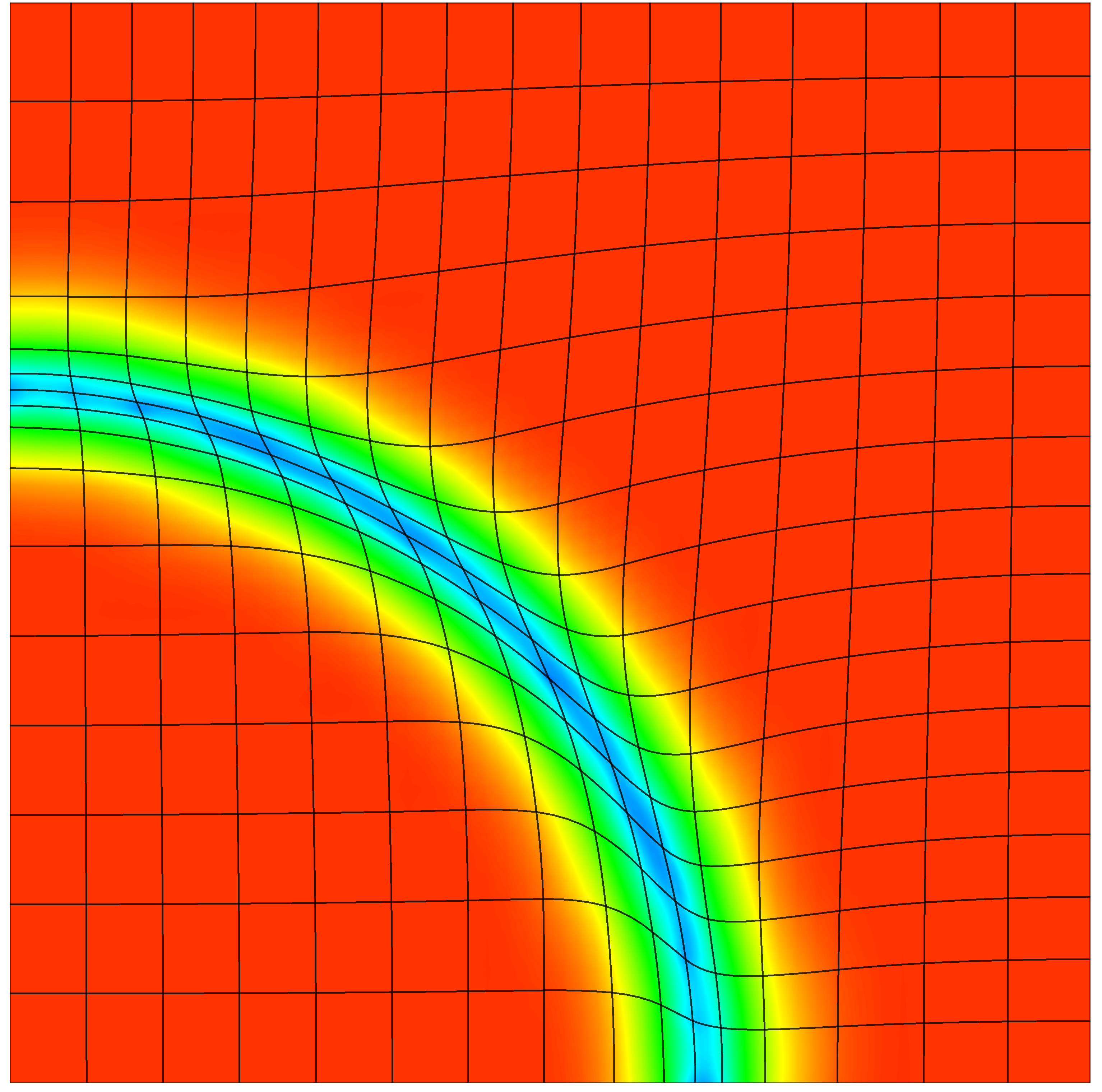} &
  \includegraphics[width=0.1\linewidth]{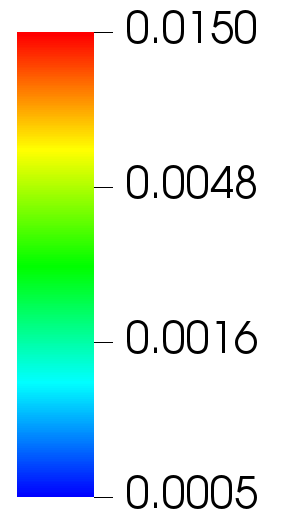}
  \end{array}$
  \end{center}
  \vspace{-4mm}
  \caption{(left) Target mesh size computed using the gradient of the solution on the original mesh. (right) The target size is mapped to the updated node positions after each iteration of $r$-adaptivity, and it is shown on the final mesh here.}
\label{fig_radapt_target}
\end{figure}

Figure \ref{fig_radapt_target} shows the target mesh size obtained using the magnitude of the gradient of the discrete solution \eqref{eq_pois_u} on the initial mesh. This target size is remapped from the initial mesh after each $r$-adaptivity iteration, and is shown on the final mesh in Figure \ref{fig_radapt_target}. Note that the mesh-to-mesh remap for $r$-adaptivity is not required to be conservative as it is solely used to drive the mesh motion.

\subsection{Closest-point projection for tangential relaxation during $r$-adaptivity} \label{sec_results_sliding}

Tangential relaxation at boundaries is important to maximize mesh quality improvement with $r$-adaptivity. However, it is not trivial to effect for curvilinear boundaries when an analytic description of the surface is not available.
We address this challenge by leveraging the methodology proposed for surface meshes in Section \ref{sec_method_surf}.

For a given area/volume mesh, we extract the surface mesh from the boundaries for tangential relaxation. Then the nodes on these boundaries are allowed to move freely during $r$-adaptivity with TMOP \cite{TMOP2019}. After each TMOP iteration, the boundary nodes are searched on the original surface mesh using the proposed method. As long as these nodes are within the bounding boxes associated with the surface mesh, the Newton's method converges to the closest point on the surface.
The boundary nodes are then moved to these closest points, and the corresponding displacement is blended into the interior of the domain using a Poisson solve \cite{mittal2019mesh}.

\begin{figure}[bt!]
  \begin{center}
  $\begin{array}{cccc}
  \includegraphics[width=0.25\linewidth]{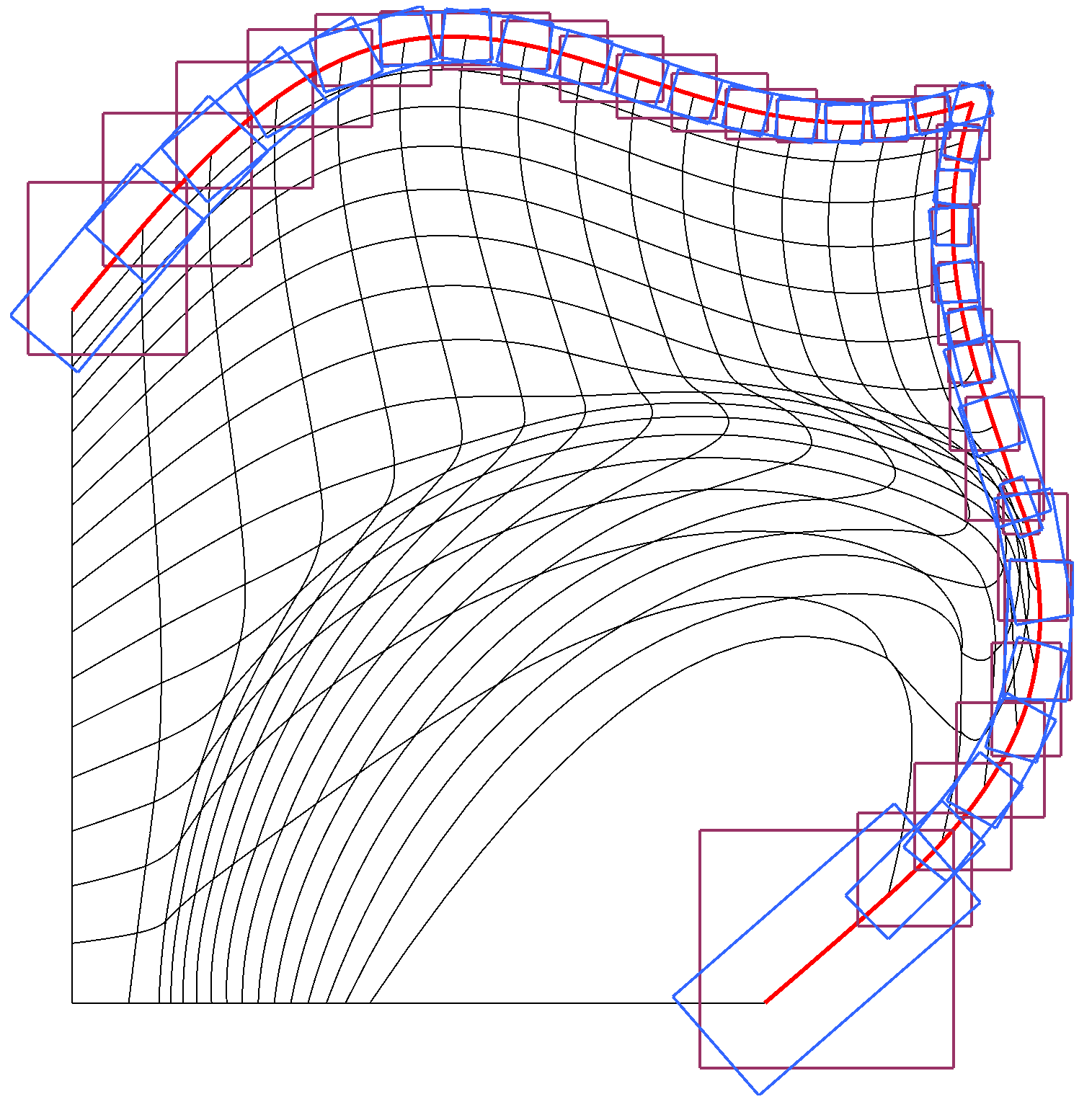} &
  \includegraphics[width=0.25\linewidth]{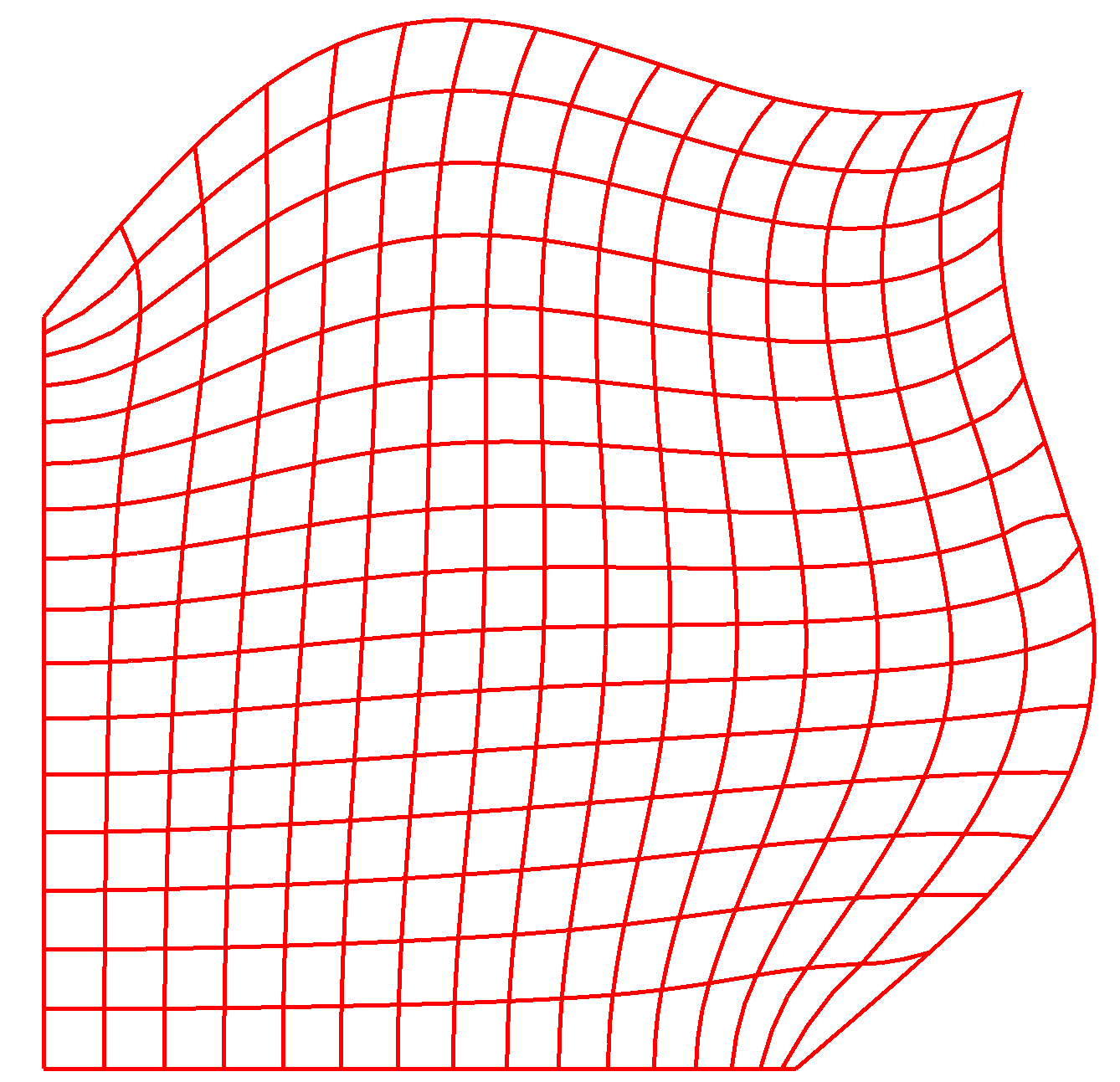} &
  \includegraphics[width=0.25\linewidth]{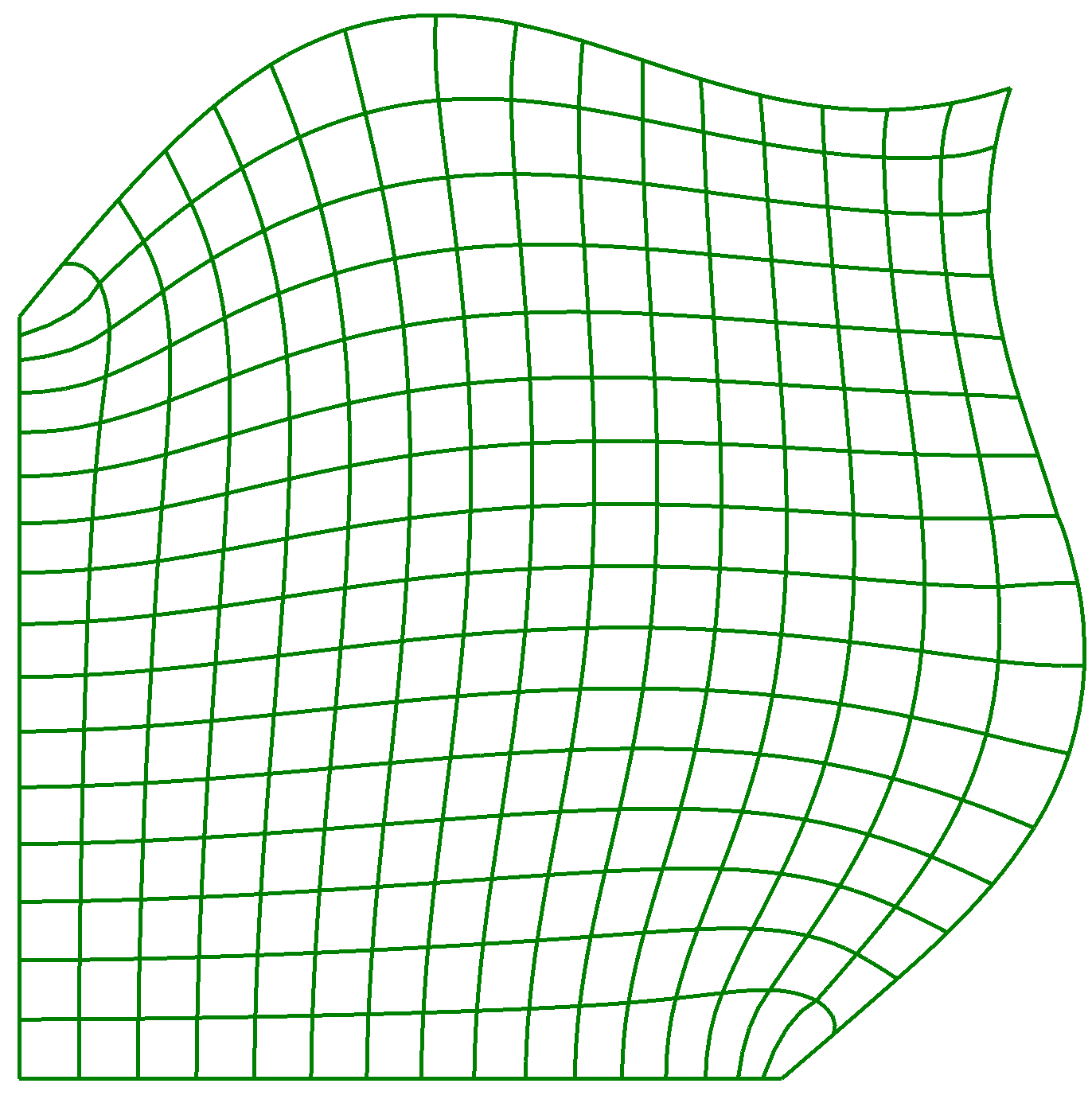} &
  \includegraphics[width=0.25\linewidth]{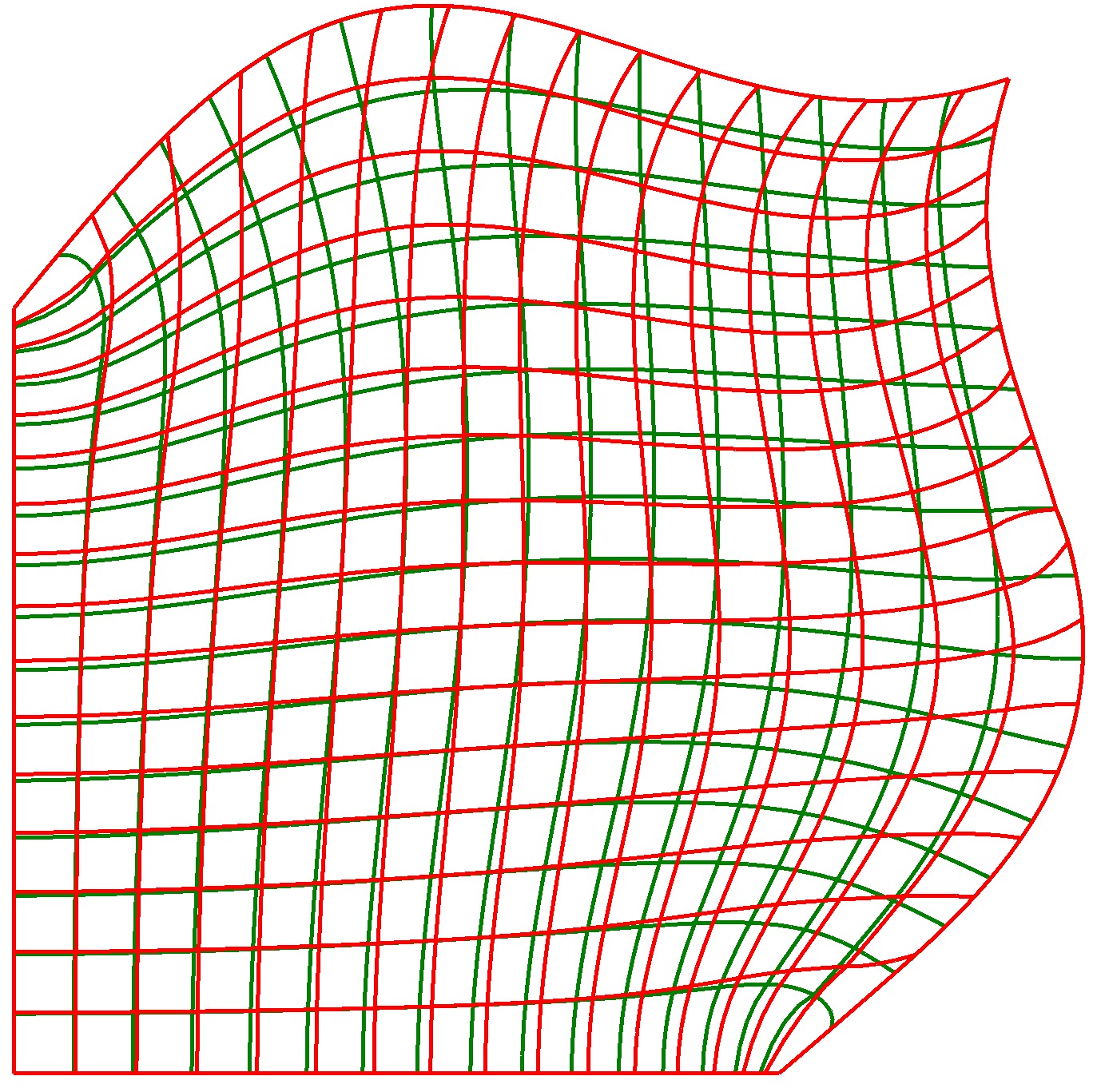} \\
  \textrm{(a)} & \textrm{(b)} &
  \textrm{(c)} & \textrm{(d)}
  \end{array}$
  \end{center}
  \vspace{-4mm}
  \caption{(a) A deformed quadratic mesh with the boundaries for tangential relaxation highlighted in red. The AABB and OBB for surface elements discretizing these boundaries are also shown. (b) Optimized mesh with nodes held fixed along the curved boundaries. (c) Optimized mesh with tangential relaxation enabled on the curved boundaries. (d) Both optimized meshes from (b) and (c) shown overlapping each other to highlight that tangential relaxation preserves the original boundary.}
\label{fig_radapt_sliding}
\end{figure}

Figure \ref{fig_radapt_sliding}(a) shows an example of a quadratic mesh deformed due to the solution of the Euler equations using the Arbitrary Lagrangian-Eulerian formulation \cite{Dobrev2012}. The surface for tangential relaxation is highlighted in red, along with the corresponding AABB and OBB for surface elements. Note that to ensure that the boundary nodes stay within the bounding boxes after each TMOP iteration, we expand the bounding boxes by $100\%$ in each direction. If the boundary nodes still move beyond these bounding boxes, we iteratively scale down the displacement produced by the mesh optimizer.
Figure \ref{fig_radapt_sliding}(b) shows the optimized mesh obtained using the TMOP-based approach where the nodes are allowed to slide on Cartesian axes-aligned boundary but remain fixed on the curved boundaries. As evident, restricting node movement on the boundary limits mesh improvement with uneven element size along the right boundary near its bottom, and kinks in element edges near its center.
Figure \ref{fig_radapt_sliding}(c) shows the optimized mesh with tangential relaxation along the curved boundaries. Both these meshes are shown overlapping each other in Figure \ref{fig_radapt_sliding}(d) for comparison. We observe that tangential relaxation preserves the original boundary and leads to better mesh quality; elements are equally spaced along the boundary, are free of kinks, and have their edges orthogonal to the boundary.

\subsection{Interdomain boundary data exchange for overlapping grids} \label{sec_results_overlapping}

\begin{figure}[bt!]
  \begin{center}
  $\begin{array}{cccc}
  \includegraphics[width=0.25\linewidth]{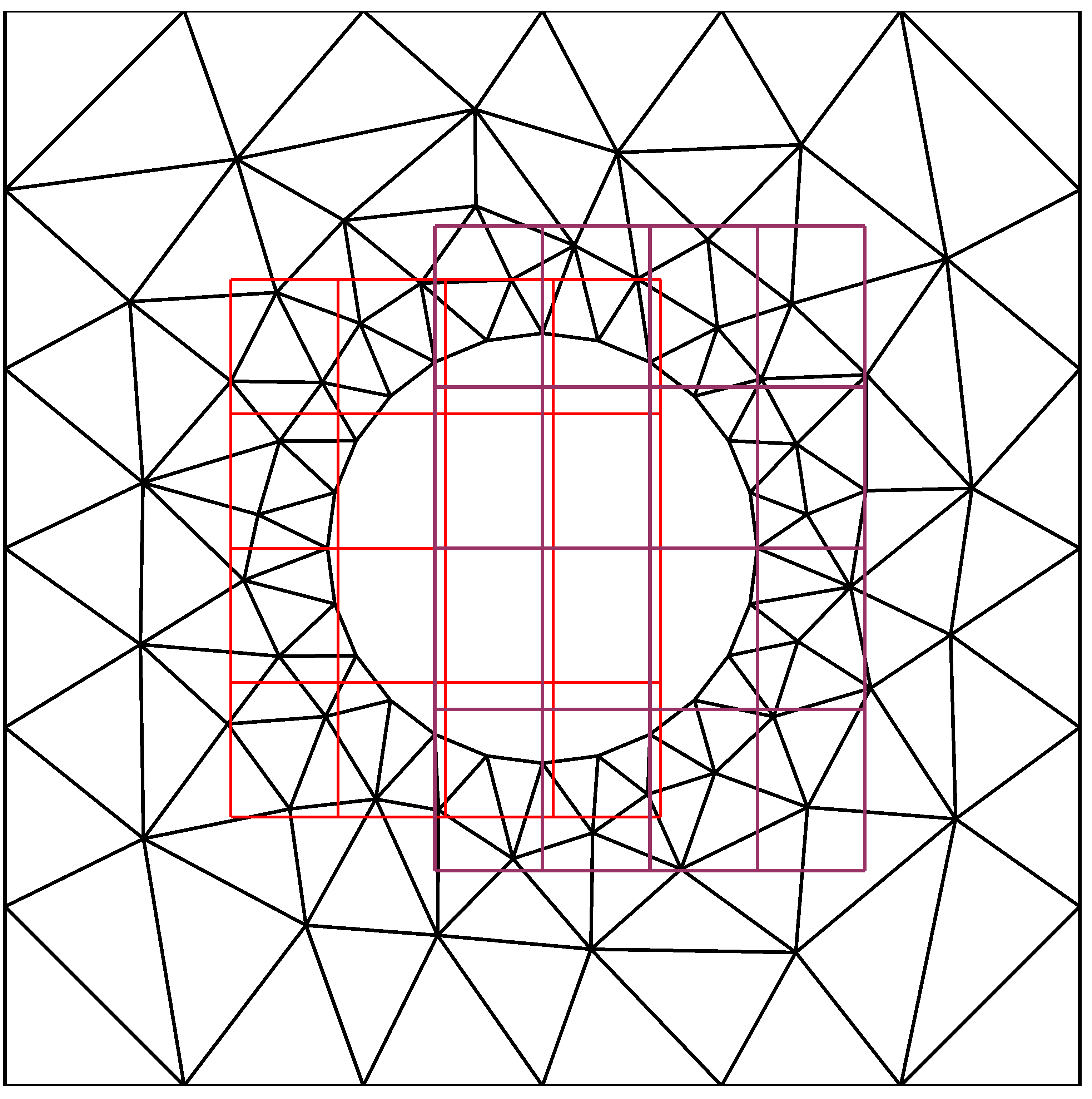} &
  \includegraphics[width=0.25\linewidth]{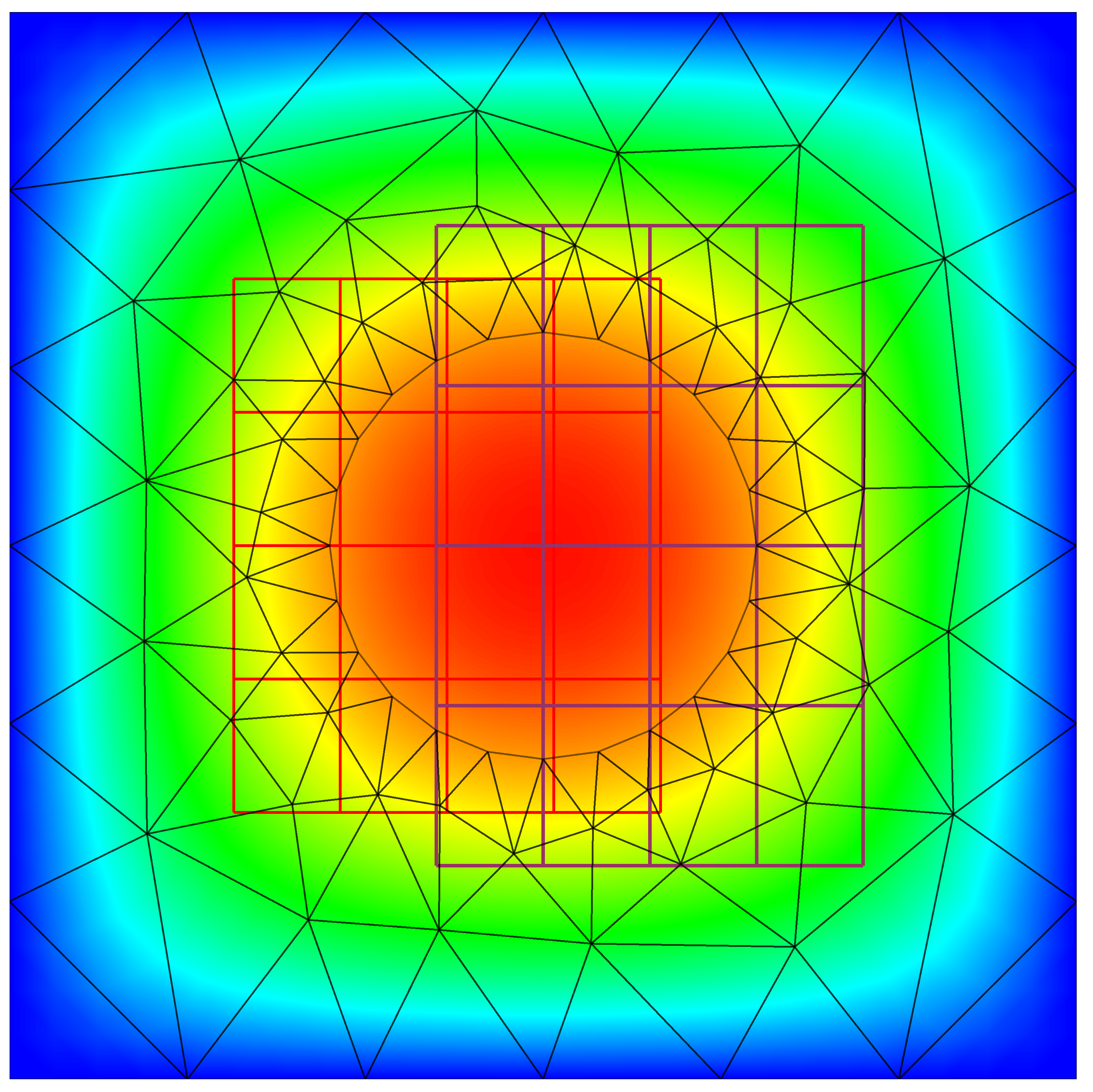} &
  \includegraphics[width=0.1\linewidth]{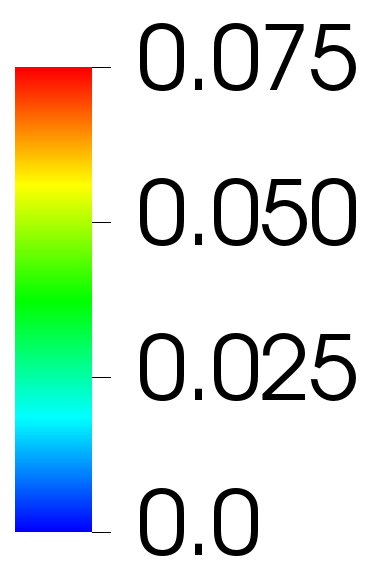}
  \end{array}$
  \end{center}
  \vspace{-4mm}
  \caption{(left) Domain $\Omega \in [0,1]^2$ modeled using three overlapping meshes. (right) The solution to the Poisson problem $-\nabla^2 u = 1$ is obtained on the overlapping grids using a predictor-corrector scheme following \cite{mittal2019highly}.}
\label{fig_schwarz}
\end{figure}

Overlapping grid based techniques have proven to be effective at circumventing the hex mesh generation problem in complex geometries by representing the domain $\Omega$ as a union of $S$ overlapping subdomains $\Omega = \bigcup_{s=1}^S \Omega_s$, $s=1\dots S$, each of which are relatively easier to mesh independently.
In \cite{mittal2019nonconforming}, we introduced a framework for solving the incompressible Navier-Stokes equations (INSE) in an arbitrary number of overlapping grids. Therein, we described a predictor-corrector scheme to solve the INSE on each subdomain independently, with the boundary condition for nodes discretizing the interdomain boundaries ($\partial \Omega_{s,I} := \partial \Omega_s \subset \bigcup_{u=1,u\neq s}^S\Omega_u$) interpolated from the overlapping grid prior to each solver iteration. In cases where an interdomain boundary is overlapped by multiple subdomains, the boundary condition is interpolated from the subdomain that minimizes the error at the boundary (Section 2.3 of \cite{mittal2019nonconforming}).
This interdomain boundary data exchange is enabled by the arbitrary point interpolation capability described in the current work, and has been used in simulation of complex flow problems such as heat transfer enhancement in a pipe with wire-coil insert \cite{mittal2019nonconforming}, thermally buoyant plumes \cite{mittal2021multirate}, resolved flow around rotating ellipsoidal particle \cite{mittal2020}, and turbulent flow over an array of shark denticles \cite{lloyd2023multi}.
Figure \ref{fig_schwarz} shows a simple example of three overlapping grids used to solve the Poisson problem (-$\nabla^2 u=1$ with $u|_{\partial \Omega}=0$) using a predictor-corrector scheme; see section 1.2 of \cite{mittal2019highly} for more details.

\subsection{Lagrangian particle tracking} \label{sec_results_lagrangian}
Lagrangian particle tracking is inherent to flow visualization \cite{dutta2018visualization}, and modeling transport of particles \cite{dutta2016large,dutta2017bulle} and aerosols \cite{fabregat2021direct} in turbulent flows using the point-particle approach. Within the Nek5000 solver, the one-way \cite{dutta2017bulle} and two-way \cite{zwick2020scalable} coupled Lagrangian particle tracking had been implemented for efficient-parallel computations on CPUs using the CPU implementation of \textit{findpts}. The GPU adaptation of \textit{findpts} has been utilized to facilitate Lagrangian particle tracking calculations.
Algorithm \ref{algo:lpm} illustrates its implementation for computing one-way coupled transport of particles.

\begin{algorithm}
        \caption{Lagrangian Particle Tracking}
        \begin{algorithmic}[1]
                \For {$time\text{-}step=N_{start},\ldots,N_{steps}$}
            \State Solve the fluid flow: FluidSolve
            \State Evaluate Eulerian fluid velocity at particle locations: \emph{Interpolate}
            \State Evaluate the aerodynamic forces on each particle: ParticleRHS
            \State Integrate the particles in time: Integrate
            \State Find rank, element and reference coords. of the particles: \emph{Find}
            \State Enforce boundary conditions on particles: ParticleBC
            \If {\# (rank non-local particles) fraction$>0.1$}
              \State Migrate particles to their overlapping ranks: Migrate
            \EndIf
                \EndFor
        \end{algorithmic}
    \label{algo:lpm}
\end{algorithm}

\begin{figure}[bt!]
  \begin{center}
  $\begin{array}{cc}
  \includegraphics[width=0.56\linewidth]{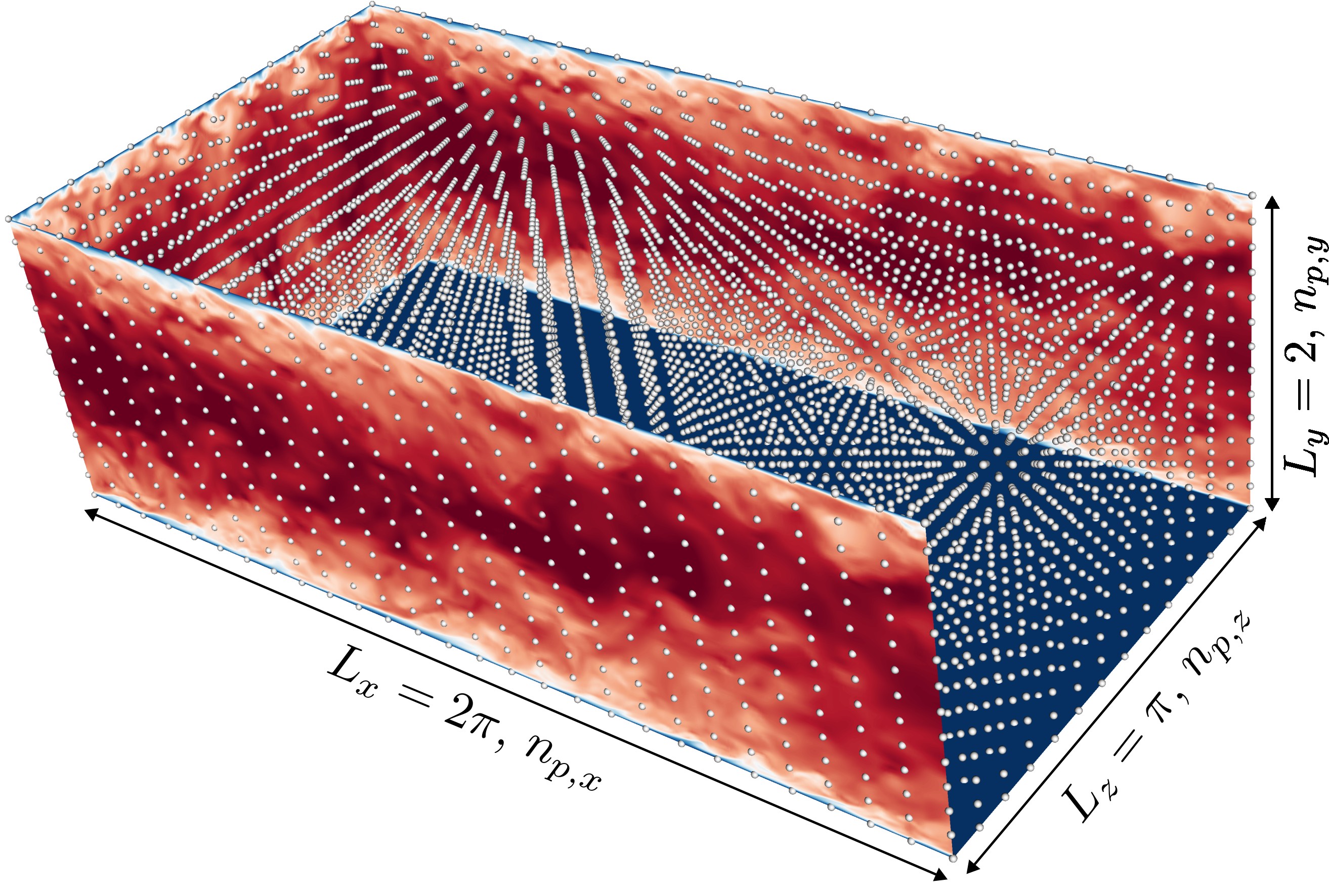} &
  \includegraphics[width=0.38\linewidth]{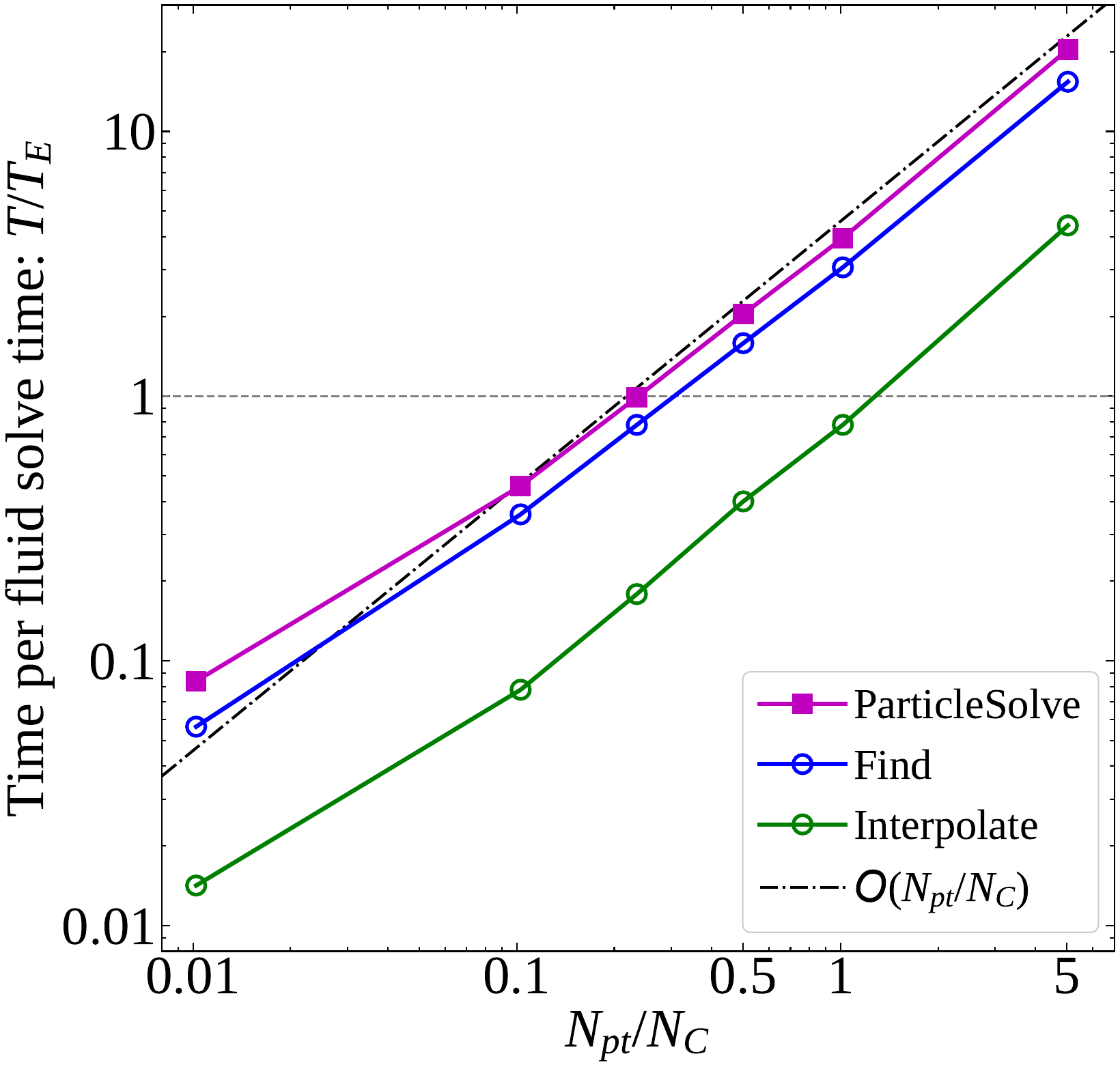}
  \end{array}$
  \end{center}
  \vspace{-4mm}
  \caption{(left) Flow and particle setup for the test case. (right)
  Solve times for the most expensive parts of particle calculation (\emph{Find} and \emph{Interpolate} steps) and total particle solve, normalized by time to solve the flow ($T /T_E$) vs. ratio of Lagrangian particles to Eulerian computational points ($N_{pt} / N_C$).}
\label{fig_lagrangian}
\end{figure}

To demonstrate the utility of the GPU implementation of \fpt\ in facilitating Lagrangian particle tracking and to demonstrate the relative speed of different parts of the algorithm, we simulate the transport of one-way coupled inertial particles by turbulent flow in a channel at $\mathit{Re}_{bulk} = 10,000$ using the gpu-optimized high-order SEM-based incompressible Navier-Stokes solver NekRS \cite{nekrs-turb}.
The turbulent channel's computational domain and the initial distribution of the particles is shown in Figure \ref{fig_lagrangian}(left).
The computational domain comprises of $N_E = 31104$ elements with a polynomial order $p = 7$, resulting in $N_C = 15.9 ~ million$ computational points.
Periodic boundary conditions is enforced in the $x$ and $z$ directions, and no-slip is enforced at the walls in the $y$ direction.
A constant mean velocity of $1$ is enforced in the x direction, and the flow without particles is simulated till $t=300.0$ time-units ($\sim 47$ flow-through) to ensure statistically steady state.
A timestep $\Delta t = 10^{-3}$ time-units ensures that CFL number remains below $0.5$ for the simulation.

The flow is restarted at time $t=300.0$, with a total of $N_{pt} = n_{pt,x}\times n_{pt,y}\times n_{pt,z}$ particles that are uniformly distributed in each dimension, according to the values of $n_{pt,x}$, $n_{pt,y}$, and $n_{pt,z}$, which are assumed to follow $n_{pt,x}: n_{pt,y}: n_{pt,z} = 2:1:1$.
If the number of particles per computational point $\tilde N_{pt} = N_{pt}/N_C$ is specified, the corresponding $n_{pt,y} = n_{pt,z}$ is the rounded up value of $(0.5\, \tilde N_{pt}\, N_C )^{1/3}$.
We assume that the particle Stokes number $\tau = 5$.
For particle velocity $\mathbf{v}$, and fluid velocity interpolated at the particle location $\mathbf{u}$, the acceleration observed by the particles under Stokes drag formulation can be described as $\dot{\mathbf{v}} = (\mathbf{u}-\mathbf{v})/\tau$.
Adam-Bashforth time-stepping scheme of the same order as the fluid flow ($2$\textsuperscript{nd} order in this case) is used to integrate the particle in time.

All test simulations are performed on $2$ nodes of Frontier supercomputer, each of which consists of a 64-core AMD “Optimized 3rd Gen EPYC” CPU and $4$ AMD MI250X GPUs. Each GPU has 2 Graphics Compute Dies (GCDs) that makes a total of 8 GCDs per node, which is equivalent of having 8 independednt GPUs per noder. Each GCD effectively has access to 64 GB of high-bandwidth memory (HBM2E).

\begin{table}[bt!]
\centering
\renewcommand{\arraystretch}{1.5} 
\resizebox{\textwidth}{!}{%
\begin{tabular}{|ll|cccccc|}
\hline
$\tilde N_{pt}= N_{pt}/N_C$ &                & 0.01   & 0.10   & 0.23   & 0.50   & 1.02   & 5.03   \\ \hline
FluidSolve $T_E$        & (sec)          & 160.90 & 166.74 & 163.49 & 166.34 & 172.05 & 168.79 \\
ParticleSolve           & ($\times T_E$) & 0.08   & 0.46   & 0.99   & 2.04   & 3.96   & 20.39  \\
-- \emph{Interpolate}   & ($\times T_E$) & 0.01   & 0.08   & 0.18   & 0.40   & 0.78   & 4.41   \\
-- ParticleRHS          & ($\times T_E$) & 0.00   & 0.00   & 0.00   & 0.00   & 0.00   & 0.01   \\
-- Integrate            & ($\times T_E$) & 0.01   & 0.01   & 0.01   & 0.01   & 0.01   & 0.03   \\
-- \emph{Find}          & ($\times T_E$) & 0.06   & 0.36   & 0.78   & 1.59   & 3.07   & 15.43  \\
-- -- \emph{FindKernel} & ($\times T_E$) & 0.04   & 0.29   & 0.63   & 1.29   & 2.49   & 12.29  \\
-- ParticleBC           & ($\times T_E$) & 0.001  & 0.001  & 0.001  & 0.001  & 0.002  & 0.004  \\
-- Migrate              & ($\times T_E$) & 0.00   & 0.01   & 0.02   & 0.05   & 0.10   & 0.51   \\ \hline
\end{tabular}
}
\caption{Breakdown of the time required to perform different parts of the Lagrangian particle tracking computation, as described in algorithm \ref{algo:lpm}. Time to solve the fluid flow ($T_E$) is given in seconds. All other times are scaled in terms of $T_E$. }
\label{table_lpm}
\end{table}

The time spent in \emph{Find} step, \emph{Interpolate} step, and total particle solve (all three non-dimensionalized with the time required to solve the fluid flow), from $t=310.0$ to $315.0$ units ($5000$ timesteps) is shown in Figure \ref{fig_lagrangian} (right).
A more detailed breakdown of the relative time required to perform all the different parts of the algorithm (\ref{algo:lpm}) is listed in table \ref{table_lpm}.
It is found that a Lagrangian particles to Eulerian computational points ratio, $N_{pt}/N_C$, of approximately $0.23$ results in equal time to solve particles and the flow.
We also note that almost the entirety of the computational costs involved in particle solve is spent in the \emph{Find} ($\sim 75\%$) and \emph{Interpolate} ($20 \%$) steps. All other particle local calculations such as the evaluations of the forcing terms, time integration, and boundary condition checks are relatively inexpensive and account for only about $5 \%$ of the total particle solve.
Thus, the scaling of particle solve costs with respect to number of particles is almost entirely dictated by the scaling characteristics of the general field evaluation algorithm, which is $\bigO(N_{pt})$ as illustrated in Figure \ref{fig_lagrangian} (right).

\section{Summary and Future Work} \label{sec_conclusion}

We have presented a robust and efficient technique for general field evaluation in large-scale high-order curvilinear meshes with quadrilaterals and hexahedra. The key components of this framework are (i) a combination of globally partitioned and processor-local maps to determine candidate elements overlapping a given point, (ii) element-wise bounding boxes to quickly determine if an element overlaps a point, (iii) Newton’s method with trust region to invert element mapping and determine the reference space coordinates corresponding to a given physical location, and (iv) specialized kernels to effect the proposed methodology on GPUs. The effectiveness of the method is demonstrated using various numerical experiments featuring large scale curvilinear meshes and various applications such as Lagrangian particle tracking and tangential relaxation.

In future work, we will improve the method proposed for bounding boxes in Section \ref{sec_method_precomp_aabb} to construct theoretically provable bounds for general high-order FEM bases. We will also explore ways to construct even tighter oriented bounding boxes compared to the existing approach in Section \ref{sec_method_precomp_obb}. Finally, we will develop techniques to construct more efficient maps for determining candidate elements that overlap a given point on a  surface mesh, as the current approach can lead to suboptimal maps (Section \ref{sec_method_surf_maps}).

\noindent \section*{Acknowledgments}
\noindent The authors would like to thank Yohann Dudouit, Malachi Phillips, Thilina Rathnayake and Yu-Hsiang Lan for the helpful discussions and input throughout this work.
S.D and P.F's involvement was supported through U.S. Department of Energy's Center for Efficient Exascale Discretization (CEED).
This work was performed under the auspices of the U.S. Department of Energy by Lawrence Livermore National Laboratory under Contract DE-AC52-07NA27344. LLNL-JRNL-872076.

\noindent \section*{In memoriam}

\noindent This paper is dedicated to the memory of Prof. Arturo Hidalgo L\'opez
($^*$July 03\textsuperscript{rd} 1966 - $\dagger$August 26\textsuperscript{th} 2024) of the Universidad Politecnica de Madrid,
organizer of HONOM 2019 and active participant in many other editions of HONOM. Our thoughts and wishes go to his wife Lourdes and his sister Mar\'ia Jes\'us, whom he left behind.

\bibliographystyle{elsarticle-num}
\bibliography{findpts}

\newcommand{\noopsort}[1]{} \newcommand{\printfirst}[2]{#1}
  \newcommand{\singleletter}[1]{#1} \newcommand{\switchargs}[2]{#2#1}
\begin{thebibliography}{10}
\expandafter\ifx\csname url\endcsname\relax
  \def\url#1{\texttt{#1}}\fi
\expandafter\ifx\csname urlprefix\endcsname\relax\def\urlprefix{URL }\fi
\expandafter\ifx\csname href\endcsname\relax
  \def\href#1#2{#2} \def\path#1{#1}\fi

\bibitem{plimpton2004parallel}
S.~J. Plimpton, B.~Hendrickson, J.~R. Stewart, A parallel rendezvous algorithm
  for interpolation between multiple grids, Journal of Parallel and Distributed
  Computing 64~(2) (2004) 266--276.

\bibitem{herring2021portage}
A.~M. Herring, C.~R. Ferenbaugh, C.~M. Malone, D.~W. Shevitz, E.~Kikinzon,
  G.~A. Dilts, H.~N. Rakotoarivelo, J.~Velechovsky, K.~Lipnikov, N.~Ray,
  et~al., {Portage: A modular data remap library for multiphysics applications
  on advanced architectures}, Journal of Open Research Software
  9~(LA-UR-20-24654) (2021).

\bibitem{ray2023efficient}
N.~Ray, D.~Shevitz, Y.~Li, R.~Garimella, A.~Herring, E.~Kikinzon, K.~Lipnikov,
  H.~Rakotoarivelo, J.~Velechovsky, Efficient kd-tree based mesh redistribution
  for data remapping algorithms, in: International Meshing Roundtable,
  Springer, 2023, pp. 25--41.

\bibitem{slattery2013data}
S.~Slattery, P.~Wilson, R.~Pawlowski, The data transfer kit: A geometric
  rendezvous-based tool for multiphysics data transfer, in: International
  conference on mathematics \& computational methods applied to nuclear science
  \& engineering (M\&C 2013), 2013, pp. 5--9.

\bibitem{mittal2019nonconforming}
K.~Mittal, S.~Dutta, P.~Fischer, Nonconforming schwarz-spectral element methods
  for incompressible flow, Computers \& Fluids 191 (2019) 104237.

\bibitem{Roca2018}
G.~Aparicio-Estrems, A.~Gargallo-Peir{\'o}, X.~Roca, Defining a stretching and
  alignment aware quality measure for linear and curved 2D meshes, Springer
  International Publishing, 2019, pp. 37--55.

\bibitem{lipnikov2023conservative}
K.~Lipnikov, M.~Shashkov, Conservative high-order data transfer method on
  generalized polygonal meshes, Journal of Computational Physics 474 (2023)
  111822.

\bibitem{lacroix2024comparative}
M.~Lacroix, S.~F{\'e}vrier, E.~Fern{\'a}ndez, L.~Papeleux, R.~Boman, J.-P.
  Ponthot, A comparative study of interpolation algorithms on non-matching
  meshes for pfem-fem fluid-structure interactions, Computers \& Mathematics
  with Applications 155 (2024) 51--65.

\bibitem{chandar2019overset}
D.~D. Chandar, On overset interpolation strategies and conservation on
  unstructured grids in openfoam, Computer Physics Communications 239 (2019)
  72--83.

\bibitem{gslib-github}
{GSLIB}, \url{https://https://github.com/Nek5000/gslib}.

\bibitem{dutta2016large}
S.~Dutta, P.~Fischer, M.~H. Garcia, Large eddy simulation (les) of flow and
  bedload transport at an idealized 90-degree diversion: Insight into
  bulle-effect, in: Proc., Int. Conf. on Fluvial Hydraulics, CRC, Boca Raton,
  FL, 2016, pp. 101--109.

\bibitem{zwick2020scalable}
D.~Zwick, S.~Balachandar, A scalable euler--lagrange approach for multiphase
  flow simulation on spectral elements, The International Journal of High
  Performance Computing Applications 34~(3) (2020) 316--339.

\bibitem{fabregat2021direct}
A.~Fabregat, F.~Gisbert, A.~Vernet, J.~A. Ferr{\'e}, K.~Mittal, S.~Dutta,
  J.~Pallar{\`e}s, Direct numerical simulation of turbulent dispersion of
  evaporative aerosol clouds produced by an intense expiratory event, Physics
  of Fluids 33~(3) (2021).

\bibitem{yang2021scalable}
Y.~Yang, S.~Balachandar, A scalable parallel algorithm for direct-forcing
  immersed boundary method for multiphase flow simulation on spectral elements,
  The Journal of Supercomputing 77 (2021) 2897--2927.

\bibitem{min2024exascale}
M.~Min, Y.-H. Lan, P.~Fischer, E.~Merzari, T.~Nguyen, H.~Yuan, P.~Shriwise,
  S.~Kerkemeier, A.~Davis, A.~Dubas, et~al., Exascale simulations of fusion and
  fission systems, arXiv preprint arXiv:2409.19119 (2024).

\bibitem{dutta2018visualization}
S.~Dutta, M.~W.~V. Moer, P.~Fischer, M.~H. Garcia, Visualization of the
  bulle-effect at river bifurcations, in: Proceedings of the practice and
  experience on advanced research computing, 2018, pp. 1--4.

\bibitem{mittal2020}
K.~Mittal, S.~Dutta, P.~Fischer, Direct numerical simulation of rotating
  ellipsoidal particles using moving nonconforming schwarz-spectral element
  method, Computers \& Fluids 205 (2020) 104556.

\bibitem{mfem-github}
{MFEM}, \url{https://github.com/mfem/mfem/}.

\bibitem{MFEM2024}
J.~Andrej, N.~Atallah, J.-P. B{\"a}cker, J.~Camier, D.~Copeland, V.~Dobrev,
  Y.~Dudouit, T.~Duswald, B.~Keith, D.~Kim, et~al., {High-performance finite
  elements with MFEM}, arXiv preprint arXiv:2402.15940 (2024).

\bibitem{lindquist2021scalable}
N.~Lindquist, P.~Fischer, M.~Min, Scalable interpolation on gpus for thermal
  fluids applications, Tech. rep., Argonne National Lab.(ANL), Argonne, IL
  (United States) (2021).

\bibitem{fischer2022nekrs}
P.~Fischer, S.~Kerkemeier, M.~Min, Y.-H. Lan, M.~Phillips, T.~Rathnayake,
  E.~Merzari, A.~Tomboulides, A.~Karakus, N.~Chalmers, et~al., Nekrs, a
  gpu-accelerated spectral element navier--stokes solver, Parallel Computing
  114 (2022) 102982.

\bibitem{nekrs-github}
{NekRS}, \url{https://https://github.com/Nek5000/nekrs}.

\bibitem{MFEM2021}
R.~Anderson, J.~Andrej, A.~Barker, J.~Bramwell, J.-S. Camier, J.~Cerveny, V.~A.
  Dobrev, Y.~Dudouit, A.~Fisher, T.~V. Kolev, W.~Pazner, M.~Stowell, V.~Z.
  Tomov, I.~Akkerman, J.~Dahm, D.~Medina, S.~Zampini, {MFEM}: a modular finite
  elements methods library, Comput. Math. Appl. 81 (2021) 42--74.
\newblock \href {https://doi.org/10.1016/j.camwa.2020.06.009}
  {\path{doi:10.1016/j.camwa.2020.06.009}}.

\bibitem{deville2002high}
M.~O. Deville, P.~F. Fischer, E.~H. Mund, High-order methods for incompressible
  fluid flow, Vol.~9, Cambridge university press, 2002.

\bibitem{novak2012rasterized}
J.~Nov{\'a}k, C.~Dachsbacher, Rasterized bounding volume hierarchies, in:
  Computer Graphics Forum, Vol.~31, Wiley Online Library, 2012, pp. 403--412.

\bibitem{TMOP2020}
V.~A. Dobrev, P.~Knupp, T.~V. Kolev, K.~Mittal, R.~N. Rieben, V.~Z. Tomov,
  Simulation-driven optimization of high-order meshes in {ALE} hydrodynamics,
  Comput. Fluids (2020).

\bibitem{TMOP2019}
V.~A. Dobrev, P.~Knupp, T.~V. Kolev, K.~Mittal, V.~Z. Tomov, The
  {T}arget-{M}atrix {O}ptimization {P}aradigm for high-order meshes, SIAM J.
  Sci. Comp. 41~(1) (2019) B50--B68.

\bibitem{mittal2019mesh}
K.~Mittal, P.~Fischer, Mesh smoothing for the spectral element method, Journal
  of Scientific Computing 78~(2) (2019) 1152--1173.

\bibitem{Dobrev2012}
V.~Dobrev, T.~Kolev, R.~Rieben, High-order curvilinear finite element methods
  for {L}agrangian hydrodynamics, SIAM J. Sci. Comp. 34~(5) (2012) 606--641.

\bibitem{mittal2021multirate}
K.~Mittal, S.~Dutta, P.~Fischer, {Multirate timestepping for the incompressible
  Navier-Stokes equations in overlapping grids}, Journal of Computational
  Physics 437 (2021) 110335.

\bibitem{lloyd2023multi}
C.~Lloyd, K.~Mittal, S.~Dutta, R.~Dorrell, J.~Peakall, G.~Keevil, A.~Burns,
  Multi-fidelity modelling of shark skin denticle flows: insights into drag
  generation mechanisms, Royal Society Open Science 10~(2) (2023) 220684.

\bibitem{mittal2019highly}
K.~Mittal, Highly scalable solution of incompressible {Navier-Stokes} equations
  using the spectral element method with overlapping grids, Ph.D. thesis,
  University of Illinois at Urbana-Champaign (2019).

\bibitem{dutta2017bulle}
S.~Dutta, Bulle-effect and its implications for morphodynamics of river
  diversions, Ph.D. thesis, University of Illinois at Urbana-Champaign (2017).

\bibitem{nekrs-turb}
{NekRS Example: Wall-resolved LES of turbulent channel flow},
  \url{https://github.com/Nek5000/nekRS/tree/master/examples/turbChannel}.

\end{thebibliography}

\end{document}